\documentclass[12pt]{article}
\usepackage{jheppub}
\pdfoutput=1
\usepackage{epstopdf}

\usepackage{mathrsfs}
\usepackage{psfrag}
\usepackage{color}
\usepackage{hyperref}

\usepackage{slashed}
\usepackage{feynmp-auto}
\usepackage{simplewick}
\usepackage{cancel}

\usepackage[normalem]{ulem}

\usepackage[table]{xcolor}

\usepackage{amsmath,bbm,array,amsfonts,graphicx,wrapfig,arydshln,lscape,float,multirow,longtable,rotating,makecell}
\usepackage{url}

\newcommand{\ab}[1]{\langle #1 \rangle}

\newcommand{\sab}[1]{s_{#1}}

\def\lra{\leftrightarrow}
\def\dlog{d\log}

\def\C{\mathcal{C}}
\def\D{\mathcal{D}}
\def\O{\mathcal{O}}
\def\Sy{\mathcal{S}}
\def\N{\mathcal{N}}

\def\I{\mathcal{I}}

\def\Li{\text{Li}}

\newcommand{\tw}[1]{\widetilde{#1}}

\def\sm{\text{-}}
\def\sp{\!\text{+}\!}

\definecolor{airforceblue}{rgb}{0.36, 0.54, 0.66}
\definecolor{bananayellow}{rgb}{1.0, 0.88, 0.21}
\definecolor{bittersweet}{rgb}{1.0, 0.44, 0.37}
\definecolor{blue(ncs)}{rgb}{0.0, 0.53, 0.74}
\definecolor{bole}{rgb}{0.47, 0.27, 0.23}
\definecolor{brass}{rgb}{0.71, 0.65, 0.26}
\definecolor{bronze}{rgb}{0.8, 0.5, 0.2}
\definecolor{brgreen}{rgb}{0.0, 0.26, 0.15}
\definecolor{burgundy}{rgb}{0.5, 0.0, 0.13}
\definecolor{cherry}{rgb}{1.0, 0.72, 0.77}
\definecolor{cocao}{rgb}{0.82, 0.41, 0.12}
\definecolor{citrine}{rgb}{0.99, 0.82, 0.07}

\newcommand{\twhite}[1]{\textcolor{white}{#1}}

\title{Bootstrapping two-loop Feynman integrals for planar $\mathbf{\N=4}$ sYM}

\author[1,2]{\!\!\!\,Johannes~Henn}
\affiliation[1]{PRISMA Cluster of Excellence, \\
Johannes Gutenberg University, 55128 Mainz, Germany}
\affiliation[2]{Max-Planck-Institute f\"ur Physik, \\
Werner-Heisenberg-Insitut, 80805 M\"unchen, Germany}
\emailAdd{henn@uni-mainz.de}

\author[3]{\!\!\!,\,Enrico~Herrmann}
\affiliation[3]{ SLAC National Accelerator Laboratory, Stanford University, Stanford, CA 94039, USA}
\emailAdd{eh10@stanford.edu}

\author[4]{\!\!\!,\,Julio~Parra-Martinez}
\affiliation[4]{Mani L. Bhaumik Institute for Theoretical Physics,\\
UCLA Department of Physics and Astronomy, Los Angeles, CA 90095, USA}
\emailAdd{jparra@physics.ucla.edu}

\preprint{MPP-2018-135, MITP/18-049}

\abstract{We derive analytic results for the symbol of certain two-loop Feynman integrals relevant for seven- and eight-point two-loop scattering amplitudes in planar $\N=4$ super-Yang--Mills theory. We use a bootstrap inspired strategy, combined with a set of second-order partial differential equations that provide powerful constraints on the symbol ansatz. When the complete symbol alphabet is not available, we adopt a hybrid approach. Instead of the full function, we bootstrap a certain discontinuity for which the alphabet is known. Then we write a one-fold dispersion integral to recover the complete result.
At six and seven points, we find that the individual Feynman integrals live in the same space of functions as the amplitude, which is described by the 9- and 42-letter cluster alphabets respectively. Starting at eight points however, the symbol alphabet of the MHV amplitude is insufficient for individual integrals. In particular, some of the integrals require algebraic letters involving four-mass box square-root singularities. We point out that these algebraic letters are relevant at the amplitude level directly starting with N$^2$MHV amplitudes even at one loop.}

\begin{document}
\setlength{\unitlength}{1mm}
\maketitle

\begin{fmffile}{2loop_draft}
%
%
%
\newpage
\section{Introduction}
%

The traditional path for computing scattering amplitudes follows a two-step process. One first constructs the Feynman \emph{integrand} either by summing Feynman diagrams, or in terms of generalized unitarity based \cite{Bern:1994zx,Bern:1994cg,Bern:2007ct} methods. In the second step, one carries out the space-time integrations to obtain the final answer.

In the present work we focus on the second step, i.e.~integrating to go from integrands to actual amplitudes. A number of powerful tools have been developed to tackle this task, including Mellin-Barnes integration \cite{Vermaseren:1998uu,Czakon:2005rk,Smirnov:2009up,Anastasiou:2013srw} as well as solving differential equations for the Feynman integrals under consideration \cite{Henn:2014qga,Kotikov:1990kg,Remiddi:1997ny,Gehrmann:1999as,Argeri:2007up,Henn:2013pwa}.  However, these techniques often meet practical limitations, especially when a large number of particles is involved.

As has been tradition in the field of scattering amplitudes, the natural place to sharpen one's toolbox is the realm of planar $\N=4$ sYM theory. To give one example, recent progress in understanding the link between Feynman integrands, multiple polylogarithms \cite{Goncharov:2001iea,Brown:2009ta,Panzer:2014gra,Brown:2015qmm} appearing in the integrated answer, and canonical forms of differential equations \cite{ArkaniHamed:2010gh,Henn:2013pwa}, originated in this theory, but is by now widely applied to  generic quantum field theories.

Planar $\N=4$ sYM theory shows a number of simplifying features; one of the most remarkable is the existence of a hidden dual conformal symmetry (DCI) \cite{Drummond:2006rz,Alday:2007hr,Drummond:2008vq} which effectively reduces the number of kinematic variables a given amplitude can depend on. One further simplification of $\N=4$ sYM is the simplicity of allowed rational prefactors; for planar MHV amplitudes, the only allowed rational prefactor in amplitudes is the Parke-Taylor tree (super-)amplitude \cite{Arkani-Hamed:2014bca}. In light of this discussion it is not so surprising that also the corresponding integrals relevant for these amplitudes have similar special properties.  

Another simplifying feature is the existence of a particular $4L$-$\dlog$-form of the four-dimensional
\emph{integrand} \cite{ArkaniHamed:2012nw}, which can be expressed in terms of on-shell diagrams, the positive Grassmannian \cite{ArkaniHamed:2012nw}, and the amplituhedron framework \cite{Arkani-Hamed:2013jha,Arkani-Hamed:2013kca}\footnotemark. At the level of individual integrals, the $\dlog$-representation makes manifest that all leading singularities~\cite{Britto:2004nc,Cachazo:2008vp} are $\{\pm1,0\}$, and conjecturally these \emph{unit leading singularity} integrals evaluate to \emph{pure} weight $2L$ polylogarithmic functions without any rational prefactors. 
\footnotetext{The initial examples satisfying this conjecture were massless, planar, finite, dual conformal integrals. It has since been generalized to more generic integrals within dimensional regularization \cite{Henn:2013pwa}. 
A number of interesting features at the integrand level has also been observed beyond the planar limit of $\N=4$ sYM theory \cite{Bern:2014kca,Bern:2015ple} as well as in $\N=8$ supergravity \cite{Herrmann:2016qea}. For a recent application to non-planar integrated amplitudes at the three-loop level, see \cite{Henn:2016jdu}.
}

Even with the simplifying features of planar $\N=4$ sYM at play, the integration of the relevant Feynman integrals still poses a considerable challenge. In order to circumvent direct integration, an alternative approach to calculating scattering amplitudes is being pursued with considerable success to very high loop order \cite{Dixon:2011nj,Dixon:2011pw,Dixon:2013eka,Golden:2014pua,Dixon:2014iba,Drummond:2014ffa,Dixon:2015iva,Caron-Huot:2016owq,Dixon:2016nkn} for six- and seven-particle amplitudes. This alternative framework has its philosophical roots in the bootstrap program of the 1960s \cite{Eden:1966dnq}, which aims at fixing scattering amplitudes from the knowledge of a few basic physical principles alone. Such a strategy resurfaced in the 1990s \cite{Bern:1994zx}, and in modern times it has been successfully applied at the level of perturbation theory in planar $\N=4$ sYM. 

In the amplitude bootstrap program, at step zero, one starts from an initial guess for the relevant analytic structure of a given amplitude and the corresponding function space. For multiple polylogarithms, this essentially amounts to knowing their singularities, which are encoded in the so-called symbol alphabet \cite{Goncharov:2010jf}. Often, this initial guess is first inspired by explicit lower loop calculations. In the context of six- and seven- particle amplitudes, it has been further motivated by the (conjectured) relation to the mathematical structure of cluster varieties \cite{Fomin:2001aa,Fock:2003aa,Fock:2005aa,Golden:2013xva,Golden:2014xqa,Harrington:2015bdt}. One alternative avenue that can inform the guess for the required singularity structures of a given amplitude comes from the study of Landau singularities that has attracted some attention recently \cite{Dennen:2015bet,Dennen:2016mdk,Prlina:2017azl,Prlina:2017tvx}. 

In this paper we focus on planar two-loop integrals in $\N=4$ sYM. While the symbols of the MHV amplitudes are known \cite{CaronHuot:2011ky}, the loop integrals contributing to them have been evaluated explicitly only up to $n=6$ particles, see \cite{Dixon:2011nj}, and in part for $n=7,8$ \cite{Drummond:2017ssj,Bourjaily:2018aeq}. The present work is motivated by two open problems. First, beyond seven particles, it is currently not known what the appropriate function space is. Second, even in cases where the amplitude is known, it is desirable to learn how the individual loop integrals contributing to it can be evaluated. As we will see, this often requires a larger function space. Here we focus on the particular interesting seven- and eight-particle cases.

A few years ago, Drummond, Trnka and one of the authors showed that such loop-integrals satisfy certain second order differential equations \cite{Drummond:2010cz}. The inhomogeneous part of these equations consists of known one-loop integrals. Solving these equations directly is generally difficult (see \cite{Caron-Huot:2018dsv} for progress in this direction). However, when combined with the bootstrap approach, the differential equations are particularly powerful. This is the strategy we follow in this paper.

Since these differential equations play such a prominent role in our work, we briefly summarize their salient features in Appendix \ref{subsec:diff_mechanisms} before applying them to explicit seven- and eight-point examples. We make contact with the heptagon cluster bootstrap for planar scattering amplitudes in $\N=4$ sYM and probe whether the heptagon function space satisfying the Steinmann conditions \cite{Caron-Huot:2016owq,Dixon:2016nkn} is sufficient to express the two-loop seven-point double-pentagon integrals. We find that the seven-point MHV-integrals we consider here do in fact live in the heptagon Steinmann space and can be expressed in terms of cluster letters. A similar approach to some of the seven-point integrals has been followed by Drummond et al.~\cite{Drummond:2017ssj} where cluster adjacency properties have been studied. 

Proceeding to eight points, we find that the two-loop MHV-alphabet from~\cite{CaronHuot:2011ky} is insufficient to describe individual Feynman integrals that contribute in a local expansion of the MHV scattering amplitude. We discuss this in the light of the analysis of Landau equations. We also point out that the need for additional alphabet letters can be anticipated from the differential equations. The reason is that the latter involve a four-mass box function that itself contains letters that are not rational in momentum-twistor variables. 

At this point we postpone the question of finding the full eight-particle alphabet, and present a tool that allows to circumvent this complication. We observe that the complicated four-mass box terms disappear in a certain discontinuity of the eight-point integral. The discontinuity we take is similar to the one recently discussed at six points in \cite{Caron-Huot:2018dsv}. We then argue that the eight-point discontinuity function can be bootstrapped using a simple rational alphabet. Using complex analysis we then recover the full function from a dispersion integral before we comment on some properties of the result.

The paper is structured as follows. In section \ref{labelsection2loopMHV}, we review the two-loop integrand representation for MHV amplitudes and introduce the necessary notation. In section \ref{sec:seven_point_bootstrap}, we use the seven-point cluster alphabet to bootstrap the seven-point double pentagon integrals appearing in the MHV amplitudes. In section \ref{sec:eightpointintro}, we discuss the appearance of new alphabet letters at eight points, and bootstrap a dispersive formula for one eight-particle double pentagon integral. This constitutes our main result. In section \ref{sec:amplitudes_bootstrap_implication}, we briefly discuss some implications of our analysis for developing a future eight-point amplitudes bootstrap program. We conclude in section \ref{sec:conclusion}. Appendix \ref{subsec:diff_mechanisms} reviews the mechanisms behind the second-order differential equations of~\cite{Drummond:2010cz}, and Appendix \ref{sec:seven_and_eight_point_differential_euqations} contains the explicit equations for the seven- and eight-point integrals computed in this paper. Appendix \ref{subsubsec:integrating_symbols} discusses an algorithm to integrate the one-fold dispersive representation.

%
%
\section{\texorpdfstring{Two-loop MHV amplitude in $\N=4$ sYM}{Two-loop MHV amplitude in N=4 sYM}}
\label{labelsection2loopMHV}
%
%
The local representation of two-loop MHV scattering amplitudes in planar $\N=4$ sYM involves a particular class of Feynman integrals \cite{ArkaniHamed:2010gh},
\begin{align}
\label{fig:chiral_2_loop_integral_general}
\raisebox{-45pt}{
\includegraphics[scale=.5,trim={0cm 3.2cm 0cm 3.2cm},clip]{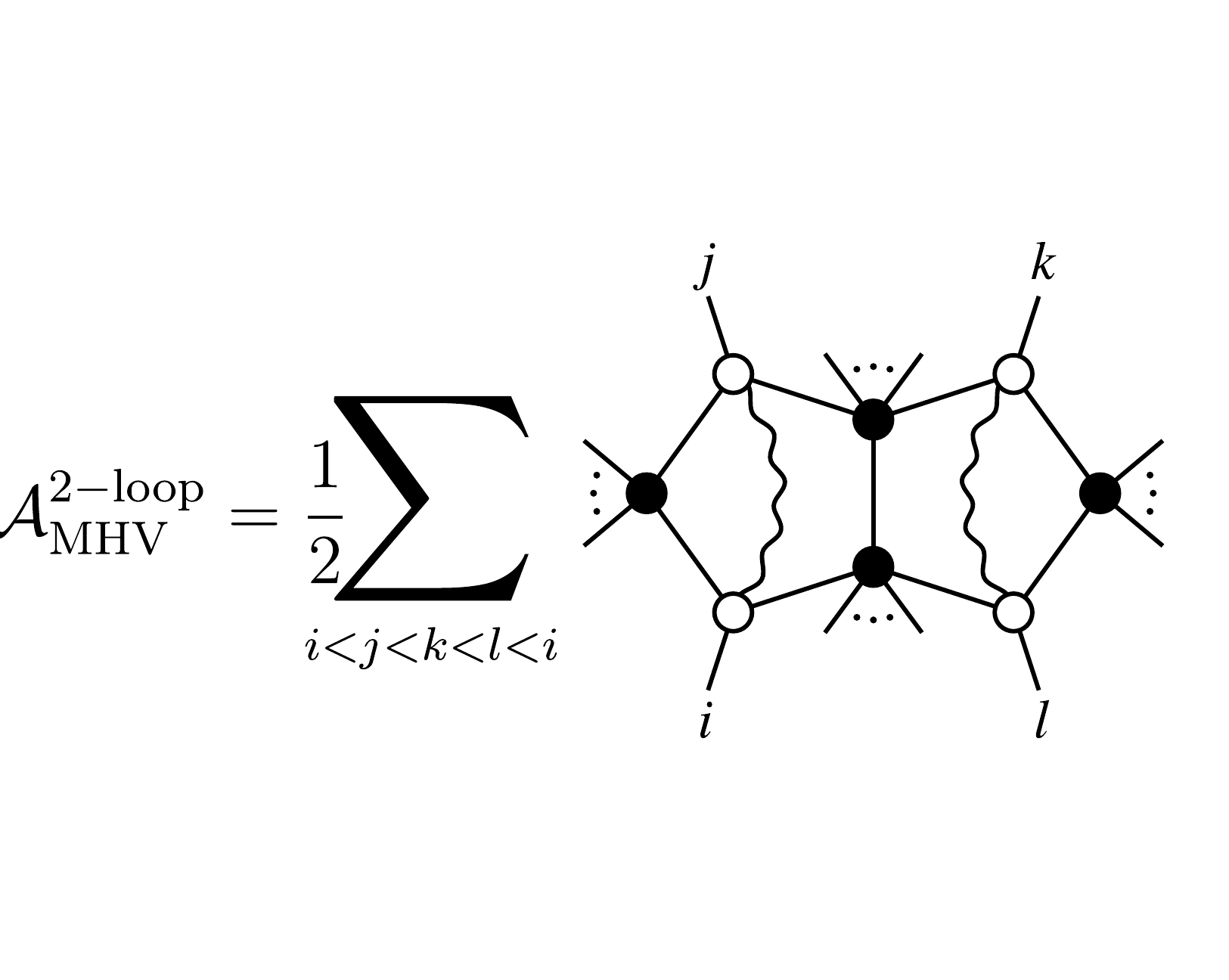}}\,,
\end{align}
where the wavy lines denote a special chiral numerator for the $n$-point integral,
\begin{align}
\label{eq:2loop_MHV_integral}
	 \iint\limits_{\ell_1,\ell_2} \!\!\!
\frac{\ab{\ell_1(i\sm 1ii\sp 1)\cap(j\sm 1jj\sp 1)}\ab{\ell_2(k\sm 1kk\sp 1)\cap(l\sm 1ll\sp 1)}\ab{ijkl}}
{\ab{\ell_1i\sm 1i}\ab{\ell_1ii\sp 1}\ab{\ell_1j\sm 1j}\ab{\ell_1jj\sp 1}\ab{\ell_1\ell_2}\ab{\ell_2k\sm 1k}\ab{\ell_2kk\sp 1}\ab{\ell_2l\sm ll}\ab{\ell_2ll\sp 1}}\,.
\end{align}
In momentum-twistor space~\cite{Hodges:2009hk} the integral is over lines $\ell_1\!=\!(AB)$ and $\ell_2\!=\!(CD)$. For further explanations on the integration measure as well as various twistor space conventions, see e.g.~\cite{ArkaniHamed:2010gh}. At the integrand level, the numerator enforces the vanishing on certain leading singularities \cite{Britto:2004nc,Cachazo:2008vp}, which renders the integrands chiral. One striking implication of their chiral nature is that the integrals are infrared (IR-) finite since all collinear- and soft-singular regions are canceled by the numerators\footnotemark.\footnotetext{For special values of the index range in (\ref{fig:chiral_2_loop_integral_general}), the degenerate configurations can have IR-singularities. For $j=i+1$ and the remaining indices generic, one such degeneration is the penta-box integral with massless legs on the left. This integral is indeed IR-singular.} With the help of residue theorems one can furthermore show that once the integral is normalized according to eq.~(\ref{eq:2loop_MHV_integral}), all leading singularities are $\{\pm1,0\}$. In this case, the integrals are commonly referred to as having \emph{unit leading singularity}. Conjecturally, these integrals evaluate to \emph{pure} functions without any rational prefactors. Taking into account the existence of a $\dlog$-form for the double-pentagon integrand \cite{ArkaniHamed:2012nw}, it is expected that all integrals (\ref{eq:2loop_MHV_integral}) evaluate to uniform-transcendental polylogarithmic functions of weight four.

Besides the chiral integrals denoted by wavy lines in (\ref{fig:chiral_2_loop_integral_general}), there are integrals with the opposite chirality numerators that will play a central role in the present work. These are represented by dashed lines 
\[	
  \Omega^{(2)}_{n}(i,j,k,l) = 
  \raisebox{-49pt}{\includegraphics[scale=.5]{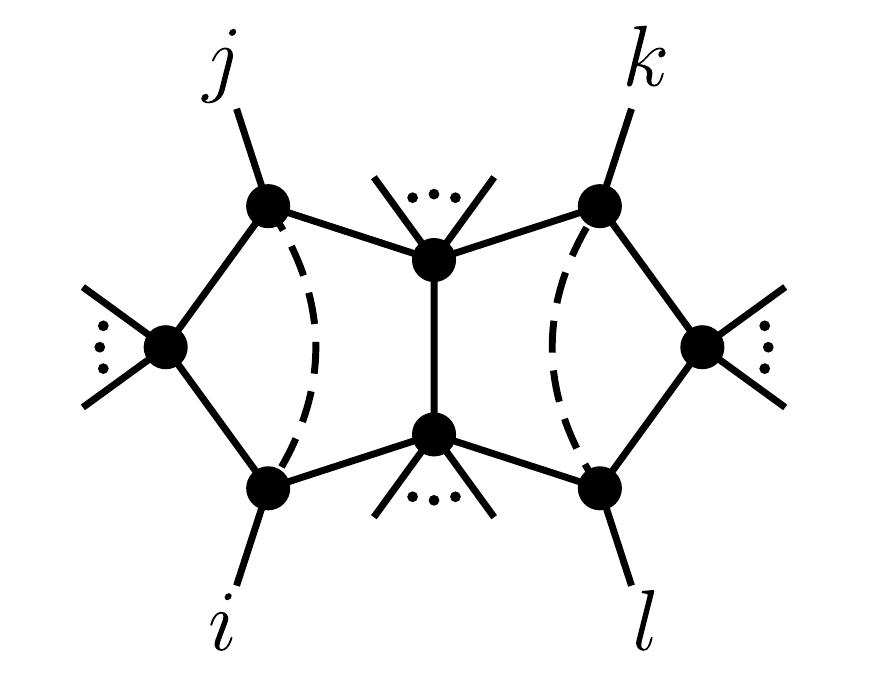}}\,,
\]
to indicate the following structure
\begin{align}
\label{eq:2loop_MHVbar_integral}
  \iint\limits_{\ell_1,\ell_2} \!\!\!
\frac{\ab{\ell_1ij}\ab{\ell_2kl}\ab{(i\sm 1ii\sp 1)\cap(j\sm 1jj\sp 1),(k\sm 1kk\sp 1)\cap(l\sm 1ll\sp 1)}}
{\ab{\ell_1i\sm 1i}\ab{\ell_1ii\sp 1}\ab{\ell_1j\sm 1j}\ab{\ell_1jj\sp 1}\ab{\ell_1\ell_2}\ab{\ell_2k\sm 1k}\ab{\ell_2kk\sp 1}\ab{\ell_2l\sm ll}\ab{\ell_2ll\sp 1}}\,.
\end{align}
At the integrand level, parity acts by interchanging wavy- and dashed numerators, and hence these are relevant for the $\overline{\text{MHV}}$-integrand. In the language of momentum-twistor space, parity corresponds to the duality between points $Z_i$ and planes $(Z_{i-1},Z_i,Z_{i+1})$. Under integration this effectively allows us to focus on the particular chirality that is most convenient in a given context, keeping in mind that we can always convert back to the opposite chirality if needed. We refer to the two-loop integrals with same-chirality numerators in both loops as $\Omega^{(2)}(u_i)$ which are pure weight four transcendental functions that only depend on dual conformal cross ratios, $u_i$. There are also the mixed chirality integrals, where one pentagon has a wavy and the other a dashed numerator. In the six-point case this double-pentagon integral, denoted as $\tw{\Omega}$, was evaluated in \cite{Dixon:2011nj}. Its solution is more complicated and involves both parity-even and parity-odd terms (parity of integrated results refers to spacetime parity acting on external data).  

A few years ago, Drummond, Trnka and one of the authors investigated a number of second order differential operators for these loop integrals \cite{Drummond:2010cz}. Schematically, the differential equations can be written as  
$$
D^{(2)} \quad  \raisebox{-37pt}{\includegraphics[scale=.4]{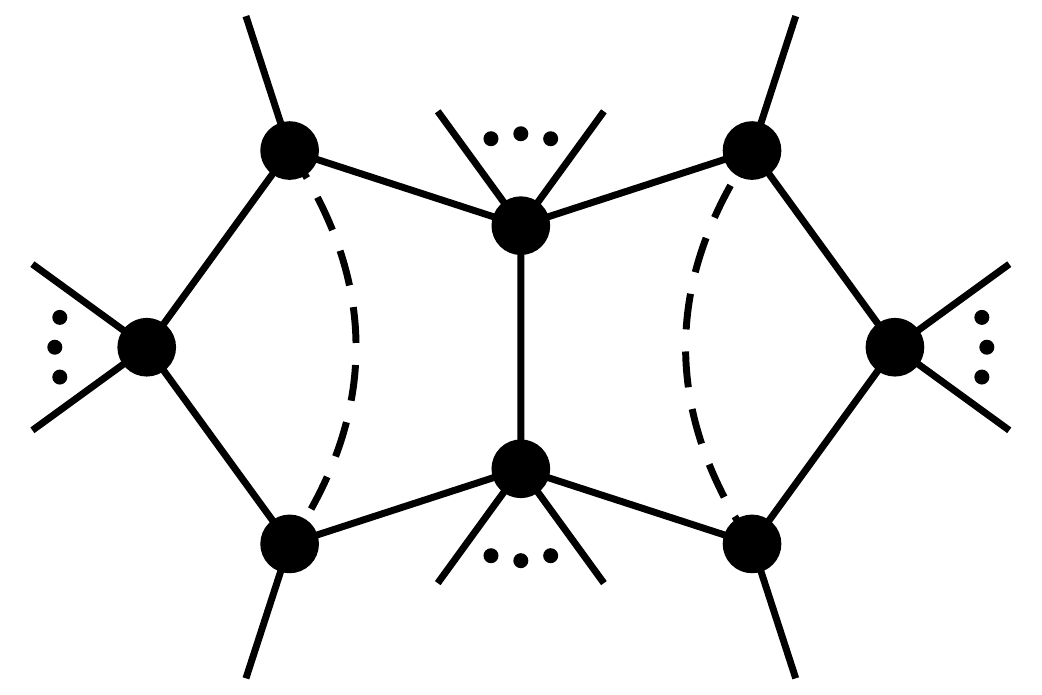}} \quad =  \quad  \raisebox{-38pt}{\includegraphics[scale=.4]{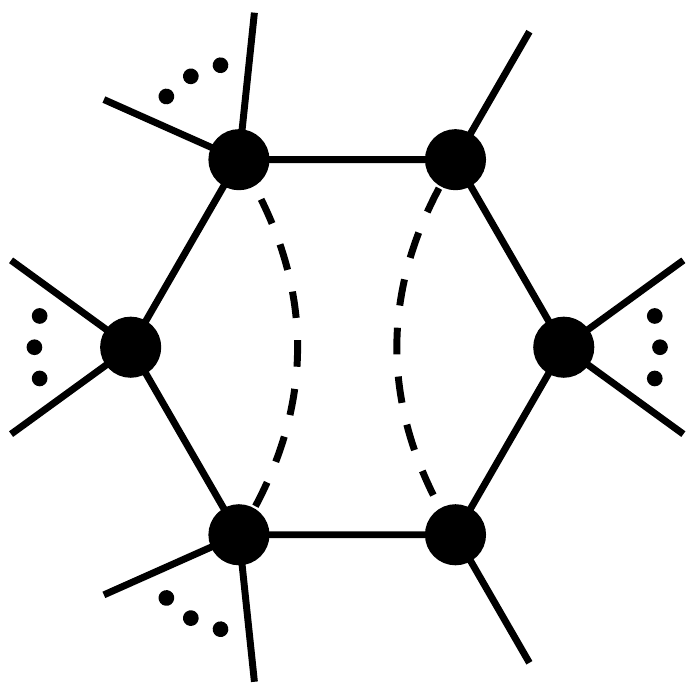}}\,,
$$
where the integral on the right hand side is a particular chiral hexagon.
We present details of the different mechanisms for obtaining differential equations in Appendix \ref{subsec:diff_mechanisms}, and then apply them to the seven- and eight-point MHV integrals in Appendix \ref{sec:seven_and_eight_point_differential_euqations}. These equations will be our main tool for generating constraints for the bootstrap approach which we pursue in the following sections.

%
%
\section{Seven-point integrals from Steinmann bootstrap}
\label{sec:seven_point_bootstrap}
%

There exist two seven-point generalizations of $\Omega^{(2)}_6$ which appear in MHV amplitudes
$$
\Omega^{(2)}_{7,a} = \raisebox{-47.5pt}{\includegraphics[scale=.5]{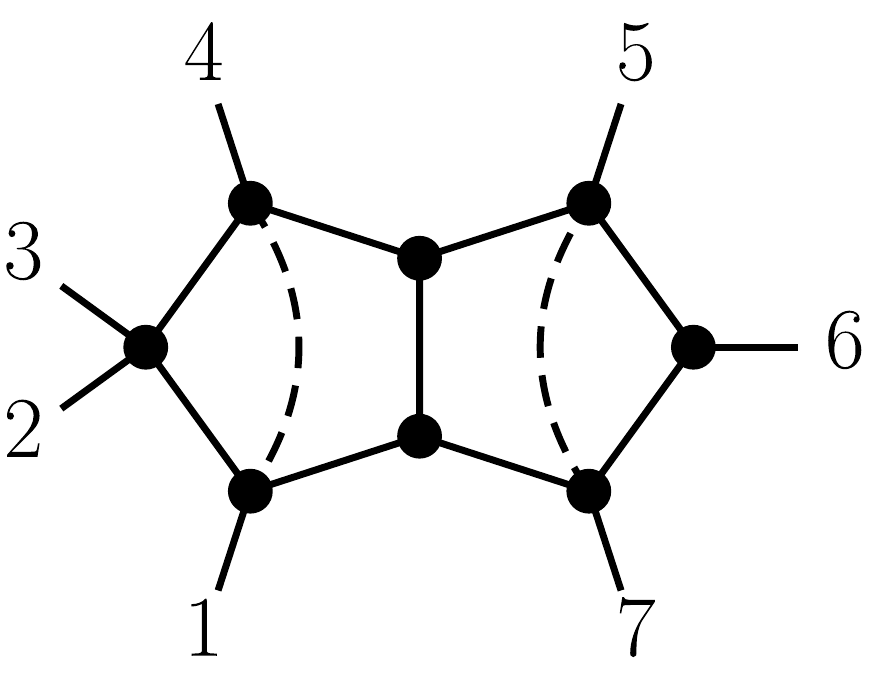}}\,,
\qquad
\Omega^{(2)}_{7,b} = \raisebox{-47.5pt}{\includegraphics[scale=.5]{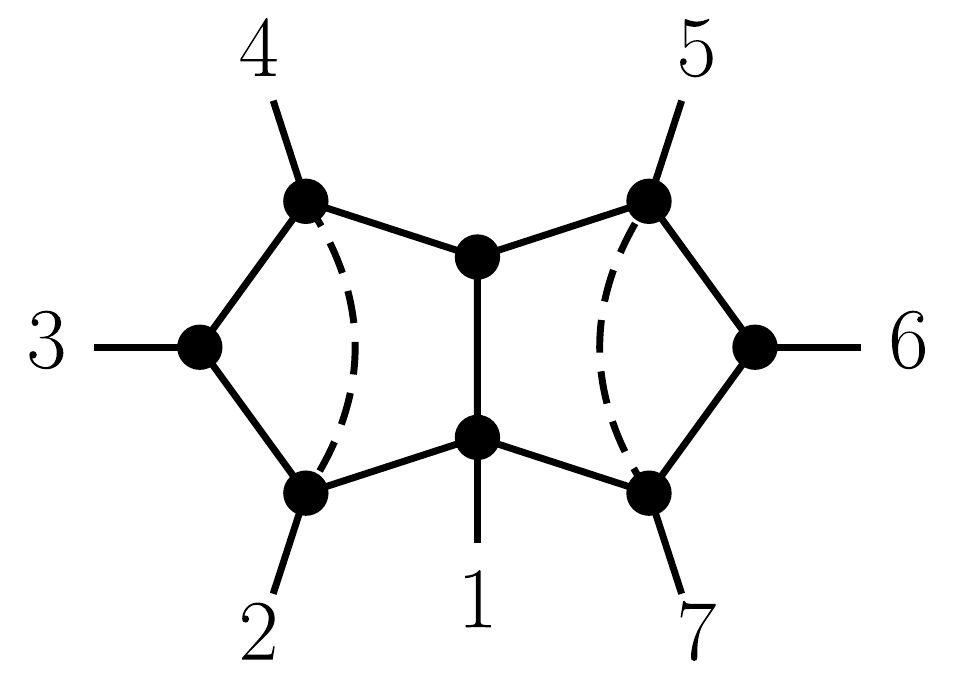}}\,,
$$
The second-order differential equations satisfied by these integrals are reviewed in Appendix \ref{sec:seven_and_eight_point_differential_euqations}. The larger dimensionality ($6=3\times7-15$) of the kinematic DCI seven-point space makes a direct integration of the second order differential equations challenging.
Instead of directly integrating up the second order partial differential equations, we turn to an alternative strategy that is inspired by the bootstrap approach for planar scattering amplitudes in $\N=4$ sYM, see e.g.~\cite{Dixon:2011pw,Dixon:2013eka,Dixon:2014iba,Golden:2014pua,Dixon:2015iva,Caron-Huot:2016owq,Drummond:2014ffa,Dixon:2016nkn}. Several of the constraints that apply in the bootstrap program for full amplitudes are relevant for Feynman integrals as well. A similar analysis has recently been pursued for nonplanar five-point integrals \cite{Chicherin:2017dob}.

The general philosophy is as follows: we make an initial ``educated guess'' about the function space the Feynman integral under consideration should live in, and then apply different constraints to isolate the desired integral within that space. Since the $\Omega$-integrals have unit-leading singularity and are given by a $\dlog$-form \cite{ArkaniHamed:2012nw}, it is natural to expect that they evaluate to pure, uniform-transcendental functions of weight four. In order to describe the essential analytic properties of these function spaces, the mathematical tools of symbol technology (see e.g.~\cite{Goncharov:2010jf,Duhr:2011zq}) prove extremely useful. In the following, we will assume a rudimentary knowledge of the relevant concepts as described in e.g.~\cite{Duhr:2011zq,Duhr:2014woa}. Equipped with these tools, we build a basis for the (symbols of the) corresponding functions through weight four and impose the differential equations from the previous section as constraints on this ansatz.

For the two-loop seven-point cases discussed below, it turns out that we have sufficiently many constraints to find unique solutions within our initial weight four function space. Any constraints not used to constrain the initial ansatz then serve as valuable cross-checks to ensure that our answer is complete (at symbol level). 

Here we note that upgrading the symbol-level expressions to full function level results is possible, but requires more work and would involve fixing certain multiple-zeta valued ambiguities (e.g. $\zeta_3 \times \log s$, etc.) that are in the kernel of the symbol map. This is beyond the scope of our current work where are content with obtaining the symbol. 

%
\subsection{Constructing the function space}
\label{subsec:steinmann_bootstrap_heptagon}
%

%
\subsection*{Alphabet}
%
%
As mentioned above, our bootstrap inspired approach to solving the differential equations for the seven-point integrals starts from an educated guess of the relevant function space. To this end, we can either look at the two-loop seven-point MHV symbol of Caron-Huot~\cite{CaronHuot:2011ky} and extract the independent symbol alphabet or focus on the existing alphabet to describe the heptagon symbols relevant for MHV and NMHV scattering \cite{Drummond:2014ffa,Dixon:2016nkn}. In both cases we obtain equivalent 42-letter alphabets. The dimension of the alphabet coincides with the dimension of the cluster algebra associated to the Grassmannian $Gr(4,7)$  \cite{Drummond:2014ffa}. 

A priori it is unclear if the symbol alphabet that is relevant for the full amplitudes is sufficient to describe individual integrals. It is logically possible that Feynman integrals depend on additional letters that drop out in the amplitude. As we will show below, this is not the case for the seven point examples we discuss in this section but becomes relevant starting at eight points. In the opposite direction, individual Feynman integrals might live in lower-dimensional subspaces of the full heptagon-space, but for uniformity of the discussion, it will be advantageous to embed the integrals into the more general setting of the six dimensional heptagon space characterized by the 42 heptagon letters introduced in \cite{Drummond:2014ffa,Dixon:2016nkn}. Here we follow the conventions of \cite{Drummond:2014ffa,Dixon:2016nkn} and denote the letters by $a_{i,j}$. 
\begin{align}
\label{eq:heptagon_alphabet}
\begin{split}
  a_{1,1} & = \frac{\ab{1234} \ab{1567} \ab{2367}}{\ab{1237} \ab{1267} \ab{3456}}\,, 
	\qquad 
	a_{2,1}  = \frac{\ab{1234} \ab{2567}}{\ab{1267} \ab{2345}} \,,\\
	a_{3,1} & = \frac{\ab{1567} \ab{2347}}{\ab{1237} \ab{4567}} \,, \twhite{\frac{\ab{1234}}{\ab{1234}}}  \hskip -.05cm
	\qquad
	a_{4,1}  = \frac{\ab{2457} \ab{3456}}{\ab{2345} \ab{4567}} \,,\\
	a_{5,1} & = \frac{\ab{1245} \ab{1367} - \ab{1267} \ab{1345}}{\ab{1234} \ab{1567}}
						= \frac{\ab{1(23)(45)(67)}}{\ab{1234} \ab{1567}} \,,	\\
	a_{6,1} & = \frac{\ab{1356} \ab{1472} - \ab{1372}\ab{1456}}{\ab{1234} \ab{1567}}
	= \frac{\ab{1(34)(56)(72)}}{\ab{1234} \ab{1567}} \,.
\end{split}
\end{align}
The remaining letters are given by the cyclic permutation of the particle labels,
\begin{align}
  a_{ij} = a_{i1}\Big|_{Z_k \mapsto Z_{k+j-1}}\,.
\end{align}
Parity transformations act nontrivially on these letters, it leaves $a_{1i}$ and $a_{6i}$ invariant and maps the other letters to cyclic images according to the rule \cite{Drummond:2014ffa,Dixon:2016nkn},
\begin{align}
	a_{11}\leftrightarrow a_{11}\,,\ 
	a_{61}\leftrightarrow a_{61}\,,\
	a_{21}\leftrightarrow a_{37}\,,\
	a_{41}\leftrightarrow a_{51}\,,
\end{align}
and cyclic images thereof. 

\subsection*{First entry condition}

Scattering amplitudes for massless particles as well as Feynman integrals with massless internal propagators can have branch cuts only when Mandelstam invariants $\sab{ij}$ or $\sab{i_1,...,i_k}$ go to zero or infinity as dictated by locality and unitarity \cite{Eden:1966dnq}. In the bootstrap literature, this is known as the branch-cut condition and imposes constraints on the allowed first entries of a symbol where only Mandelstam invariants can appear. In the example of the heptagon alphabet this branch cut condition imposes that only the letters $a_{1,j}$ can appear as valid first entries \cite{Drummond:2014ffa,Dixon:2016nkn},
\begin{align}
\label{eq:a11first_entry_letter}
a_{1,1} & = \frac{\ab{1234} \ab{1567} \ab{2367}}{\ab{1237} \ab{1267} \ab{3456}}  = \frac{\sab{23}\sab{67}\sab{712}}{\sab{12}\sab{17}\sab{45}}\,,
\end{align}
as $a_{1,1}$ and its cyclic images are the only letters which exclusively involve Mandelstam invariants. 

%
\subsection*{Steinmann conditions}
%
%
Feynman integrals in a local Lorentz-invariant quantum field theory are subject to analyticity constraints derived from locality. The Steinmann conditions \cite{Steinmann,Steinmann2,Cahill:1973qp}, which were originally derived in the context of axiomatic quantum field theory, are some of the key constraints that recently experienced a revived interest concerning scattering amplitudes for massless particles~\cite{Caron-Huot:2016owq,Dixon:2016nkn} (see also e.g.~\cite{Bartels:2008ce,Bartels:2008sc} for the use of Steinmann conditions in the multi-Regge-limit). At a practical level, the Steinmann conditions state that scattering amplitudes (and Feynman integrals) can not have double-discontinuities in overlapping kinematic channels. For a more detailed discussion we refer the reader directly to \cite{Caron-Huot:2016owq,Dixon:2016nkn}. For the bootstrap program, disallowed double-discontinuities are associated with certain exclusions on consecutive letters in the symbol.

At the level of multi-particle invariants in the heptagon case, this means that symbols of Feynman integrals can not involve adjacent letters that are associated with overlapping three-particle channels. Since the structure of the 42-letter heptagon alphabet (\ref{eq:heptagon_alphabet}) is such that each of the 7 three-particle invariants is associated to a unique letter, the Steinmann conditions are easily imposed at the symbol level. For example from (\ref{eq:a11first_entry_letter}) we see that $a_{1,1}\leftrightarrow\sab{712}$ which indicates that $a_{1,1}$ can not be adjacent to $a_{1,2}, a_{1,3},$ and $a_{1,6}$ (and cyclic images thereof). For further details, we refer directly to \cite{Drummond:2014ffa,Dixon:2016nkn}. 

\subsection*{Integrability}

Besides these two more detailed constraints that rely on the analytic structure of Feynman integrals, there is a general constraint imposed on symbols coming from the fact that partial differentials ought to commute for well defined functions. This constraint is known as the \emph{integrability condition}. Starting from the symbol of a weight $n$ function $F$ written in a factorized form
\begin{align}
  \mathcal{S}( F ) = \sum_{i,j} \mathcal{S}(F^{i,j}) \otimes \log \alpha_i, \otimes \log \alpha_j \,,
\end{align}
where the sum $i,j$ is over all letters of the alphabet $\alpha_{i,j}$ and $F^{i,j}$ are weight $n-2$ functions. Equipped with these definitions, we can write the integrability conditions as follows,
\begin{align}
\label{eq:integrability_condtions_ffas}
(\partial_r\partial_k -\partial_k \partial_r)F = 0 \Leftrightarrow 
\sum_{i,j}\left[\mathcal{S}(F^{i,j})-\mathcal{S}(F^{j,i})\right] \left(\frac{\partial \log \alpha_i}{\partial x_r}\right)\left(\frac{\partial \log \alpha_j}{\partial x_k}\right) = 0\,,
\end{align}
for any combination of partial derivatives with respect to the underlying variables $x_r,\ x_k$.  These integrability equations give constraints on the antisymmetric combinations, $\mathcal{S}(F^{i,j})-\mathcal{S}(F^{j,i})$, which can then be imposed iteratively in weight to construct integrable symbols. Further details can be found e.g.~in \cite{Drummond:2014ffa,Dixon:2016nkn}.  

\noindent
Following the discussion above, one can count the number of integrable symbols that satisfy,
\begin{itemize}
	\item first entry condition (only $a_{1j}$ can appear in first entry)
	\item Steinmann constraints (no overlapping discontinuities in incompatible channels)
	\item integrability
\end{itemize}
at weights $(1,2,3,4)$ there are $(7,28,97,322)$ symbols satisfying the above conditions in agreement with the numbers given in \cite{Dixon:2016nkn}. For our bootstrap inspired approach to solving the differential equations for the seven-point double-pentagon integrals, the 322-dimensional Steinmann heptagon space is going to be our initial starting point.

%
\subsection{Constraining the ansatz}
\label{subsec:bootstrap_constraints}
%
%
\subsection*{Second order differential equations and final entry condition}
%

The constraints on the 322-dimensional Steinmann-heptagon ansatz from the second order differential equations can conceptually be split into two categories. First, we only use the fact that the inhomogeneous term in the differential equation is a uniform-transcendental weight-two function but apart from that do not require any details about its functional form. Acting with a second order differential operator on a generic weight four function typically results in a mix of weight-three and weight-two terms due to the fact that the second derivative can act on the rational prefactor generated by the first derivative. A simple toy example of this phenomenon is given by,
\begin{align}
\label{eq:diff_eq_toy}
  \partial_y \partial_x \left[\mathcal{S}(f)\otimes y \otimes (1+xy) \right]
  = \frac{1}{(1 + x y)^2}\, \mathcal{S}(f)\otimes y + \frac{1}{1+xy}\,\mathcal{S}(f)\,.
\end{align}
where $\mathcal{S}(f)$ is the symbol of an arbitrary weight-two function. The first term on the right hand side of~(\ref{eq:diff_eq_toy}) is of weight three and gets generated when the $\partial_y$ derivative acts on the rational prefactor $\frac{y}{1+xy} = \partial_x \log (1+xy)$. Demanding that all weight-three terms cancel (we know that the inhomogeneity is weight two) therefore imposes \emph{homogeneous constraints} on the initial ansatz. Since we are matching zero on the right hand side, these constraints are very easy to implement, and sometimes they are even sufficient to fix the answer up to an overall scale.

Alternatively, we can phrase these homogeneous weight-drop conditions as a constraints on the final entries of the weight-four symbol,
\begin{align}
\label{eq:symbol_ansatz_w4}
	\Sy(\omega^{(4)}) = \sum^{n_f}_{i=1} \sum^{n_{\omega^{(3)}}}_{j=1} c_{ij} \ \Sy(\omega^{(3)})_j \otimes \alpha_i\,,
\end{align}
where the sums run over the $n_f$ final entries as well as the number of linearly independent weight-three symbols $n_{\omega^{(3)}}$. Acting with the first derivative on the weight-four symbol leads to rational prefactors, $\frac{\partial \log \alpha_i}{\partial x_k}$, times weight-three symbols. For generic final entries, $\alpha_i$, the second derivative now acts on the rational prefactors (just like in our toy example (\ref{eq:diff_eq_toy})) without decreasing the transcendental weight further. Demanding that all such terms cancel therefore imposes constraints on the possible final entries of the symbol. Concretely, demanding that the second order differential operators $\D^{(2)}$ annihilate the final entries independent of the weight-three basis element, gives constraints on the allowed final entries of the integral
\begin{align}
 \sum^{nf}_{i=1} c_{ij} \D^{(2)}\log \alpha_i = 0 \,, \text{for all }j = (1,...,n_{\omega^{(3)}})\,.
\end{align} 

In the second category we can use the \emph{detailed knowledge} of the \emph{inhomogeneous} weight-two terms on the right-hand side of the differential equations. The inhomogeneities are one-loop integrals which are either known, it is straightforward to integrate them directly, or we can express them in terms of the standard one-loop basis of integrals using generalized unitarity \cite{Bern:1994zx,Bern:1994cg,Bern:2007ct}. Knowing the precise form of the inhomogeneous terms then gives additional constraints on the undetermined coefficients in our symbol ansatz, $c_{ij}$, by plugging the weight-four ansatz (\ref{eq:symbol_ansatz_w4}) and the known inhomogeneities into the differential equations and matching both sides,
\begin{align}
\label{eq:full_differential_constraints}
 \sum^{n_f}_{i=1} \sum^{n_{\omega^{(3)}}}_{j=1} c_{ij} \  \D^{(2)} \left[\Sy(\omega^{(3)})_j \otimes \alpha_i \right]= \text{weight-two source term}\,.
\end{align}
In principle we could have obtained all constraints from the differential directly from (\ref{eq:full_differential_constraints}), but we find it conceptually enlightening to phrase things in terms of the two-step process outlined above to get insights into the structure of allowed final entries of the integrals. 

%
\subsection*{Further constraints}
%

Besides the differential equations, we can use the knowledge of discrete symmetries of the integral or certain kinematic limits. In the seven-point case the Feynman integrals should smoothly degenerate to the known hexagon answer in the appropriate soft- or collinear limits. Implementing these limits is straightforward at the level of twistor expression, see e.g.~\cite{Dixon:2016nkn}, and we do not discuss them in detail here. 

%
\subsection{Result for \texorpdfstring{$\Omega^{(2)}_{7,a}$}{Omega(2)7b}  }
\label{subsec:omega_7b}
%

As outlined above, we start from the \emph{assumption} that $\Omega^{(2)}_{7,a}$ lives inside the 322-dimensional weight-four heptagon Steinmann space. We impose the homogeneous conditions that the three differential operators identified in eq.~(\ref{eq:diff_ops_dp_b}) lead to a weight drop by two. Imposing the weight-drop constraint leaves 23 unfixed parameters in our ansatz.

In the next step, we use the knowledge of the inhomogeneities on the right hand side of the differential equations to fix additional degrees of freedom. Looking at the first operator and matching to the weight-two chiral hexagon function (\ref{eq:ULS_chi_hex_7pt_23_massive}) fixes 15 additional parameters, leaving 8 unfixed degrees of freedom. Surprisingly, the two remaining operators $\O_2$ and $\O_3$ (\ref{eq:diff_ops_heptagon_7b}) do not fix any additional free parameters but serve as valuable consistency checks.  We are able to match the inhomogeneous terms of these differential equations as well. 

So far we have not imposed any discrete symmetries of the integral on our general weight-four Steinmann-symbol ansatz. Implementing the $2\lra3, 1\lra4, 5\lra7$ discrete symmetry and demanding that the ansatz be even under this flip, fixes 4 additional parameters.

In the final step, we demand that the seven-point integral smoothly matches the known six-point result if we consider the $2||3$ collinear limit. This collinear limit fixes all remaining degrees of freedom in our ansatz and we are left with a unique result which is attached in an ancillary file. We summarize the number of unfixed parameters after various constraints in tab.~\ref{tab:constraints_summary_7pt}.

\noindent
The final result for the symbol of $\Omega^{(2)}_{7,a}$ shows some interesting features that are worth pointing out:
\begin{enumerate}
  \item As expected this simple integral does not live in the full six-dimensional heptagon kinematic space. This is related to the fact that there are two commuting first-order differential operators that annihilate this integral, $O_{12}$ and $O_{43}$, which implies that the kinematic dependence of this integral is reduced to a four-dimensional subspace of the full heptagon space. Consistent with this expectation, we find that our result indeed only depends on four variables.
 \item One special feature of this integral is that the symbol has only four independent final entries, the same as the number of kinematic variables. These four independent final entries can be chosen as special combinations of the heptagon letters,
 \begin{align}
 \begin{split}
 g_1 & = \frac{a_{1,2}}{a_{2,6}a_{3,5}} = \frac{\ab{1347}\ab{4567}}{\ab{1467}\ab{3457}}\,, \quad
 g_2  = \frac{a_{3,2}}{a_{2,6}a_{2,7}} = \frac{\ab{1345}\ab{4567}}{\ab{1456}\ab{3457}}\,,\\
 g_3 & = \frac{a_{1,3}}{a_{2,7}a_{3,6}} = \frac{\ab{1245}\ab{1567}}{\ab{1257}\ab{1456}}\,,\quad
 g_4  = \frac{a_{2,3}}{a_{3,5}a_{3,6}} = \frac{\ab{1247}\ab{1567}}{\ab{1257}\ab{1467}}\,.
 \end{split}
 \end{align}  
so that any projective first-order differential operator for this integrals should only involve four independent directions.
\item In tune with the simpler kinematic dependence, this integral only depends on 16 independent letters out of the 42 possible heptagon letters. Furthermore, there are only four independent first entries, again reduced from the 7 possible first entries in the full heptagon space. Inspired by the result for the hexagon integrals \cite{Caron-Huot:2018dsv}, we have analyzed the discontinuity in the $\sab{567}$-channel where the six-point counterpart led to a simplification of the alphabet to very high weight, when boxes are inserted between the two pentagons along this channel. Here we also observe a simplification of the alphabet with a dimension drop from 16 to 11 on the discontinuity. Since we are still at relatively low weight, it would be interesting to investigate this point further for higher loop examples and determine the influence of the differential equations on this dropout. 
 \item Relation to six-dimensional scalar one-mass hexagon integral, $\Phi^{(23)}_7\!:$

Now that we have the symbol-level expression for  $\Omega^{(2)}_{7,a}$ one can ask the question if there is a first-order differential operator relating this symbol to the symbol of the scalar $6d$ hexagon with one mass,
\begin{align}
  D^{(1)} \Sy(\Omega^{(2)}_{7,a}) \stackrel{?}{=} \Sy(\Phi^{(23)}_7)\,.
\end{align}
Since we have both symbols in terms of heptagon letters, it is easy to implement a generic first-order differential operator on the last letter and then try to find a solution for $D^{(1)}$. This is further simplified by the fact that $\Omega^{(2)}_{7,a}$ only has four independent final entries, the same as the number of variables. If we now choose the final entries as our independent coordinates, the first-order differential operator drastically simplifies and one can show that
\begin{align}
  \left[\frac{\partial}{\partial \log g_2} - \frac{\partial}{\partial \log g_4}\right] \Sy(\Omega^{(2)}_{7,a} )=\Sy( \Phi^{(23)}_7)\,.
\end{align}
We have checked that this first-order differential operator is independent from the factorized first-order differential operators in (\ref{eq:scaled_diff_ops_dp_b}) so that the relation between the scalar $6d$ hexagon and the double pentagon integral is not as straightforward as in the six-point case. 
In this particular case, the differential operator above only acts along a single direction in the four-dimensional kinematic space of $\Omega^{(2)}_{7,a}$, but we do have access to the full three-dimensional boundary data of the 6pt integrals so that the combined knowledge of the first-order differential equation and the boundary data is sufficient to fix the answer. We leave a detailed discussion on solving the differential equations more directly to future work.
\end{enumerate}

%
\subsection{Result for  \texorpdfstring{$\Omega^{(2)}_{7,b}$}{Omega(2)7a}}
\label{subsec:omega_7a}
%
For the double pentagon $\Omega^{(2)}_{7,b}$, we found four second-order differential operators (\ref{eq:diff_eqs_omega_7a}). Following the bootstrap strategy as before, we apply these second-order differential operators to the 322-dimensional heptagon Steinmann space of weight four. For this particular integral, only using the homogeneous constraints implied by the weight drop condition for the four differential operators is sufficient to fix \emph{all} coefficients (up to an overall scale) so that we uniquely identify the symbol for $\Omega^{(2)}_{7,b}$.  We have fixed the absolute scale and cross-checked our result by checking the inhomogeneous terms of the second-order differential operators acting on our weight-four $\Omega^{(2)}_{7,b}$-symbol against the symbol of the weight two chiral hexagon. The result written in terms of the heptagon alphabet is attached in an ancillary file. Note that this seven-point integral has been discussed from a similar perspective in \cite{Drummond:2017ssj}.

\begin{table}
\centering
\begin{tabular}{|c|c|c|}
\hline
constraints & $\Omega^{(2)}_{7,a}$ & $\Omega^{(2)}_{7,b}$ \\
\hline
ansatz				& 322	& 322 \\
hom.~diff.~eqs.		 	& 23 		& 1 \\
inhom.~diff.~eqs.		& 8		& 0	\\
symmetries			& 4		& -	\\
collinear limit			& 0		& -	\\
\hline
\end{tabular}
\caption{Number of unconstraint coefficients in the heptagon Steinmann ansatz after applying various constraints for both two-loop heptagon integrals.}
\label{tab:constraints_summary_7pt}
\end{table}


\noindent
\newline

\noindent
Before we go on to discuss the bootstrap approach to eight-point integrals, let us comment on the structure of the symbol of $\Omega^{(2)}_{7,b}$. First of all, it is interesting to note that this double-pentagon integral depends on 38 out of the 42 possible heptagon letters. The fact that individual integrals depend on a smaller number of letters might not be too surprising as a given integral only represents one particular cyclic configuration of external legs. 


In comparison to the other seven-point double-pentagon integral, $\Omega^{(2)}_{7,a}$, which we discussed in the previous subsection, $\Omega^{(2)}_{7,b}$ has a more complicated kinematic dependence. This was expected as this integral depends on seven dual points in comparison to the six dual points before. In comparison to the four first entries for $\Omega^{(2)}_{7,a}$, the current integral depends on all seven possible first entries of the heptagon space. An even more prominent difference is in the final entries. For $\Omega^{(2)}_{7,a}$ there were only four independent final entries which coincided with the number of variables this integral depended on. The more complicated configuration, $\Omega^{(2)}_{7,b}$, on the other hand has eight independent final entries, exceeding the number of kinematic variables by two. This more complicated kinematic dependence also affects the parallel work of \cite{Bourjaily:2018aeq} where a direct Feynman-parameter integration approach is pursued.

\begin{table}[ht!]
\centering
\begin{tabular}{|c|c|c|c|c|c|c|}
\hline
integral 					&  variables 	&  letters & $1^{\text{st}}$ entries & $2^{\text{nd}}$ entries & $3^{\text{rd}}$ entries & $4^{\text{th}}$ entries \\
\hline
$\Omega^{(2)}_{7,a}$		& 4			& 16		& 4	& 10	& 15	& 4 	\\
$\Omega^{(2)}_{7,b}$		& 7			& 38		& 7   & 21	& 33	& 8	\\ 
\hline
\end{tabular}
\caption{Summary of properties of both two-loop heptagon integrals $\Omega^{(2)}_{7,a}$ and $\Omega^{(2)}_{7,b}$. We list the number of variables, the number of independent letters in total and the number of independent letters in each entry of the symbol.}
\label{tab:property_summary_7pt}
\end{table}

%
\section{Eight-point bootstrap and algebraic letters}
\label{sec:eightpointintro}
%
Inspired by the success of the bootstrap for both the hexagon- and heptagon integrals, one is tempted to pursue the same approach for the octagons as well. There are four different types of eight-point double-pentagon integrals, that contribute in a local expansion of MHV scattering amplitudes in planar $\N=4$ sYM \cite{ArkaniHamed:2010gh}.
\begin{align}
  &\Omega^{(2)}_{8,a} =\raisebox{-47.5pt}{\includegraphics[scale=.5, trim=16pt 0 0 0]{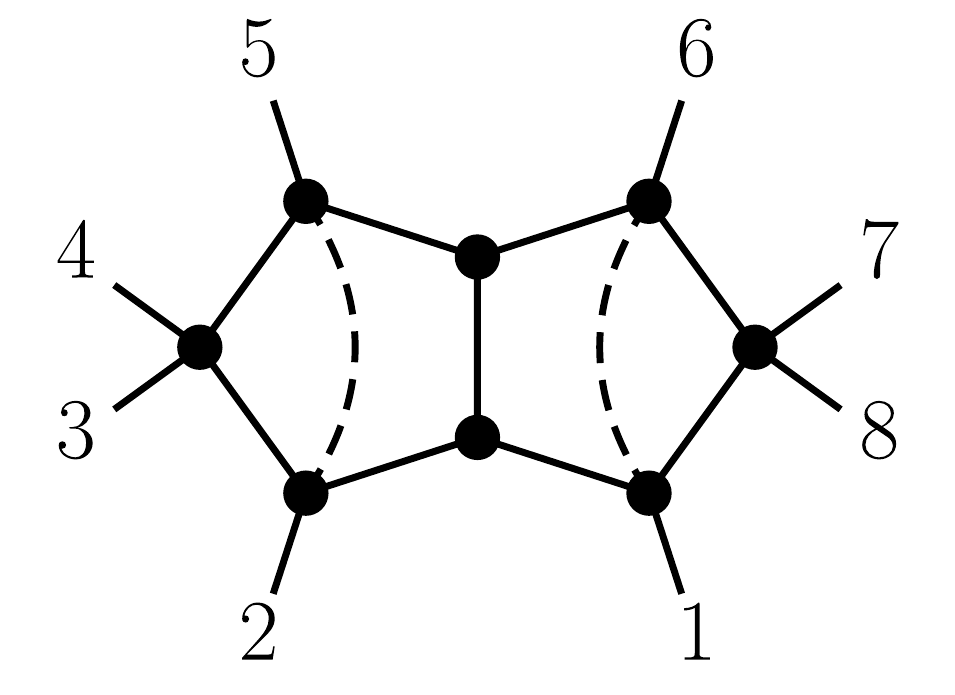}}\,, \qquad
  \Omega^{(2)}_{8,b} = \raisebox{-47.5pt}{\includegraphics[scale=.5, trim=5pt 0 0 0]{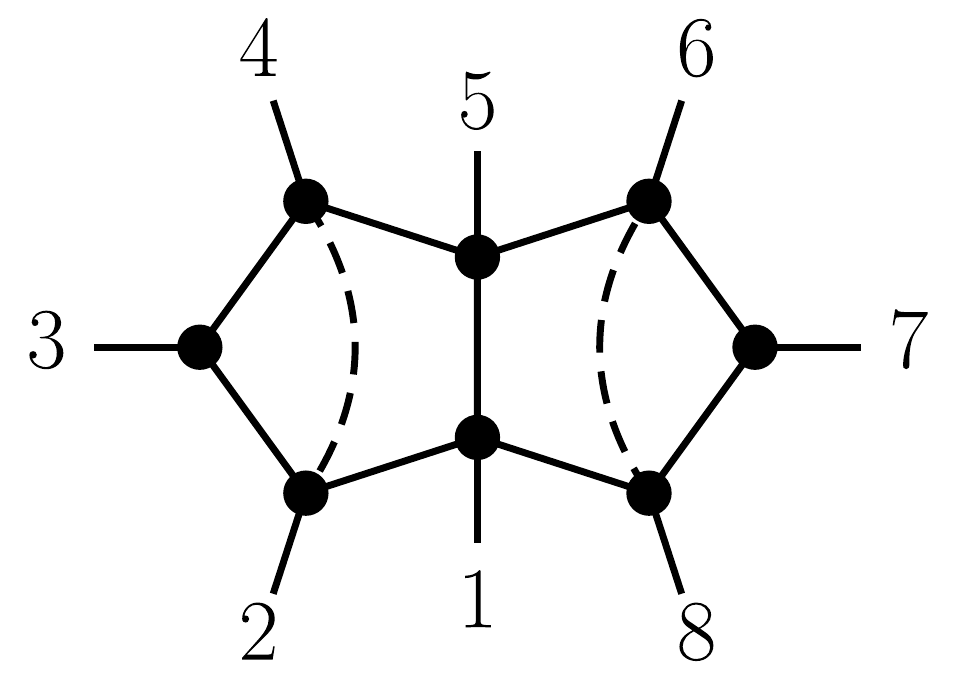}}\,,\\
 &\Omega^{(2)}_{8,c} =
 \raisebox{-47.5pt}{\includegraphics[scale=.5]{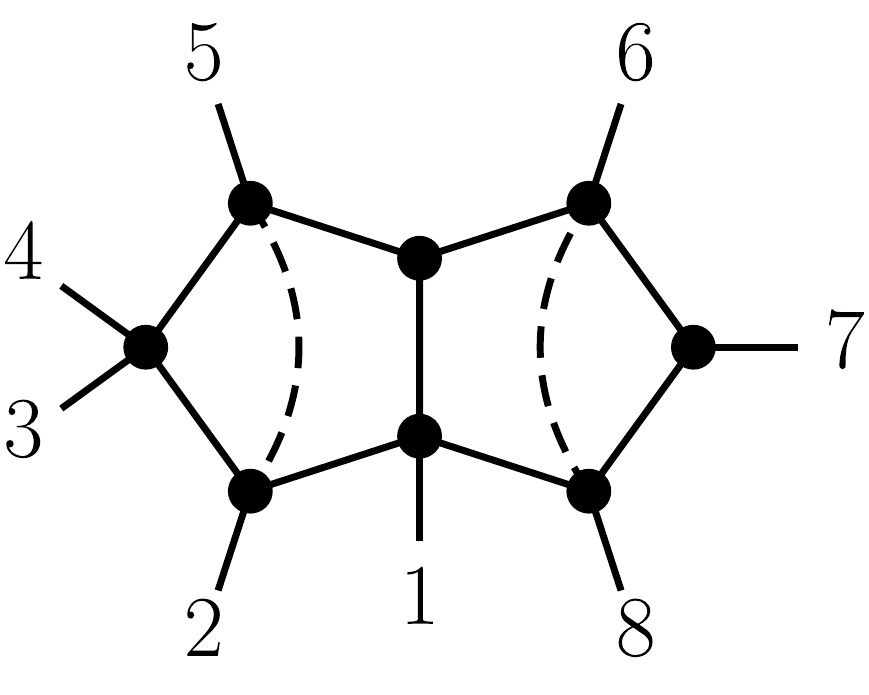}}\,, \qquad \,\,\,\,
\Omega^{(2)}_{8,d} =
\raisebox{-47.5pt}{\includegraphics[scale=.5]{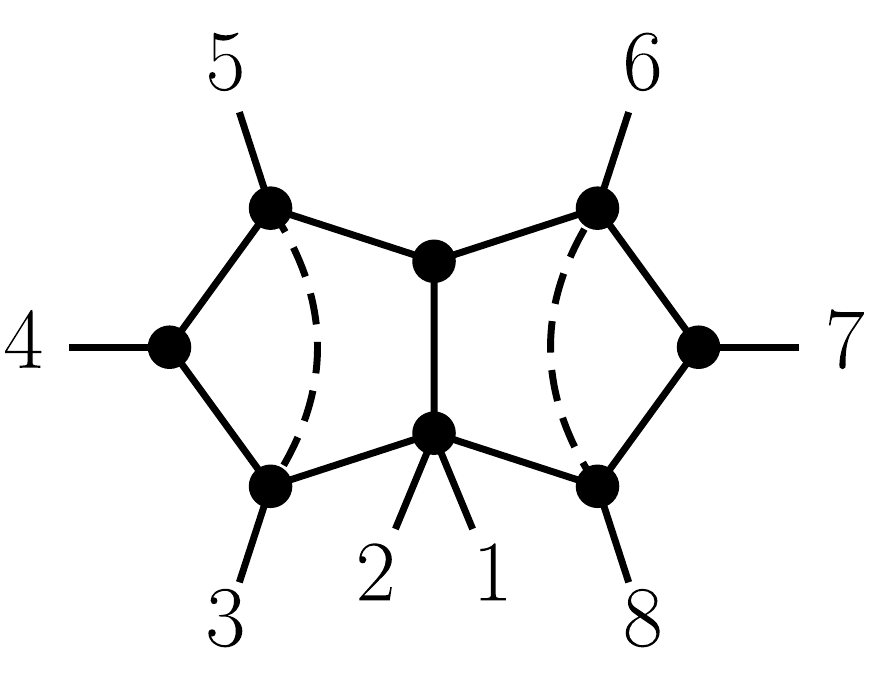}}\,.
\end{align}
Here we focus on $\Omega^{(2)}_{8,a}$, which is a generalization of the seven-point integral $\Omega^{(2)}_{7,a}$ of the previous section. This integral is structurally simpler than some of the other cases. It only depends on a restricted set of kinematic variables, but as we will see, it already shows a number of interesting new features. In comparison to $\Omega^{(2)}_{7,a}$, there is one additional massive corner and hence one can easily count five independent dual-conformal cross ratios.

%
%
\subsection{Need for algebraic letters}
%
%
The first and most important question to answer is about the relevant symbol alphabet of these integrals. The naive hope would be to find the eight-point integrals in the function space determined by the respective two-loop MHV-alphabet extracted from~\cite{CaronHuot:2011ky}. Compared to the 9 letters at six point, 42 letters at seven point, one finds 108 independent letters in the two-loop eight-point MHV amplitude of~\cite{CaronHuot:2011ky}. All of those letters are given in terms of \emph{rational functions} of twistor four-brackets. However, while this alphabet describes the two-loop eight-point MHV amplitude, there are several indications that this alphabet is insufficient to describe the double-pentagon integrals contributing to those amplitudes.

A very direct indication comes from the differential equations. To see this, let us focus our attention on the second-order differential equation for integral $\Omega^{(2)}_{8,a}$ (\ref{eq:double_pent_8pt_b_diff_eq}). There, the inhomogeneous term is the four-dimensional chiral hexagon $\I^{(1),\chi\text{-hex}}_{(34)(78)}$ (\ref{eq:chi_hex_8_34_78_massive}), whose unitarity decomposition contains the following four-mass box,
\begin{equation}
  	\I^{4m}_{\text{box}}(u,v) = \raisebox{-50pt}{\includegraphics[scale=.5]{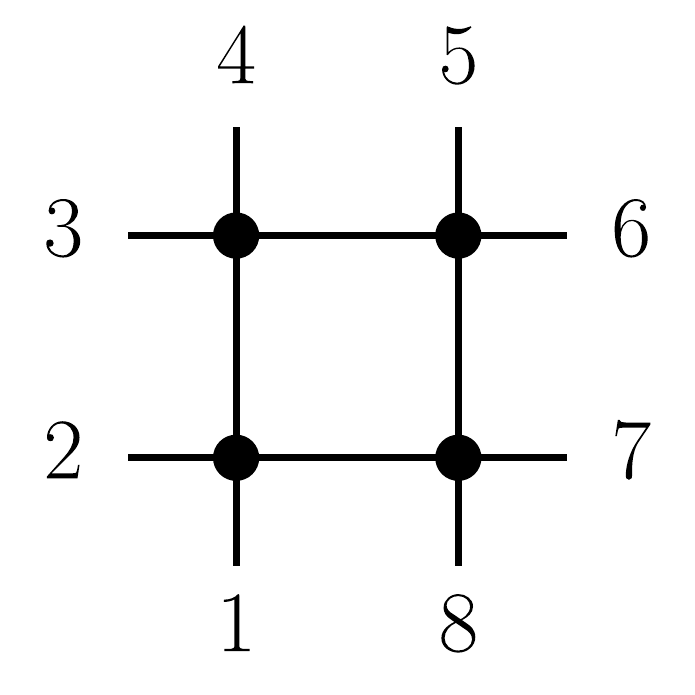}}
   	\leftrightarrow
  	u  = \frac{\ab{8123}\ab{4567}}{\ab{8145}\ab{2367}}\,, 
 	 \quad 
  	v = \frac{\ab{2345}\ab{6781}}{\ab{8145}\ab{2367}}\,
\label{eq:fourmassbox}
\end{equation}
which evaluates to
\begin{align}
\label{eq:four-mass-box-integral}
\begin{split}
  \I^{4m}_{\text{box}}(u,v) &  =  
 		 \Li_2\left[\frac{1}{2}\left(1\!+\!u\!-\!v\!+\!\sqrt{\Delta_4} \right)\right]
 		\!+\! \Li_2\left[\frac{1}{2}\left(1\!-\!u\!+\!v\!+\!\sqrt{\Delta_4} \right)\right] \!-\! \zeta_2\\
		&+\frac{1}{2} \log u \log v 
		  \! -\!  \log\left[\frac{1}{2}\left(1\!+\!u\!-\!v\!+\!\sqrt{\Delta_4} \right)\right] \log\left[\frac{1}{2}\left(1\!-\!u\!+\!v\!+\!\sqrt{\Delta_4} \right)\right]\,.
\end{split}
\end{align}
The letters of this integral involve a square root of the Gram determinant of the box, which in terms of the cross ratios is written as,
\begin{equation}
	\sqrt{\Delta_4}=\sqrt{(1-u-v)^2-4uv}\,.
  	\label{eq:gramdet4}
\end{equation}
The important observation here is that $\sqrt{\Delta_4}$ is \emph{not rationalized} by the twistor brackets. We call letters containing such roots \emph{algebraic}. In addition, unlike in the 6- and 7-point examples, one of the leading singularities for this integral also involves  $\sqrt{\Delta_4}$ which makes this integral non-pure. When integrating the differential equation, this opens the possibility that the two-loop integral can now depend on additional algebraic letters of the form,  
\begin{align}
 \frac{a_i(u_j) + \sqrt{\Delta_4}}{a_i(u_j) - \sqrt{\Delta_4}}
\end{align}
different from the ones that appear in the four-mass box. 

The Landau equations \cite{Landau:1959fi} provide an alternative approach to exploring the alphabet of the double pentagon integrals. This was pursued in \cite{Dennen:2015bet}, where such integrals were analyzed through the sub$^3$-leading Landau order. To this order, only rational branch points appeared as solutions to the Landau equations. The authors furthermore found that double-pentagon integrals in principle admit branch points that do not appear in the full symbol of the two-loop MHV amplitude. At the sub$^4$-leading order the following Landau diagram appears which has not been analyzed in \cite{Dennen:2015bet}
\begin{equation}
  \raisebox{-50pt}{\includegraphics[scale=0.6]{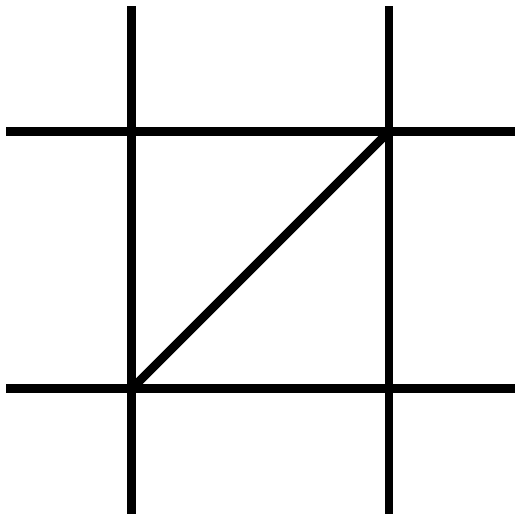}}\,.
  \label{eq:landaudiagram}
\end{equation}
Although we have not solved the Landau equations in this case, we suspect that this Landau singularity would reveal the square-root branch point contained in the algebraic letters. The reason for this is that the same Landau diagram describes a sub$^2$-leading Landau singularity of the double box integral, whose letters are known to contain the same four-mass box square root.

The reader might suspect that this effect is special to integral $\Omega^{(2)}_{8,a}$, due to its kinematic configuration, and one might wonder about other eight-point configurations. Although not described here, the differential equation for integral $\Omega^{(2)}_{8,c}$ also contains a four-mass box in the inhomogeneous term, so one expects the presence of algebraic letters there too. On the other hand, the differential equations for integrals $\Omega^{(2)}_{8,b}$ and $\Omega^{(2)}_{8,d}$ do not involve a four-mass box in the inhomogeneous term on the right hand side, nor algebraic prefactors. However, under closer inspection, both integrals contain the suspicious Landau diagram (\ref{eq:landaudiagram}) as subtopology. In fact, $\Omega^{(2)}_{8,b}$ contains two inequivalent versions of this diagram related by a cyclic transformation. To fully exclude the possibility that the MHV alphabet is sufficient to describe these integrals, we took the 1372 dimensional symbol space at weight four that is constructed from the 108-letter alphabet and is consistent with integrability and the Steinmann conditions. Once we impose the homogeneous equations from the four differential operators for integral $\Omega^{(2)}_{8,b}$, the homogeneous constraint alone fix 1368 parameters.  However, the solution only has  two independent last entries.  This is of course inconsistent with any smooth soft- or collinear limit onto the known seven-point integrals we discussed in sections \ref{subsec:omega_7b} and \ref{subsec:omega_7a} that have more complicated kinematic dependence. Not surprisingly, we find the need for additional letters to describe $\Omega^{(2)}_{8,b}$ as well. A similar conclusion can be reached for  $\Omega^{(2)}_{8,d}$.

In conclusion, we see that the eight-particle alphabet needs to be augmented in order to bootstrap the double pentagon integrals. One can check that adding only four independent four-mass box letters to the 108-letter MHV alphabet is still insufficient to describe the $\Omega^{(2)}_8$-integrals. The above arguments suggest including additional letters of the form $a \pm \sqrt{\Delta_4}$, where $a$ is some rational function of the kinematics. It is therefore interesting to ask what are the most general $a$ needed. The latter cannot be arbitrary, as we already know all possible singularities from the Landau analysis. Therefore they should be such that $a^2 -  \Delta_4$ factorizes over the rational letters of the alphabet. Analyzing which $a$ are allowed would be one possibility of continuing with the bootstrap approach.

Here, we follow a different route. Instead of bootstrapping the whole integral in one step, we will focus on bootstrapping one of its discontinuities, and will then use the latter to recover the full function via a dispersive integral. The advantage of this is, as we will argue, that the discontinuity is described by a smaller alphabet that is easier to handle.

%
%
\subsection{Bootstrap for the \texorpdfstring{$s_{2345}$}{s2345} discontinuity of \texorpdfstring{$\Omega^{(2)}_{8,a}$}{Omega(2)8b}}
\label{subsec:bootstrap_disc_omega_2_8b}
%
%

In the previous section we explained that a bootstrap for $\Omega^{(2)}_{8,a}$ necessarily involves an alphabet that goes beyond the one needed for MHV amplitudes. In principle, we could attempt to construct the required alphabet and proceed as with the seven-point integrals. Here we describe a shortcut that allows one to overcome the problem of knowing the number and the precise structure of the algebraic letters. 
So far we have only made use of the differential equations for the double-pentagon integrals which allowed us to bootstrap the six- and seven-point MHV integrals. 
At the level of symbols, the differential equations act on the final entry of the symbol and generically drop the transcendental weight. In this section, we are going to illustrate that it is advantageous, to also consider discontinuities of the higher point $\Omega$-integrals. \footnotetext{We thank Lance Dixon for encouraging us to study a particular discontinuity of $\Omega^{(2)}_{8,a}$. See~\cite{Caron-Huot:2018dsv} for related ideas.} 

In~\cite{Caron-Huot:2018dsv} , the authors noticed a peculiarly simple structure of the hexagon-integrals (with many box insertions between the two pentagon-caps) once one takes a discontinuity in the channel carrying momentum along the ladders where the relevant alphabet was reduced to five letters from the original nine hexagon letters. The important observation for us is the following;  if we look at the equivalent four-particle discontinuity in the eight-point case, corresponding to the $\sab{2345}$-channel, 
\begin{align}
  \underset{\sab{2345}}{\text{Disc}}\ \Omega^{(2)}_{8,a} =  \raisebox{-47.5pt}{\includegraphics[scale=.5]{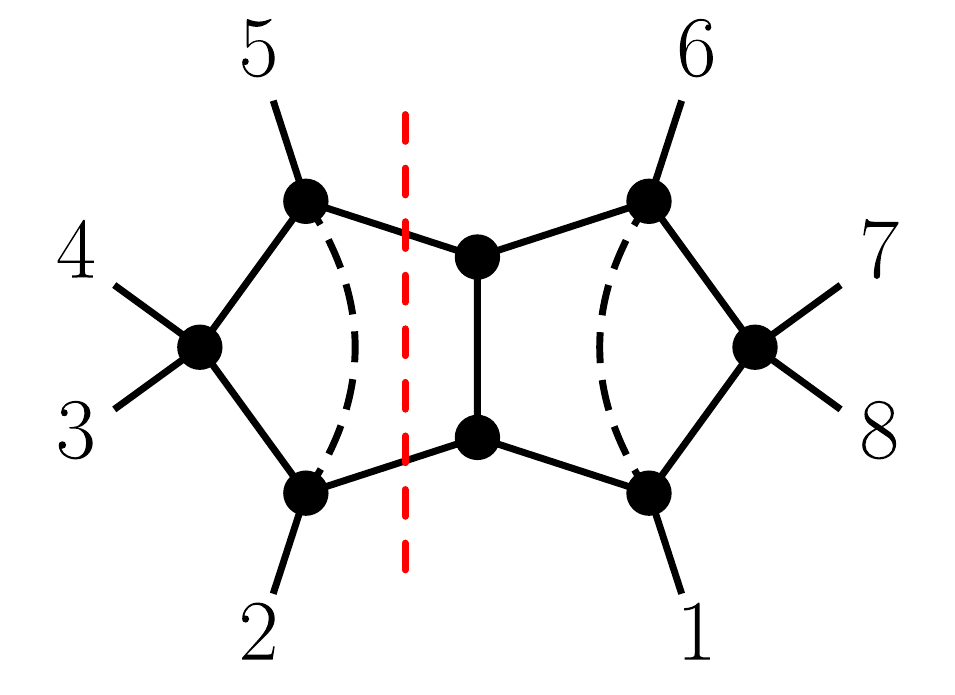}} 
  										     + \raisebox{-47.5pt}{\includegraphics[scale=.5]{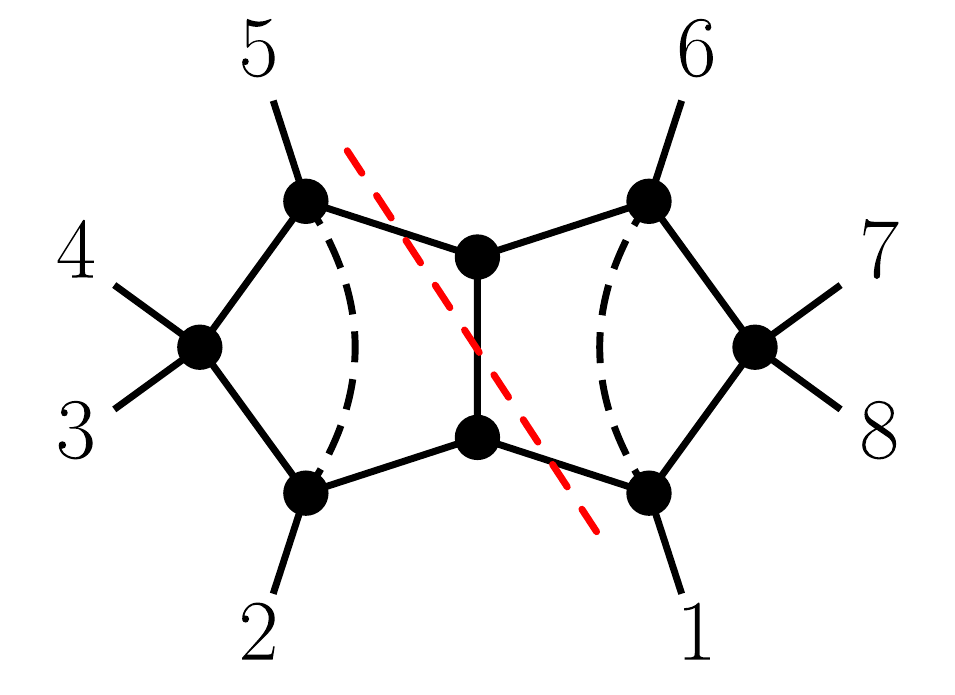}}
										     + \cdots
\end{align}
and combine this with the differential equation (\ref{eq:double_pent_8pt_b_diff_eq}), we realize that the four-mass box configuration in the decomposition of the inhomogeneous term of (\ref{eq:double_pent_8pt_b_diff_eq}) does not contribute. Furthermore, focussing on this specific discontinuity opens up these propagators in the Landau diagram, so that the dangerous four-mass box type configuration (\ref{eq:landaudiagram}) is also ruled out. From these arguments it is now clear that all algebraic letters should drop out on this discontinuity so that we are sidestepping the problem of guessing the correct algebraic letters. 

Now our proposal to obtain $ \underset{\sab{2345}}{\text{Disc}}\ \Omega^{(2)}_{8,a}$ starts from a rational set of letters required to describe the weight three discontinuity function. These rational letters can be directly given by suitable projectivizations of the rational branch points studied in the Landau-singularity analysis. We find a set of 124 independent letters composed of
\begin{itemize}
\item the original 108 MHV letters
\item eight cyclic images of $\frac{\ab{(123)\cap(245),(781)\cap(856)}}{\ab{2345} \ab{5681} \ab{7812}}$ 
\item eight inequivalent dihedral images of $\frac{\ab{(123)\cap(245),(567)\cap(681)}}{\ab{1245} \ab{2367} \ab{5681}}$.
\end{itemize}
For the particular $\Omega^{(2)}_{8,a}$-integral we can even simplify things further by going directly to the relevant five-dimensional kinematic subspace. In this subspace, the number of multiplicatively independent letters decreases to 25. 


Equipped with these 25 independent letters, we build the space of integrable symbols of weight three. Unlike for the full integrals, we do not have the first entry conditions at our disposal which leads to an increased number of admissible symbols, we find $25,\ 377$ and $4441$ at weights one, two and three respectively. These 4441 integrable weight-three symbols are our starting point for the bootstrap of the $\sab{2345}$-discontinuity of $\Omega^{(2)}_{8,a}$. As for the full six- and seven-point integrals, we can again use the differential equations (\ref{eq:double_pent_8pt_b_diff_eq}) to give us constraints on our ansatz. Instead of the full differential equations, we now look at the corresponding discontinuity of the differential equations. As for the complete integrals, we find that the differential equations are extremely powerful in constraining the ansatz; imposing both homogeneous and inhomogeneous parts of the differential equations fixes 3774 degrees of freedom, leaving us only with 667 unconstrained parameters. If we furthermore impose the discrete symmetries that are respected by the discontinuity, only 24 free parameters remain. In order to fix those, one can either match double-discontinuities or match onto the discontinuities of the seven point integrals by taking a soft limit. Doing so, we find a unique result from our originally 4441-dimensional ansatz. 


Here we note that the discontinuity function is surprisingly simple and only 13 independent letters remain in the final result. This is a dramatic simplification in comparison to the 25 letters we started with. We also note that there are only five independent final entries and six independent first entries in the weight three discontinuity function. 

%
%
\subsection{Dispersion representation for \texorpdfstring{$\Omega^{(2)}_{8,a}$}{Omega(2)8b}}
%
%
It is well known that Feynman integrals can be reconstructed from the knowledge of its discontinuities via dispersion integrals \cite{Remiddi:1981hn,Bauberger:1994hx,Remiddi:2016gno,Abreu:2014cla}. More concretely, the knowledge of the discontinuity in a single channel is sufficient to recover the full result as a one-fold integral. For a review and some examples see e.g.~\cite{Abreu:2014cla} and references therein. Dispersive representations for Feynman integrals have also been discussed recently to represent massive loop integrals in ${\mathcal{N}=4}$ super Yang-Mills \cite{Caron-Huot:2014lda}, and in the context of simplifying the analysis of differential equations \cite{Remiddi:2016gno}.

At the heart of the dispersion relations for Feynman integrals is once again Cauchy's theorem. 
One can begin by writing the full function as
\begin{align}
\label{eq:dispersion_integral_10}
F(s,t_i) = \frac{1}{2\pi i}\int \limits_\Gamma \!\!\!\frac{ds'}{s'-(s+i\epsilon)} \, F(s',t_i) \end{align}
where the contour $\Gamma$ wraps the pole at $s'=s$. By a contour deformation, assuming there is no contribution from the arc at infinity, we can write the same function as
\begin{align}
\label{eq:dispersion_integral_0}
F(s,t_i) = \frac{1}{2\pi i}\int \limits_\C \!\!\!\frac{ds'}{s'-(s+i\epsilon)} \underset{s}{\text{Disc}}\, F(s,t_i) \Big |_{s=s'}\,,
\end{align}
where the new contour $\C$ goes along the branch cut corresponding to the discontinuity.
If is often the case that the convergence at infinity is not guaranteed. In case function behaves poorly as $|s'|\to\infty$ and a pole at infinity contributes, one can improve the convergence properties by introducing so-called \emph{subtraction terms}. This is done by adding extra factors of $(s-s^\ast)$ in the denominator, where $s^\ast$ is a point where the value of the integral is known. After the contour deformation, these additional poles introduce the subtraction terms and require the knowledge of the full function on these special points $F(s^\ast,t_i)$, yielding
\begin{align}
\label{eq:dispersion_integral}
F(s,t_i) = \frac{1}{2\pi i}\int \limits_\C \!\!\!\frac{ds'}{s'-(s+i\epsilon)}\frac{s-s^\ast}{(s'-s^\ast)} \underset{s}{\text{Disc}}\, F(s,t_i) \Big |_{s=s'} + F(s^\ast,t_i) \,.
\end{align}
It is wise to judiciously choose subtraction terms where the behavior of the function $F(s^\ast,t_i)$ is simple. From a practical point of view, these subtraction terms can be chosen to be for example non-singular terms in the normalization of the integral itself or points that correspond to nonsingular soft- or collinear limits if these are related to the relevant dispersive channel, $s$. It is interesting to remark that the procedure of fixing the ambiguity in the dispersion integral with a subtraction term is very similar to fixing boundary terms when solving differential equations. In a different context, this is also related to the problem of reconstructing tree-level scattering amplitudes in effective field theories from BCFW-like recursion relations~\cite{Cheung:2015cba,Cheung:2015ota,Cheung:2016drk}, where it is possible to trade the poles at infinity for the knowledge of the universal soft-behavior of amplitudes at finite locations. 

An equation such as (\ref{eq:dispersion_integral}) is a very useful representation of the answer. Being a one-fold integral, one can extract its symbol algorithmically. We review this, and provide a pedagogical example in Appendix \ref{subsubsec:integrating_symbols}.

In subsection \ref{subsec:bootstrap_disc_omega_2_8b}, we bootstraped the discontinuity of $\Omega^{(2)}_{8,a}$ in the $\sab{2345}$ channel. Following the dicussion above, we are now in a position to write down the corresponding dispersion integral for $\Omega^{(2)}_{8,a}$. Before doing so, we introduce the relevant five cross ratios 
\begin{align}
\label{eq:cross_ratios_omega_2_8b_for_discontinuity}
\begin{split}
& 	u_1 = \frac{\ab{2356}\ab{8145}}{\ab{2345}\ab{8156}}\,,\quad
 	u_2 = \frac{\ab{1245}\ab{2367}}{\ab{1267}\ab{2345}}\,,\quad
 	u_3 = \frac{\ab{4567}\ab{8123}}{\ab{2367}\ab{8145}}\,, \\ 
&\qquad u_4 = \frac{\ab{1267}\ab{2345}\ab{8156}}{\ab{1245}\ab{2356}\ab{8167}}\,, \quad
        u_5 = y^{-1}   = \frac{\ab{1245}\ab{2356}}{\ab{1256}\ab{2345}}\,.
\end{split}
\end{align}
of which only $y$ depends on $\sab{2345}$. Therefore taking the discontinuity in $y$  is equivalent to taking the discontinuity in the $\sab{2345}$ channel. The associated dispersion integral now reads,
\begin{align}
\label{eq:dispersion_omega_2_8b}
\Omega^{(2)}_{8,a}(u_i,y) = \int \limits_0^\infty \frac{y\ dy'}{(y'-y)\ y'} \, \underset{y}{\text{Disc}}\ \Omega^{(2)}_{8,a}\Big|_{y=y'} 
					= \int \limits_0^\infty \dlog\left(\!\frac{y'-y}{y'}\!\right) \, \underset{y}{\text{Disc}}\ \Omega^{(2)}_{8,a}\Big|_{y=y'}
\end{align}
where we have included a subtraction term at $y=0$. This subtraction term is related to the normalization of the integral (\ref{eq:norm_dp_2_8b}) which is $\bar{N}_{b} \sim \ab{1256} \sim \sab{2345} \sim y$. Since $\sab{2345}=0$ does not correspond to any soft or collinear region, the integral is finite in this limit and in fact vanishes along with the normalization. We therefore get zero contribution from the corresponding residue in the dispersion formula (\ref{eq:dispersion_integral}). This can be checked explicitly in the analogous limits of the six- and seven-point double pentagons, for which similar dispersion integrals exist, and the addition of new masses does not spoil this behavior. 

Let us discuss some of the features of the formula (\ref{eq:dispersion_omega_2_8b}). By construction, the dispersive representation has the correct discontinuity in $y$ that we determined in the previous section within our bootstrap approach. It is easy to compared this to the discontinuity of $\Omega^{(2)}_{8,a}$ as obtained by the parallel computation in~\cite{Bourjaily:2018aeq}. From the cross ratios defined in (\ref{eq:cross_ratios_omega_2_8b_for_discontinuity}) we immediately recognize that the dispersion variable, $y$, does not depend on twistors $7$ and $8$. This allows us to directly take the smooth soft limit that relates this eight-point integral to a relabelling of $\Omega^{(2)}_{7,a}$. This dispersion representation for  $\Omega^{(2)}_{7,a}$, and the analog for  $\Omega^{(2)}_{6}$ can be easily integrated with full analytic dependence using the algorithm explained in Appendix~\ref{subsubsec:integrating_symbols}, finding complete agreement with the know answers. The eight-point case is slightly more complicated.  Setting up the integration requires us to rationalize the six-dimensional square-root of the Gram determinant of the six dual kinematic points, which has the following structure,
\begin{align}
\sqrt{\Delta_6} \sim \sqrt{y^2\ a_2(u_i) + y\ a_1(u_i) + a_0(u_i)}\,.
\end{align}
This can be rationalized by a suitable change of variables $y \mapsto f(\alpha, u_i)$ which introduces another square root, $\sqrt{a_0(u_i)}$. This is the four-mass box square root that we expect to find in the final answer. 
Besides the four-mass square roots, several spurious square roots are generated in intermediate steps by the change in the integration contour and the partial fractioning in the integration algorithm. Still, the integration can be carried out, since none of these square roots depend on the new integration variable $\alpha$. We can simplify the analysis by restricting to a simpler kinematic setting where the spurious square-roots disappear. In this case, we find complete agreement with \cite{Bourjaily:2018aeq}. 

Having convinced ourselves of the validity of (\ref{eq:dispersion_omega_2_8b}), and having checked consistency with the result of \cite{Bourjaily:2018aeq}, we wish to make a few further comments about the symbol answer from our point of view. For this purpose, we investigated in detail the symbol obtained in \cite{Bourjaily:2018aeq} and make the following observations
\begin{itemize}
\item All first entries of the symbol of $\Omega^{(2)}_{8,a}$ are in agreement with the expected branch cut structure and only cross ratios appear.
\item As explained in section \ref{sec:seven_point_bootstrap}, the second-order differential equations, combined with the bootstrap, predict that only certain final entries, $e_{f_i}$, are allowed. In agreement with this analysis, we find that only the following five last entries appear in the symbol of $\Omega^{(2)}_{8,a}$. They can be chosen as
\begin{align}
\label{eq:final_entries_omega_2_8b}
\begin{split}
  &	e_{f_1}\!=\!\frac{\ab{1456}\ab{2567}}{\ab{1567}\ab{2456}}\,, \quad
  	e_{f_2}\!=\!\frac{\ab{1236}\ab{1257}}{\ab{1235}\ab{1267}}\,, \hspace{-5pt} \quad
	e_{f_3}\!=\!\frac{\ab{1245}\ab{2356}}{\ab{1235}\ab{2456}}\,, \\
  &	e_{f_4}\!=\!\frac{\ab{1236}}{\ab{1267}}\! \left(\!\frac{\ab{1256}\ab{1678}}{\ab{1268}\ab{1356}}\!+\!\frac{\ab{1567}\ab{2456}}{\ab{1456}\ab{2356}}\!\right)\!\,, \ \ 
	e_{f_5}\!=\!\frac{\ab{1267}\ab{1456}\ab{2356}}{\ab{1236}\ab{5671}\ab{2456}}\,.
\end{split}
\end{align}
We find it interesting to remark that the number of last entries is equal to the number of independent variables.
\item We already mentioned the powerful simplification that occurs when taking the $\sab{2345}$ discontinuity of $\Omega^{(2)}_{8,a}$, which allowed us to bootstrap the weight three discontinuity function in this particular channel. We found that only 13 independent letters remain for the discontinuity function.  This can be contrasted with the $\sab{1234}$ discontinuity, where such a significant simplification does not occur. After taking this discontinuity, there are still 23 independent letters.
\item Out of the $26$ letters, $21$ are rational, $2$ are new algebraic letters appearing in the four-mass box integral; the six-dimensional hexagon with two massive corners does not provide further relevant letters, so that there are $3$ completely new letters. They take the form $a_i + \sqrt{\Delta_4}$, with the $a_{i}$ being some rational factors.
\end{itemize}

%
\section{Prospects for the eight-point amplitude bootstrap}
\label{sec:amplitudes_bootstrap_implication}
%
%

The bootstrap program has been extremely successful in the last years and generated explicit results for six- and seven point scattering amplitudes in planar $\N=4$ sYM up to very high loop order \cite{Dixon:2011nj,Dixon:2011pw,Dixon:2013eka,Golden:2014pua,Dixon:2014iba,Drummond:2014ffa,Dixon:2015iva,Caron-Huot:2016owq,Dixon:2016nkn}. An integral part of the bootstrap program is an adequate choice of symbol alphabet. Once the analytic structure encoded in the alphabet is fixed one can construct the relevant function spaces of generalized polylogarithms or their symbols. A judicious choice of an alphabet, augmented by the use of the Steinmann relations, symmetries and relevant physical limits, allowed for the construction of MHV and NMHV six- and seven-point amplitudes up to six-loops. In addition, the observation that the letters of the six- and seven point alphabets correspond to cluster coordinates \cite{Fock:2003aa,Goncharov:2010jf,Golden:2013xva} has brought nontrivial insight into the bootstrap program. The results of this paper could be interpreted as an indication that cluster coordinates might not be enough to capture the full symbol of even the polylogarithmic part of the amplitude at eight-points and beyond. Interestingly, starting at eight points the cluster-algebras associated to the kinematic space of dual-conformal invariant scattering become infinite and it seems a priori unclear on which of the infinitely many cluster variables amplitudes should depend on, or whether cluster variables are in fact sufficient to describe scattering amplitudes at all. 

In the present paper, we analyzed the alphabet relevant for individual Feynman integrals. We would like to emphasize that simplifications can occur at amplitude level. On the other hand, one might argue that such simplifications are due to special helicity configurations, and hence not generic. A well-known example of this occurs at six points and two loops. The result for the MHV amplitude famously can be written in terms of classical polylogarithms only \cite{Goncharov:2010jf}, while the double pentagon integrals responsible for the bulk of the answer contain non-classical polylogarithms \cite{Dixon:2011nj}. The fact that the latter disappear in the MHV amplitude is due to an identity when taking the cyclic sum of the double pentagon integrals. On the other hand, in the NMHV amplitude, the same double pentagon integrals appear dressed with certain prefactors, so that that cancellation no longer occurs.

Similarly, we see that the final result for two-loop MHV amplitudes of \cite{CaronHuot:2011ky}, written in momentum twistor variables does not contain any algebraic letters, while we find five such letters in $\Omega^{(2)}_{8,a}$. This again illustrates that certain cancellations occur when combining the integrals to form the MHV amplitude. Do similar cancellations hold also for non-MHV amplitudes, and is the absence of algebraic letters generic at the amplitude level?

The Landau analysis applied to the full amplitude \cite{Dennen:2016mdk,Prlina:2017azl,Prlina:2017tvx} can shed some light on this question. The latter allows one to understand intuitively why the algebraic letters cancel in the two-loop MHV amplitudes. A simple Grassmann $\eta$ counting shows that the cut corresponding to the Landau diagram (\ref{eq:landaudiagram})
\begin{equation}
  \raisebox{-45pt}{\includegraphics[scale=0.6]{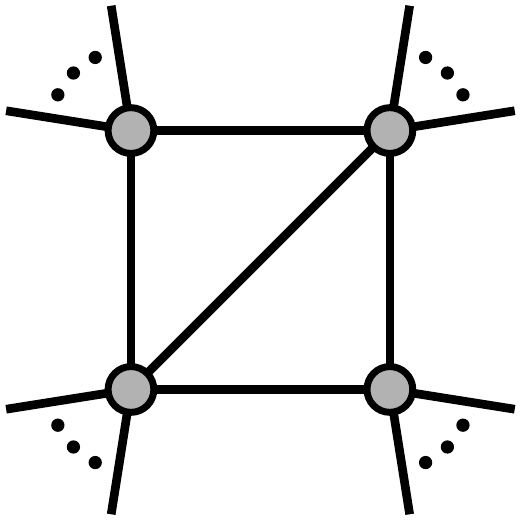}}
  \label{eq:s4llscut}
\end{equation}
does not have support on MHV amplitudes, and hence the corresponding letters should not appear in their symbol. On the other hand this is a perfectly fine NMHV cut, so one expects that algebraic letters appear in the two-loop NMHV amplitudes beyond seven points, in agreement with the proposal in \cite{Prlina:2017tvx}. However, our findings suggest that letters of the form $a_i + \sqrt{\Delta_4}$ could appear that go beyond the proposal of \cite{Prlina:2017tvx}. The knowledge of such letters will play a crucial role in any attempt to bootstrap the amplitudes and so warrants further research. 

In order to find algebraic letters, we do not necessarily have to go to higher loops. Even certain one-loop amplitudes in the N$^2$MHV sector already lie outside the cluster-letter space and are associated with the appearance of four-mass box Feynman integrals that involve square roots (\ref{eq:four-mass-box-integral}). This is easy to see by analyzing the following component-amplitude relevant for the scattering of a particular configuration of scalars in the $\N=4$ supermultiplet 
\begin{equation}
  \mathcal{A}^{(1)}_{8}(\phi^{12},\phi^{23},\phi^{23},\phi^{34},\phi^{34},\phi^{41},\phi^{41},\phi^{12})\,.
  \label{eq:scalaramp}
\end{equation}
A brief look at the Feynman rules of the theory reveals that the four-mass box
\begin{equation}
  \raisebox{-50pt}{\includegraphics[scale=0.5]{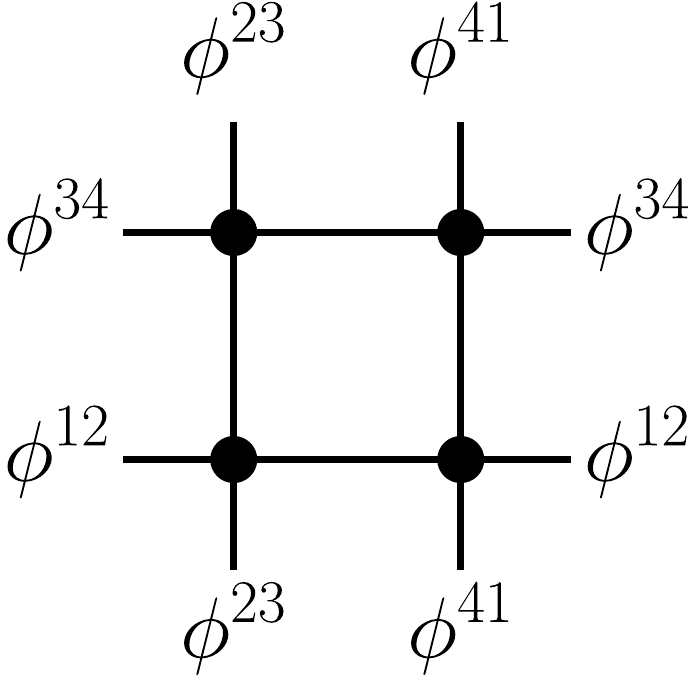}}
  \label{fig:scalarsbox}
\end{equation}
is the only diagram that contributes. Therefore one is led to conclude that the non-cluster letters are needed for this amplitude. 

Having realized that algebraic letters certainly appear at eight points in one-loop N$^2$MHV and two-loop NMHV amplitudes, one could worry that they also play a role in three-loop MHV amplitudes. It would be very interesting to understand wether this is really the case. 

%
\section{Conclusion}
\label{sec:conclusion}
%

The bootstrap approach to scattering amplitudes crucially relies on the knowledge of the appropriate function space. In this paper we have initiated the study of the relevant function space for seven- and especially eight-particle planar integrals relevant for amplitudes in $\N=4$ sYM. We showed that integrals contributing to the eight-point MHV amplitude depend on previously unknown algebraic letters. We argued that such letters may be needed not only for individual integrals, but even at the amplitude level, when going beyond the simplest MHV configuration.

One might worry that, with an increased number of legs, alphabets with a large number of letters could make it prohibitively complicated to pursue the bootstrap approach. While this does pose challenges when constructing the function space ansatz, we find that the constraints imposed by the differential equations satisfied by the integrals, are extremely powerful. For instance, the differential equations severely restrict the allowed final entries of the symbol. Our analysis, e.g. at seven points, showed that typically only a small number of free parameters survive in the ansatz after imposing such constraints. Another aspect of the differential equations is that they can provide valuable information about the function space itself. By studying their detailed properties, we could predict the necessity for previously unknown symbol letters. Finally, applying the differential constraints to a restricted bootstrap for the discontinuities of the integrals, we were able to recover the full result via dispersion relations. 

As mention in the text, the direct solution of the second-order differential equations can be quite difficult in general. It is worth pointing out however, that sometimes this can be achieved by a judicious choice of variables. See~\cite{Chicherin:2018ubl,Caron-Huot:2018dsv} for recent examples. In the context of the dual conformal integrals considered here and their higher-point, higher-loop generalizations, the novel twistor parameterizations proposed in \cite{Bourjaily:2018aeq} seem like a natural starting point to revisit a direct solution of the differential equations. We leave this for future work.

In this paper, we focused on massless, dual conformal integrals, but very similar questions arise in generic massless scattering amplitudes. There, the five-particle two-loop case is described by a $26$-letter planar alphabet \cite{Gehrmann:2015bfy}, and conjecturally by a $31$-letter non-planar alphabet \cite{Chicherin:2018ubl}. These functions are relevant for $2$ to $3$ scattering amplitudes in QCD. It would be interesting to apply the ideas discussed here to such integrals.

%
%
\subsection*{Acknowledgements}
%
%
We are especially grateful to Jake Bourjaily, Andrew McLeod, Matt von Hippel and Matthias Wilhelm for sharing with us some of their results and stimulating discussions prior to the publication of their parallel work \cite{Bourjaily:2018aeq}. We also thank Lance Dixon for a number of interesting comments and pointing out to us the simplicity of certain discontinuities of the six-particle double-pentagon integrals. Finally, we thank Zvi Bern and Michael Enciso for useful discussions and comments on the manuscript. E.H. thanks Falko Dulat and Jaroslav Trnka for stimulating conversations. E.H. is grateful to the Mani L. Bhaumik Institute for Theoretical Physics at UCLA for the kind hospitality during various stages of this project. J.M.H., E.H. and J.P.-M. thank the KITP for hospitality during the program `Scattering Amplitudes and Beyond', where this work was started. E.H. and J.P.-M. thank the theoretical high energy physics group at JGU Mainz for hospitality during the initial stages of this work. The research of E.H. was supported by the US Department of Energy under contract DE-AC02-76SF00515. J.P.-M. was supported by the U.S. Department of State through a Fulbright Scholarship, and by the Mani L. Bhaumik Institute for Theoretical Physics at UCLA. The work of J.M.H. was supported in part by the PRISMA Cluster of Excellence at Mainz university. This project has received funding from the European Research Council (ERC) under the European Union's Horizon 2020 research and innovation programme (grant agreement No 725110), ``Novel structures in scattering amplitudes''.

\newpage
\appendix

%
\section{Review of second-order differential operators}
\label{subsec:diff_mechanisms}
%

In this section, we summarize the relevant second-order differential equations for the Feynman integrals under consideration. This section is included as a convenient and concise reference for the relevant formulae 	pertinent for the remainder of our work. A comprehensive account with all the technical details can be found in~\cite{Drummond:2010cz}. Here we focus on the momentum twistor representation of the differential operators. Most of these operators are based on a careful analysis of one-loop pentagons which we turn to in the following. As a cross-check one can test the relevant differential equations on the known answer for the chiral pentagon integral~\cite{ArkaniHamed:2010gh}.

%
\subsection*{First mechanism}
\label{subsec:diff_mech_1}
%
The first mechanism for second-order differential operators is based on the chiral pentagon integral with a massless corner (4) between the two ``chiral''~legs (3 and 5),
$$
\overline\Psi^{(1)}(u,v) = 
\raisebox{-53pt}{
\includegraphics[scale=.5,trim={0cm 0cm 0cm 0cm},clip]{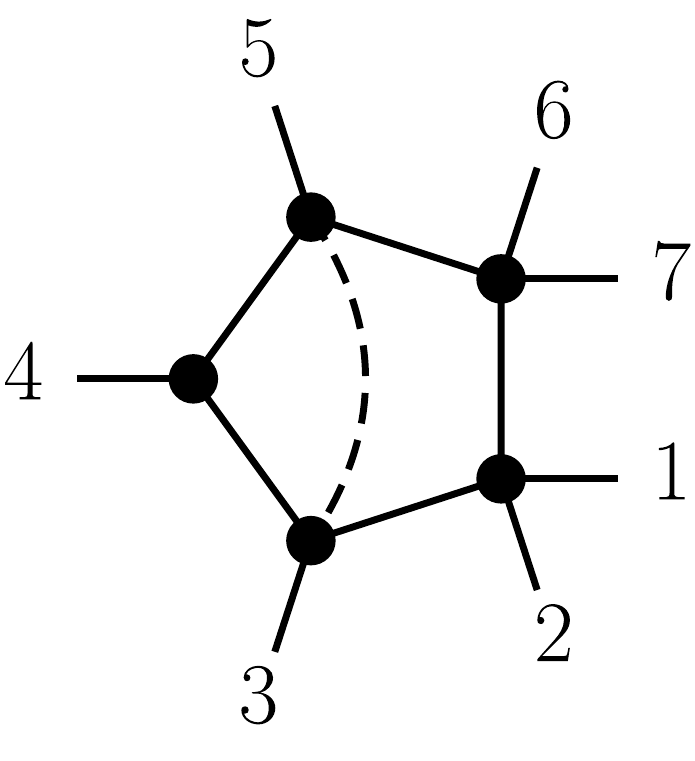}
}
= \int\limits_{(AB)} \frac{\ab{AB35}\ab{71(234)\cap(456)}}{\ab{AB71}\ab{AB23}\ab{AB34}\ab{AB45}\ab{AB56}}\,.
$$
This integral evaluates to a pure weight-two function that depends on two non-vanishing conformal cross-ratios $u$ and $v$,
\begin{align}
 u=\frac{\ab{7123}\ab{3456}}{\ab{7134}\ab{2356}},\qquad 
 v=\frac{\ab{7156}\ab{2345}}{\ab{7145}\ab{2356}}\,.
\end{align}
One can write down a Laplace-operator that acts on the dual point associated with the external line $(71)$. Taking into 
account some delta-function ``anomaly'' term, this translates to the following second-order partial differential equation 
for $\Psi^{(1)}(u,v)$,
\begin{align}
\label{eq:pent_anomaly}
	uv\partial_u\partial_v \Psi^{(1)}(u,v)=1\,.
\end{align}
This operator will play a crucial role in a moment. For later convenience, let us introduce a shorthand notation for first order differential operators in momentum twistor language,
\begin{align}
 O_{ij} \equiv Z_i \cdot \frac{\partial}{\partial Z_j}\,,
\end{align}
which acts on four-brackets according to the simple rule $O_{ij}\ab{jklm} = \ab{iklm}$. First, we look at one particular part of the pentagon \emph{integrand}, $\frac{\ab{AB35}}{\ab{AB23}\ab{AB34}\ab{AB45}}$ and act with $O_{23}$. The $\ab{AB23}$ factor is inert because it only depends on the line $(23)$ which is annihilated by the twistor derivative $O_{23}\ab{AB23}=\ab{AB22}=0$. The remaining part is more interesting,
\begin{align}
O_{23}\frac{\ab{AB35}}{\ab{AB34}\ab{AB45}} 
		 &= \frac{1}{\ab{AB45}}O_{23}\frac{\ab{AB35}}{\ab{AB34}}
		 =\frac{\ab{AB23}}{\ab{AB34}^2}\,,
\end{align}
where we used the Schouten identity. This expression now depends on twistor 4 only in the combination of line $(34)$. Acting with another differential operator $O_{34}$ now naively annihilates the expression. For the pentagon integral at hand, this analysis suggests to consider the following operator
\begin{align}
	\tw{O}_{234}\equiv N_{p1}O_{34}O_{23}N^{-1}_{p1},\qquad N_{p1}=\ab{71(234)\cap(456)}=\ab{456[1}\ab{7]234}
\end{align} 
that annihilates the pentagon integrand,
\begin{align}
 \tw{O}_{234} \frac{\ab{AB35}\ab{71(234)\cap(456)}}{\ab{AB71}\ab{AB23}\ab{AB34}\ab{AB45}\ab{AB56}} = 0\quad  (\text{naive})
\end{align}
Writing the twistor operator in terms of the dual cross ratios, one finds
\begin{align}
 \tw{O}_{234}\Psi^{(1)}(u,v) = - \frac{\ab{7135} N_{p1}}{\ab{7145}\ab{7134}\ab{3456}} 
													\underbrace{u\partial_u v\partial_v \Psi^{(1)}(u,v)}_{=1,\ \text{eq}.~(\ref{eq:pent_anomaly})}\,.
\end{align}
Stripping off the normalization factor, this equation can now be used whenever there is a pentagon subintegral with a massless corner inbetween the chiral legs,
\begin{align}
 O_{34}O_{23} \int\limits_{(AB)} \frac{\ab{AB35}}{\ab{AB71}\ab{AB23}\ab{AB34}\ab{AB45}\ab{AB56}}
				= -  \frac{\ab{7135}}{\ab{7145}\ab{7134}\ab{3456}} \,.
\end{align}
%
\subsection*{Second mechanism}
\label{subsec:diff_mech_2}
%
The second mechanism that gives rise to second-order differential equations is based on the chiral pentagon integral where the corner between the two chiral legs (3 and 6) is massive ($4+5$),
\begin{align}
\label{fig:1loop_pentagon_massive_corner}
\overline\Psi^{(1)}(u,v,w) = 
\hskip -.3cm
\raisebox{-55pt}{
 \includegraphics[scale=.5,trim={0cm 0cm 0cm 0cm},clip]{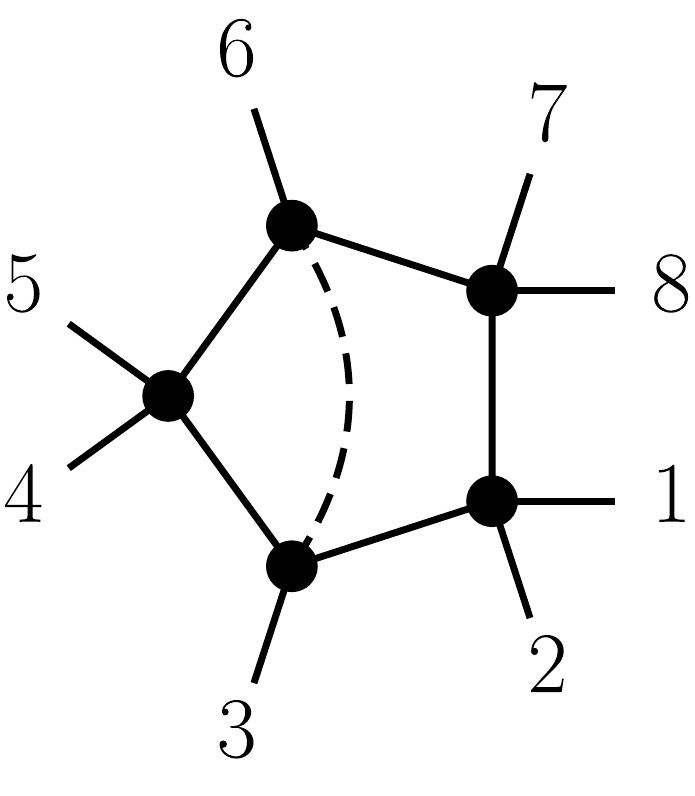}
}
\hskip -.3cm
= \int\limits_{(AB)} \!\!\!\! \frac{\ab{AB36}\ab{81(234)\cap(567)}}{\ab{AB81}\ab{AB23}\ab{AB34}\ab{AB56}\ab{AB67}}
\end{align}
Compared to $\overline\Psi^{(1)}(u,v)$ of the previous subsection, this integral has one additional scale and depends on the three cross ratios denoted by,
\begin{align}
	u = \frac{\ab{8123}\ab{3467}}{\ab{8134}\ab{2367}}\,, \quad
	v = \frac{\ab{8167}\ab{2356}}{\ab{8156}\ab{2367}}\,, \quad
	w = \frac{\ab{2367}\ab{3456}}{\ab{2356}\ab{3467}}\,.
\end{align}
Defining the modified normalization factor, $N_{p2}=\ab{81(234)\cap(567)} = \ab{567[8}\ab{1]234}$, there are two twistor operators that annihilate the integrand,
\begin{align}
 N_{p2}O_{24}O_{42}N^{-1}_{p2}\,, \qquad 
 N_{p2}O_{75}O_{57}N^{-1}_{p2}\,.
\end{align}
For this mechanism, there is no anomaly associated to the differential operators. Let us emphasize that this operator can only be used for higher loop integrals in specific situations. $O_{42}$ depends on twistor 2 so that the mechanism discussed here only applies if the integral under consideration does \emph{not} depend on twistor 2 anywhere else. For two-loop integrals, this is only true if there are \emph{at least} two additional legs attached at the joint of the double-pentagon integral. In particular, this mechanism can not be used in any of the seven-point examples and would only be required for a very special eight-point example where a massive leg is attached to the joint of the six-point double-pentagon. We leave the discussion of this particular leg-configuration to future work.

%
\subsection*{Third mechanism}
\label{subsec:diff_mech_3}
%
The third mechanism is also based on the same chiral pentagon integral from the previous subsection, where the corner between the two chiral legs is massive (\ref{fig:1loop_pentagon_massive_corner}). In contrast to the case described in (\ref{subsec:diff_mech_2}), the differential equation we introduce now is not subject to any leg-range restrictions at the joint of higher-loop integrals. 
\begin{align}
	O_{24}O_{75} \frac{\ab{AB36}\ab{81(234)\cap(567)}}
										{\ab{AB81}\ab{AB23}\ab{AB34}\ab{AB56}\ab{AB67}} 
		= N_{p2} \frac{\ab{AB36}}{\ab{AB81}\ab{AB34}^2\ab{AB56}^2}
\end{align}
This differential operator, $O_{24}O_{75}$, cancels two propagators and commutes with the normalization $N_{p2}=\ab{81(234)\cap(567)}$. The resulting integral is finite and dual conformal invariant. Since it only depends on three dual points for which there are no conformal cross ratios, it has to evaluate to a rational function. One can compute this rational function to find,
\begin{align}
O_{24}O_{75}\overline{\Psi}^{(1)}(u,v,w) \propto N_{p2} \frac{\ab{3681}}{\ab{8134}\ab{8156}\ab{3456}} 
\end{align}
which can be checked by acting with $O_{24}O_{75}$ on the known pentagon expression \cite{ArkaniHamed:2010gh}.

%
%
%

%
\section{Seven- and eigth-point double pentagon integrals
}
\label{sec:seven_and_eight_point_differential_euqations}
%

In earlier work~\cite{Drummond:2010cz}, some of the differential equations described in the previous section were used to obtain the six-point double-pentagon integral $\Omega^{(2)}_6$ directly. We will not discuss this well understood six-point example in detail, but it is worth recalling that the identification of the $6d$ hexagon integral as pure weight 3 function in the intermediate step of the chain of first order differential equations was extremely helpful
$$
  \raisebox{-47pt}{\includegraphics[scale=.5]{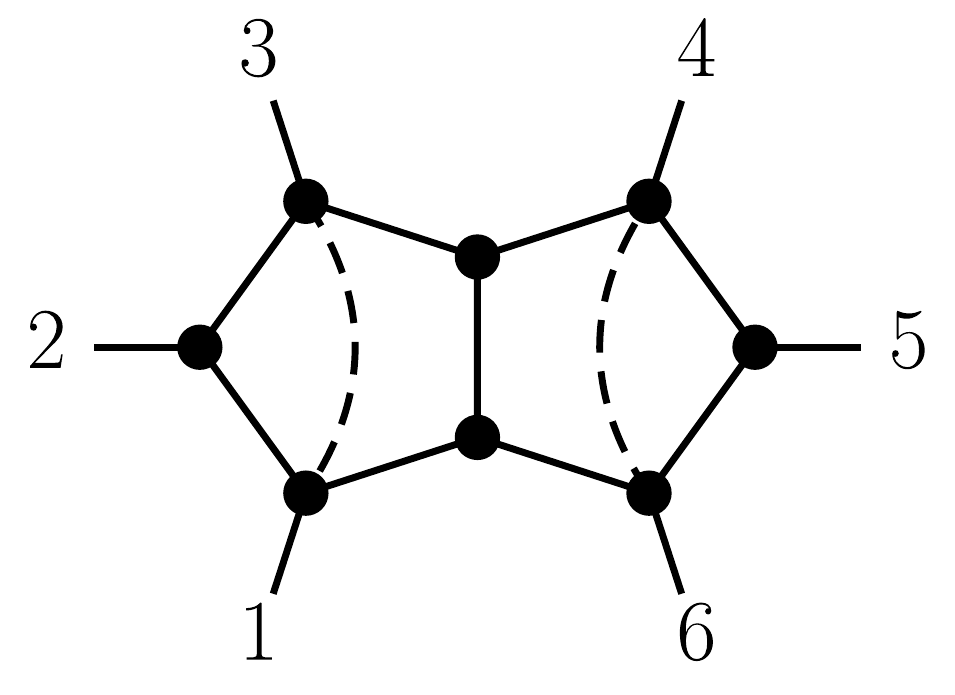}} 
   \quad \overset{D^{(1)}_1}{\longrightarrow} \quad  \raisebox{-46pt}{\includegraphics[scale=.5]{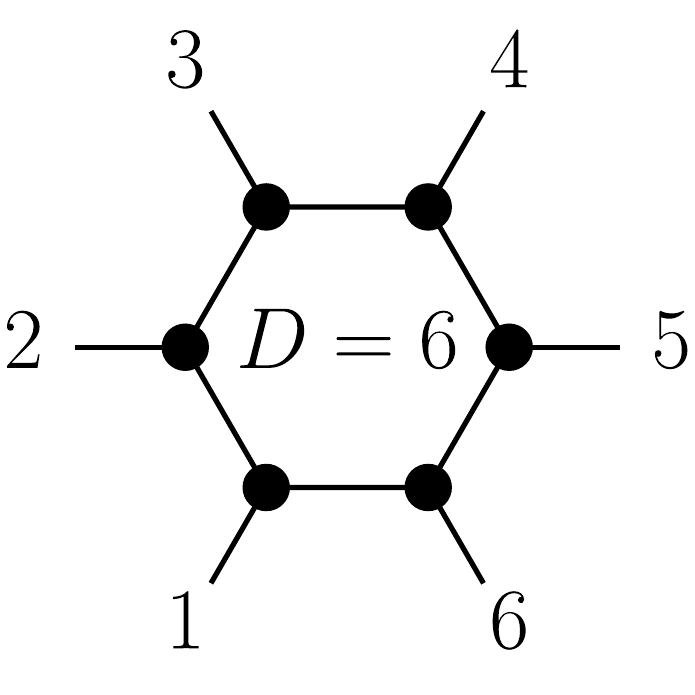}}
   \quad \overset{D^{(1)}_2}{\longrightarrow} \quad  \raisebox{-46pt}{\includegraphics[scale=.5]{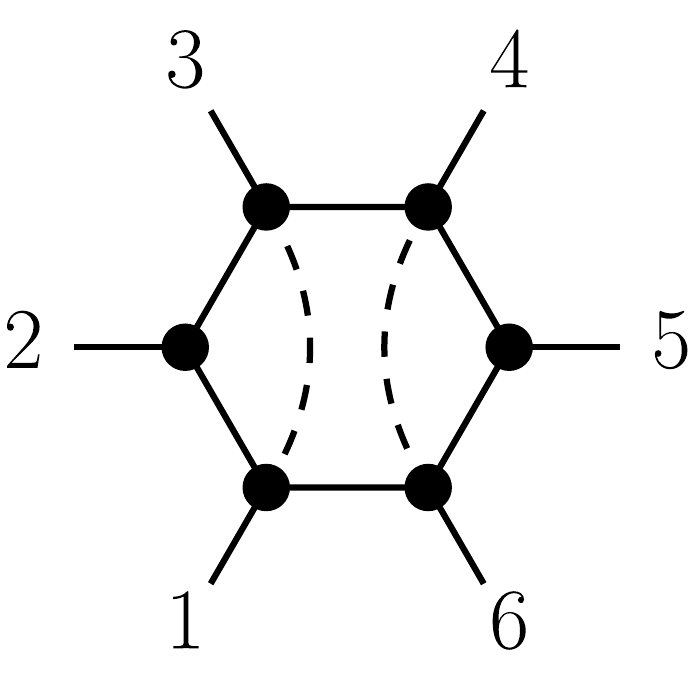}}
 $$
For higher points, the factorization of the twistor differential operators still holds, however, the direct connection to the scalar hexagon integrals does not survive generically. This severely complicates the direct solution of the differential equation. In most higher point cases, we do not have access to sufficiently many simple differential equations to control all directions of the kinematic space. This lack of differential knowledge can in principle be counterbalanced by the knowledge of sufficient boundary data. This however, requires good analytic control of the intermediate weight three function which proofs difficult in the absence of a local integral representation like the one available at six points.
%
%
\subsection{Differential equations for seven-point integrals}
%
%
In comparison to the three dual-conformal invariant cross-ratios $u,v$ and $w$ relevant for six points, the generic kinematic dependence for seven particles is six-dimensional ($6=3\times7-15$). However, a particular Feynman integral might not live in the most general heptagon-space but on a certain lower-dimensional subspace. This restricted kinematic dependence can be related to the existence of a certain number of independent, commuting, first-order differential operators that annihilate the integral \cite{Dixon:2013eka}.
The two seven-point generalizations of $\Omega^{(2)}_6$ relevant for MHV amplitudes are
$$
\Omega^{(2)}_{7,a} = \raisebox{-47.5pt}{\includegraphics[scale=.5]{./figures/2loop_7pt_dp_conf_b.pdf}}\,,
\qquad
\Omega^{(2)}_{7,b} = \raisebox{-47.5pt}{\includegraphics[scale=.5]{./figures/2loop_7pt_dp_conf_a.pdf}}\,.
$$
%
%
\subsubsection*{Differential equations for \texorpdfstring{$\Omega^{(2)}_{7,a}$}{Omega(2)7a}}
%
%
We begin by discussing the seven-point double-pentagon, $\Omega^{(2)}_{7,a}$. The variable count for this integral is relatively straightforward: it only depends on six dual points, of which five consecutive ones are lightlike separated. Starting from the three-variable $\Omega^{(2)}_6$ integral, the fourth kinematic variable is associated with the mass of the off-shell leg. Alternatively, one can start from the full six dimensional space and realize that this integral can depend on twistors 2 and 3 only via the lines $(12)$ and $(34)$ respectively. This statement is equivalent to the statement that the two independent first order differential operators $O_{12}$ and $O_{43}$ annihilate the integral which leads to a restriction on the kinematic dependence to $4=6-2$ variables.
The chiral numerator of $\Omega^{(2)}_{7,a}$ is given by \vskip -.75cm
\begin{align}
  N^{\text{dp}}_{7,a} = \ab{AB14}\ab{CD57} \overbrace{\ab{(712)\cap(345),(456)\cap(671)}}^{
	=-\ab{4571}\ab{3456}\ab{6712}\equiv \bar{N}}\,.
\end{align}

Making use of the one-loop differential equations described in Appendix~\ref{subsec:diff_mechanisms} for individual sub-pentagon integrals, we can derive one second-order differential equation (operator $\O_1$) that follows mechanism 3, and two differential equations (represented by operators $\O_2$ and $\O_3$) utilizing mechanism 1
\begin{align}
\label{eq:diff_ops_heptagon_7b}
        \O_1 = \bar{N} O_{72}O_{53} \bar{N}^{-1}\,, \qquad 
	\O_2 = \bar{N} O_{76}O_{17} \bar{N}^{-1}\,, \qquad
	\O_3 = \bar{N} O_{56}O_{45} \bar{N}^{-1}\,.
\end{align}
Note that the normalization factor $\bar{N}$ commutes with $O_{72}O_{53}$, so it could be dropped in the definition of $\O_1$. The differential equations look schematically as follows
\begin{align}
\label{eq:double_pent_7pt_a_diff_eq}
D^{(2)} \quad  \raisebox{-47.5pt}{\includegraphics[scale=.5]{./figures/2loop_7pt_dp_conf_b.pdf}} \quad= \quad  \raisebox{-47.5pt}{\includegraphics[scale=.5]{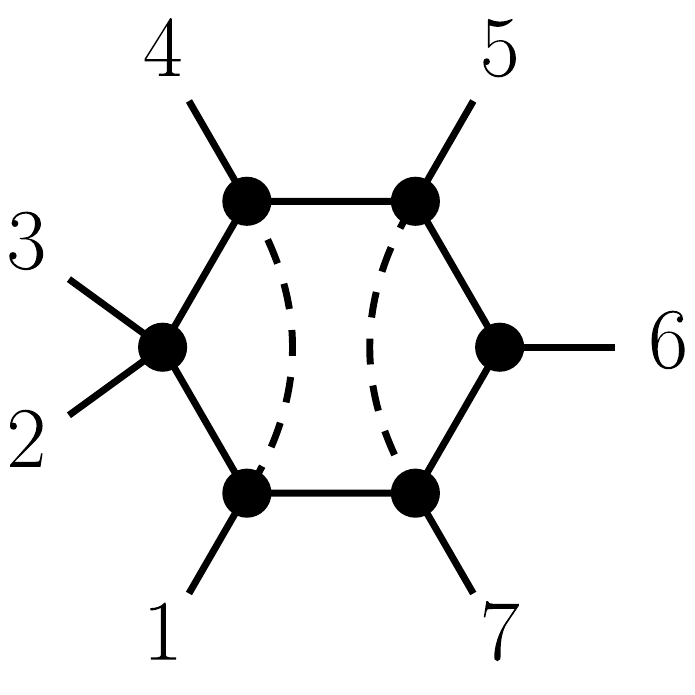}}\,.
 \end{align}
All three three differential operators on the double-pentagon $\Omega^{(2)}_{7,a}$ always produce the following unit-leading singularity chiral hexagon,
\begin{align}
\label{eq:ULS_chi_hex_7pt_23_massive}
\I^{(1),\text{ULS}}_{\chi\text{-h},(23)} = \,\,
\hskip -.3cm
\raisebox{-47.5pt}{
\includegraphics[scale=.5]{./figures/1loop_chiral_hex_4d_23_massive.pdf}}
\ = \int\limits_{(AB)}\!\!\! \frac{\ab{3456}\ab{6712}\ab{AB14}\ab{AB57}}
{\ab{AB12}\ab{AB34}\ab{AB45}\ab{AB56}\ab{AB67}\ab{AB71}}\,,
\end{align}
dressed with some rational prefactor, as inhomogeneous term:
\begin{align}
\label{eq:diff_ops_dp_b}
	\O_1\Omega^{(2)}_{7,a}\!\!  =\! \sm\frac{\ab{4571}}{\ab{1234}} \I^{(1),\text{ULS}}_{\chi\text{-h},(23)}\,,  \
	\O_2\Omega^{(2)}_{7,a}\!\! =\! \sm\frac{\ab{4571}}{\ab{4567}} \I^{(1),\text{ULS}}_{\chi\text{-h},(23)}\,, \
	\O_3\Omega^{(2)}_{7,a}\!\! =\! \sm\frac{\ab{4571}}{\ab{5671}} \I^{(1),\text{ULS}}_{\chi\text{-h},(23)}\,.
\end{align}
We can bring the rational prefactor to the left-hand side and define some modified differential operators which are projectively invariant in all momentum twistors.
In addition all second-order operators remain their factorized form in terms of a product of two projectively invariant first order differential operators. 
\begin{align}
\label{eq:scaled_diff_ops_dp_b}
\widetilde{\O}_1& = - \frac{\ab{1234}}{\ab{4571}} \O_1 = - \left(\frac{\ab{1234}}{\ab{7134}}O_{72}\right)\left(\frac{\ab{7134}}{\ab{7145}}O_{53}\right)\,,\\
\widetilde{\O}_2& = - \frac{\ab{4567}}{\ab{4571}} \O_2 = -\left(\frac{\ab{67X1}\ab{3456}}{\ab{67X3}\ab{4571}}O_{76}\right)\left(\frac{\ab{67X3}\ab{4567}}{\ab{67X1}\ab{3456}}O_{17}\right)
\end{align}
The third operator $\widetilde{\O}_3$ is related to $\widetilde{\O}_2$ by the flip symmetry of the integral. In order to reach this projectively invariant form we have multiplied by a special combination
\begin{align}
\label{eq:insert_unity_7pt_diff_op}
1 = \frac{\ab{67X1}}{\ab{67X3}} \frac{\ab{67X3}}{\ab{67X1}}
\end{align}
in which each factor commutes with $O_{67}$ for an arbitrary twistor $X$ and guarantees that both first order differential operators are neutral under rescaling the twistors (taken into account that $O_{ij}\equiv Z_i\cdot \frac{\partial}{\partial Z_j}$). For practical purposes it seems reasonable to choose $X=2$ so that all brackets appearing in (\ref{eq:insert_unity_7pt_diff_op}) are of the form $\ab{ii\sp1jj\sp1}$.
The advantage of rewriting the twistor differential operators in a projectively invariant way is that one can now express them in terms of more suitable coordinates. These can be conformal cross-ratios or a minimal parametrization of the momentum twistors.

%
%
\subsubsection*{Differential equations for \texorpdfstring{$\Omega^{(2)}_{7,b}$}{Omega(2)7b}}
%
%
The second integral we are going to discuss here is denoted by $\Omega^{(2)}_{7,b}$ and corresponds to the kinematic configuration, where the additional leg is attached to the joint of the two-loop double-pentagon. This integral depends on seven dual points and lives in the full six-dimensional heptagon kinematic space. 
The chiral numerator of this integral is given by \vskip -.75cm
\begin{align}
  N^{\text{dp}}_{7,b} = \ab{AB24}\ab{CD57} \overbrace{\ab{(123)\cap(345),(456)\cap(671)}}^{\equiv \bar{N}}
\end{align}
Employing the general mechanisms described in the previous section, there are four second-order differential operators that are related by the symmetry of the integral and follow from section \ref{subsec:diff_mech_1},
\begin{align}
\label{eq:diff_eqs_omega_7a}
 \O_1 \! =\!  \bar{N} O_{23}O_{12} \bar{N}^{-1}\!\!\!\!\,, \hskip .3cm
 \O_2 \! =\!  \bar{N} O_{76}O_{17} \bar{N}^{-1}\!\!\!\!\,, \hskip .3cm
 \O_3 \! =\!  \bar{N} O_{43}O_{54} \bar{N}^{-1}\!\!\!\!\,, \hskip .3cm 
 \O_4 \! =\!  \bar{N} O_{56}O_{45} \bar{N}^{-1}
\end{align}
We look at the action of operator $\O_1$ on the left pentagon,
\begin{align}
 O_{23}O_{12}\!\!\! \int\limits_{(AB)}\!\!\! \frac{\ab{AB24}}{\ab{AB12}\ab{AB23}\ab{AB34}\ab{AB45}\ab{ABCD}}
	= \frac{-\ab{CD24}}{\ab{CD34}\ab{CD23}\ab{2345}}
\end{align}
so that $\O_1$ on $\Omega^{(2)}_{7,b}$ is
\begin{align}
\O_1 \Omega^{(2)}_{7,b} = - \frac{\bar N}{\ab{2345}} 
					\int\limits_{(CD)} \frac{\ab{CD24}\ab{CD57}}
					{\ab{CD23}\ab{CD34}\ab{CD45}\ab{CD56}\ab{CD67}\ab{CD71}}
\end{align}
The equations have the following structure
\begin{align}
\label{eq:double_pent_7pt_b_diff_eq}
D^{(2)} \quad  \raisebox{-47.5pt}{\includegraphics[scale=.5]{./figures/2loop_7pt_dp_conf_a.pdf}} \quad= \quad  \raisebox{-48.5pt}{\includegraphics[scale=.5]{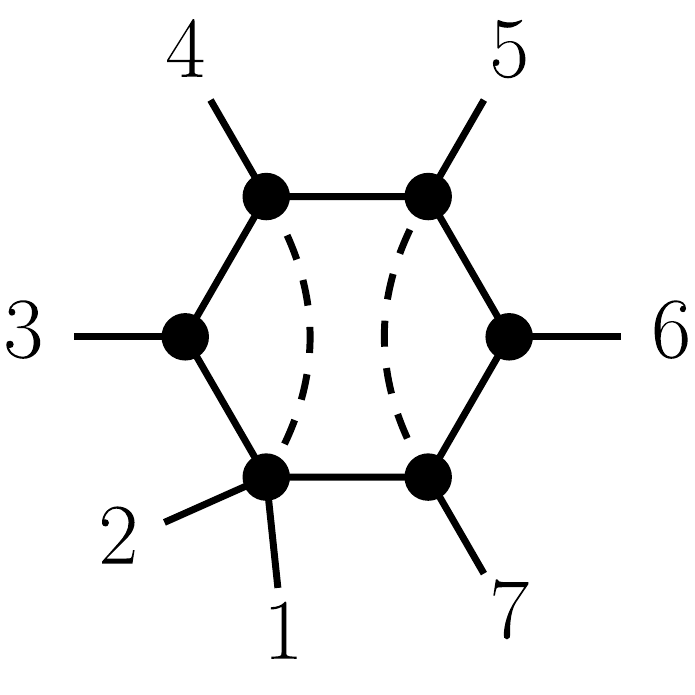}}\,.
\end{align}
The inhomogeneous term can be identified with a particular chiral hexagon integral, 
$$
I^{(1)}_{\chi\text{-h},(12)}\! =\! \!
\raisebox{-55pt}{
\includegraphics[scale=.55]{./figures/1loop_chiral_hex_4d_12_massive.pdf}
}
\!\!
=\!\int\limits_{(CD)} \!\!\!\frac{\ab{CD24}\ab{CD57}}
					{\ab{CD23}\ab{CD34}\ab{CD45}\ab{CD56}\ab{CD67}\ab{CD71}}
$$
which is not yet correctly normalized and carries little group weight in the individual twistors. Furthermore, one can check that this chiral hexagon, even with an additional normalization factor, does not have unit leading singularities. Instead, the leading singularities are proportional to different rational prefactors. With the definition of $I^{(1)}_{\chi\text{-h},(12)}$ above, we find three different leading singularities,
\begin{align}
\label{eq:chi_hex_12_massive_LS_def}
 \text{LS}_A\! =\! \frac{1}{\ab{7136}\ab{4536}}\,, 
 \text{LS}_B\! =\! \frac{\ab{4526}}{\ab{45(716)\!\cap\!(623)}\ab{4536}}\,,
 \text{LS}_C\! =\! \frac{\ab{7126}}{\ab{71(456)\!\cap\!(623)}\ab{7136}}
\end{align}
which satisfy one simple linear relation,
\begin{align}
\label{eq:chi_hex_12_massive_LS_rel}
	\text{LS}_A = \text{LS}_B+\text{LS}_C
\end{align}
In order to obtain the \emph{integrated} result for this one-loop chiral hexagon one can use generalized unitarity to express it in terms of a box expansion (parity-odd terms that integrate to zero are dropped). 
\begin{align*}
&\raisebox{-50pt}{
\includegraphics[scale=.5]{./figures/1loop_chiral_hex_4d_12_massive.pdf}
} = \text{LS}_A 
   \raisebox{-42pt}{\includegraphics[scale=.5, trim= 5 0 15 0]{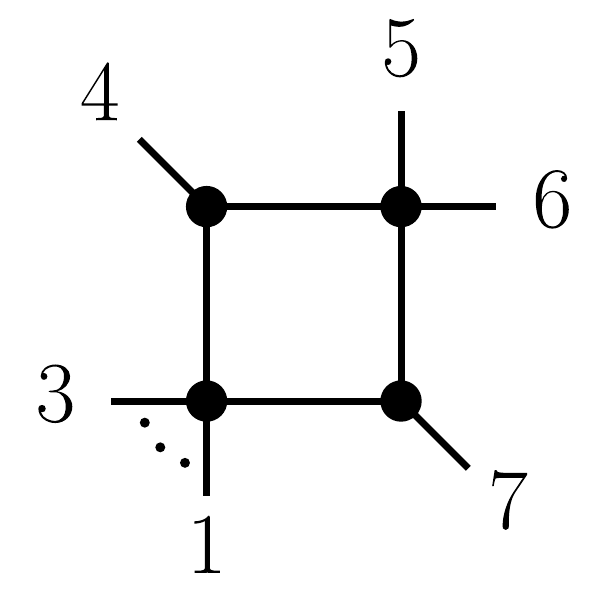}}
  + \text{LS}_A \raisebox{-42pt}{\includegraphics[scale=.5, trim= 5 0 25 0]{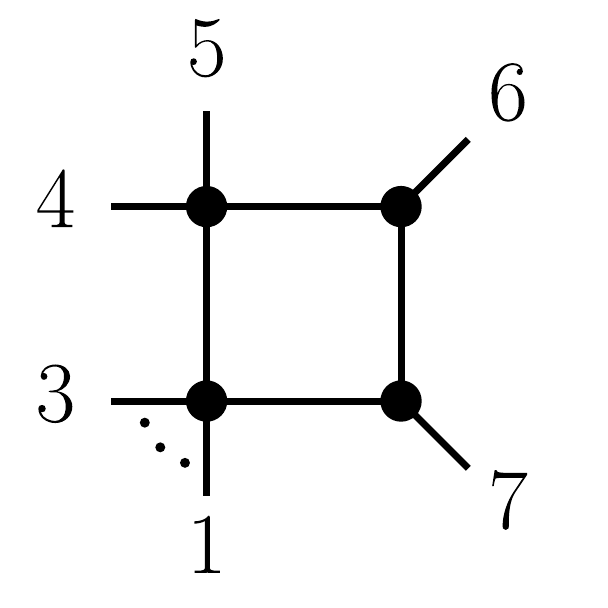}}
  +\text{LS}_A \raisebox{-42pt}{\includegraphics[scale=.5, trim= 5 0 25 0]{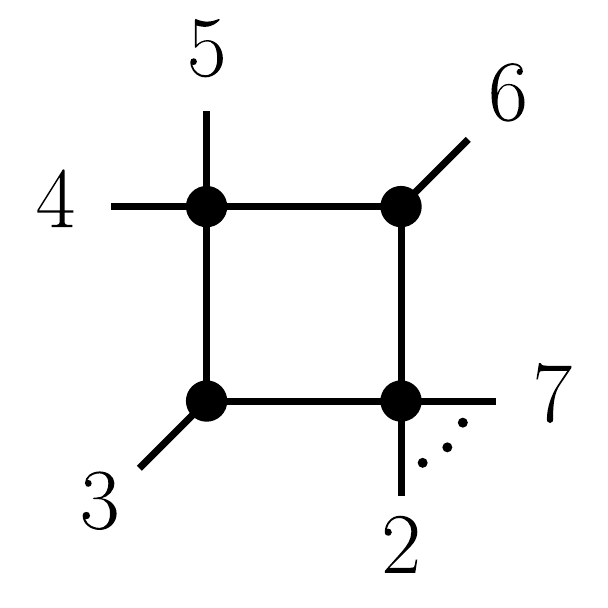}}
\\[-10pt] &  
\hspace{5pt}+ \text{LS}_A\raisebox{-42pt}{\includegraphics[scale=.5, trim= 5 0 25 0]{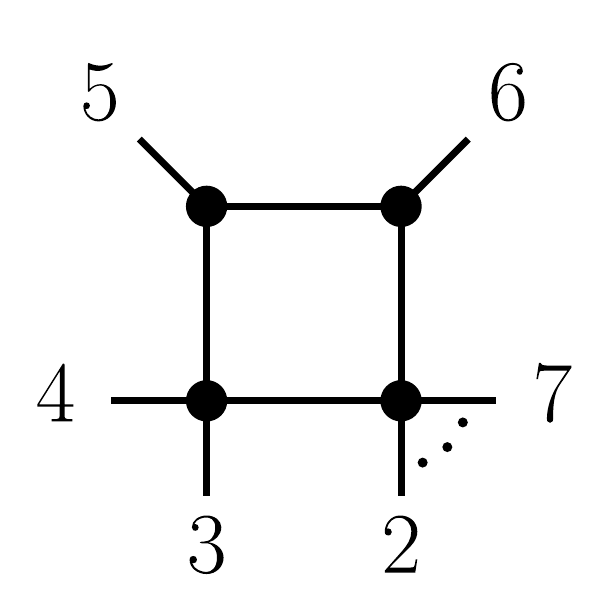}} 
  - \text{LS}_A\raisebox{-42pt}{\includegraphics[scale=.5, trim= 5 0 25 0]{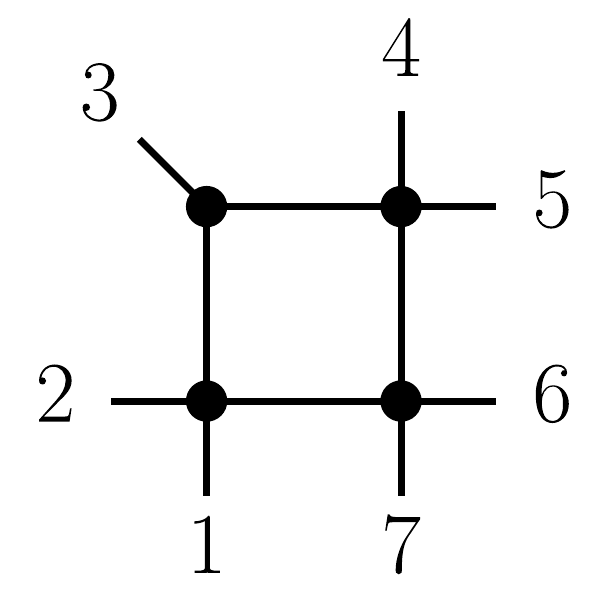}}
  - \text{LS}_A\raisebox{-42pt}{\includegraphics[scale=.5, trim= 5 0 25 0]{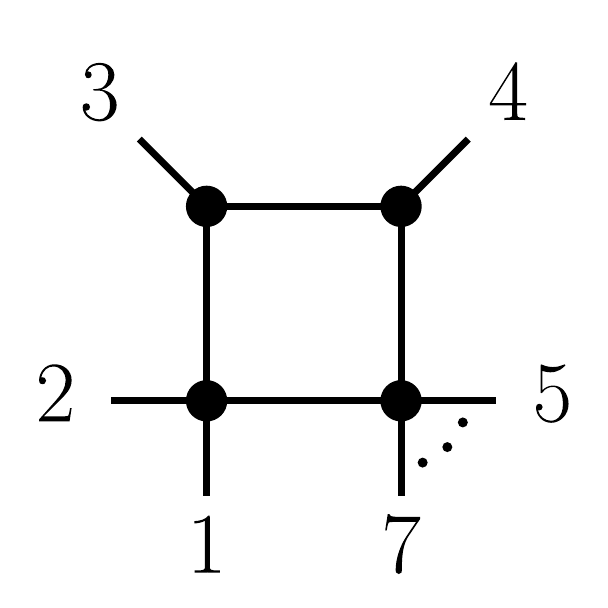}}
+\text{LS}_B 
   \raisebox{-42pt}{\includegraphics[scale=.5, trim= 5 0 25 0]{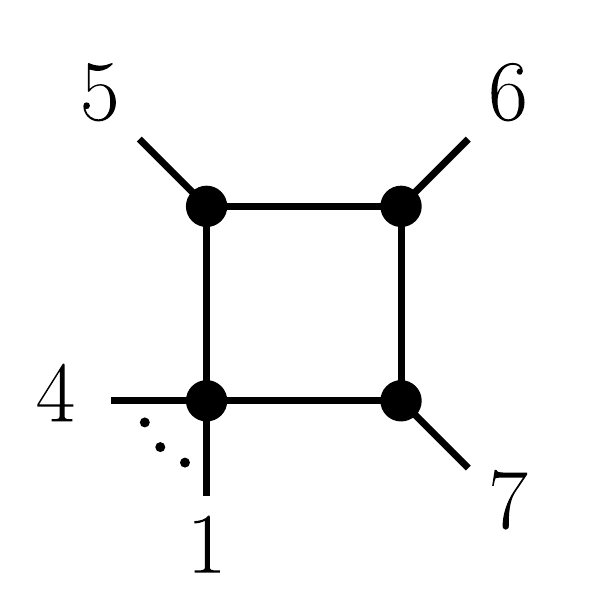}}
 \\
&\hspace{5pt}+\text{LS}_B 
  \raisebox{-42pt}{\includegraphics[scale=.5, trim= 5 0 25 0]{figures/1loop_7pt_box_34_5_6_712.pdf}}
  -\text{LS}_B 
\raisebox{-42pt}{\includegraphics[scale=.5, trim= 5 0 25 0]{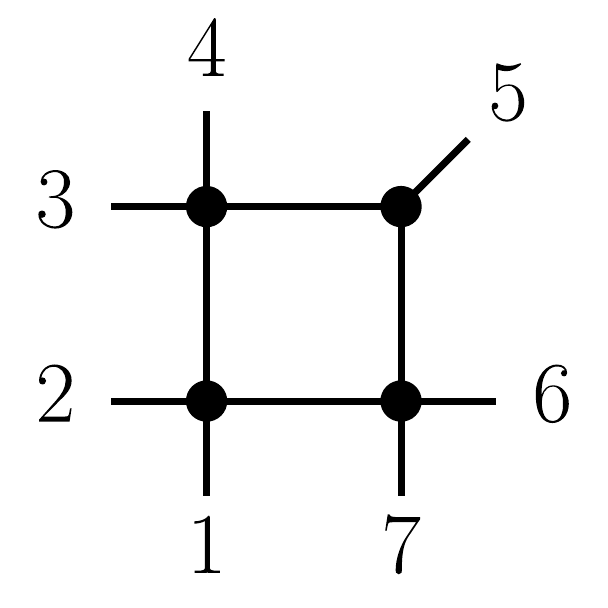}} 
 +\text{LS}_C 
   \raisebox{-42pt}{\includegraphics[scale=.5, trim= 5 0 25 0]{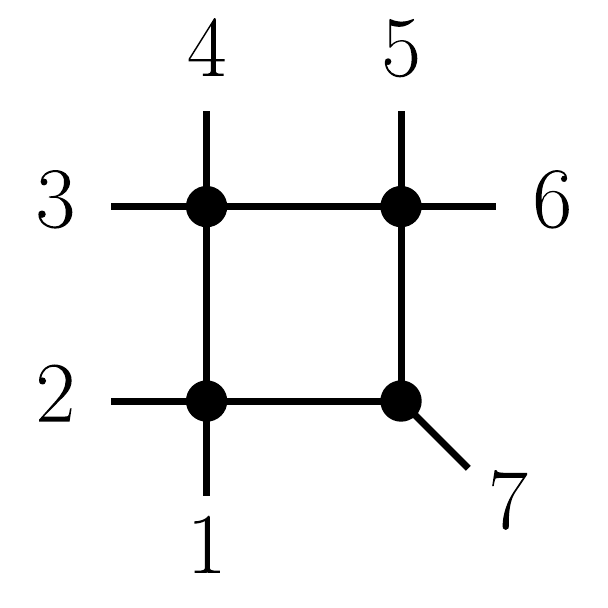}}
 +\text{LS}_C 
  \raisebox{-42pt}{\includegraphics[scale=.5, trim= 5 0 25 0]{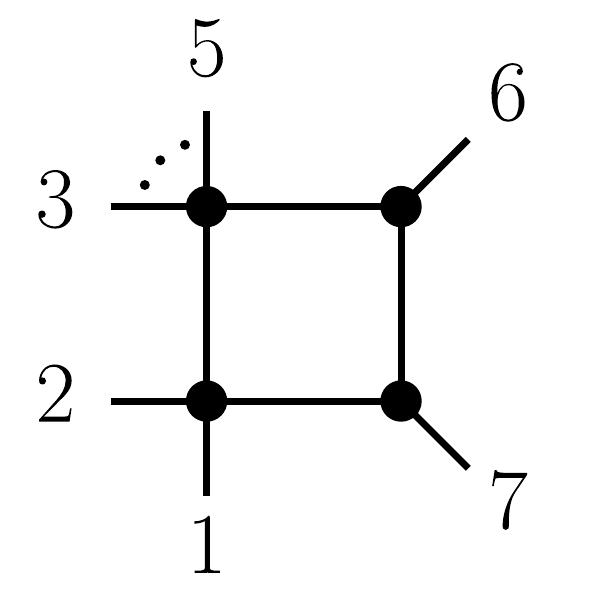}}
\end{align*}
One can now use relation (\ref{eq:chi_hex_12_massive_LS_rel}) between the leading singularities to eliminate 
$\text{LS}_A$ and define two pure weight two functions,
\begin{align}
	I^{(1)}_{\chi\text{-h},(12)}= \text{LS}_B\ \omega^{(2)}_{B} + \text{LS}_C\ \omega^{(2)}_{C}
\end{align}
where $\omega^{(2)}_{B}$ and $\omega^{(2)}_{C}$ are pure weight two functions. We insert the known expressions of the 1-loop box integrals (we find it convenient to use the DCI-regulated expressions of \cite{Bourjaily:2013mma}) into the unitarity decomposition in order to obtain analytic results for $\omega^{(2)}_{B,C}$. Their symbols, written in terms of the heptagon alphabet introduced in section \ref{subsec:steinmann_bootstrap_heptagon}, are included in ancillary files attached to this submission.

%
\subsection{Differential equations for eight-point integrals}
%
%
The two eight-point integrals we will that we will be analyzing in detail are the following
\begin{align}
  &\Omega^{(2)}_{8,a} =\raisebox{-47.5pt}{\includegraphics[scale=.5, trim=16pt 0 0 0]{./figures/2loop_8pt_dp_conf_b.pdf}}\,, \qquad
\Omega^{(2)}_{8,b} = \raisebox{-47.5pt}{\includegraphics[scale=.5, trim=5pt 0 0 0]{./figures/2loop_8pt_dp_conf_a.pdf}}\,.
\end{align}
%
%
\subsubsection*{Differential equations for \texorpdfstring{$\Omega^{(2)}_{8,a}$}{Omega(2)8a}}
%
%

First we focus on the integral $\Omega^{(2)}_{8,a}$ where the additional legs provide two new massive corners. 
The chiral numerator for this double-pentagon integral is,
 \begin{align}
 \label{eq:norm_dp_2_8b}
 N^{\text{dp}}_{8,a} = \ab{AB25}\ab{CD61} \overbrace{\ab{(123)\cap(456),(567)\cap(812)}}^{\equiv \bar{N}_b=-\ab{1256}\ab{8123}\ab{4567}}
 \end{align}
This integral depends on five kinematic variables. 
There are four second-order differential operators for $\Omega^{(2)}_{8,a}$ with the schematic form
\begin{align}
\label{eq:double_pent_8pt_b_diff_eq}
D^{(2)}_2 \quad  \raisebox{-47.5pt}{\includegraphics[scale=.5]{./figures/2loop_8pt_dp_conf_b.pdf}} = \quad  \raisebox{-47.5pt}{\includegraphics[scale=.5]{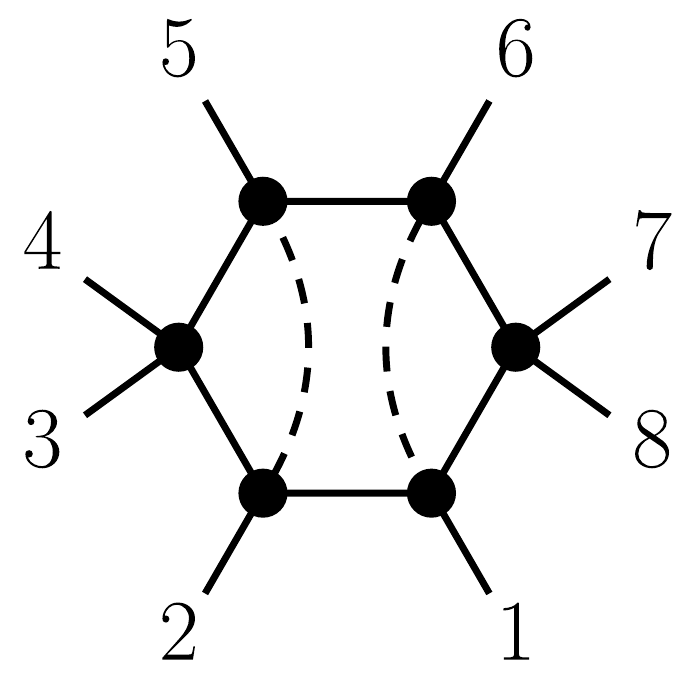}}\,.
 \end{align}
There are two second-order differential operators acting on this integral, which are related by the $Z_2$-symmetry of the integral. The differential operators follow from the one-loop discussion in subsection \ref{subsec:diff_mech_3},
 \begin{align*}
  \O_1 = O_{13}O_{64}\,, \qquad 
  \O_2 = O_{28}O_{57}\,.
 \end{align*}
Both differential operators commute with the numerator factor $\bar{N}_b$. Focussing on $\O_1$ for concreteness, the second-order differential equation reads,
\begin{align}
  \O_1 \Omega^{(2)}_{8,a} = -\frac{\ab{1256}}{\ab{2345}} \int \limits_{(CD)} \frac{\ab{CD25}\ab{CD61}\ab{8123}\ab{4567}}
 									           {\ab{CD81}\ab{CD12}\ab{CD23}\ab{CD45}\ab{CD56}\ab{CD67}}\,,
\end{align}
which can be written in a factorized form where both first order differential operators are individually little group neutral,
\begin{align}
\label{eq:diff_eq_omega_2_8b}
 \left[\frac{\ab{2345}}{\ab{1245}} O_{13}\right]\left[\frac{\ab{1245}}{\ab{1256}}O_{64}\right] \Omega^{(2)}_{8,a} = - \I^{(1),\chi\text{-h}}_{(34)(78)}\,. 
\end{align}
The chiral hexagon integral on the right hand side of the differential equation (\ref{eq:double_pent_8pt_b_diff_eq})
\begin{align}
\label{eq:chi_hex_8_34_78_massive}
\I^{(1),\chi\text{-h}}_{(34)(78)} & = \raisebox{-50pt}{\includegraphics[scale=.5]{./figures/1loop_chiral_hex_34_78_massive.pdf}}
				     =  \int \limits_{(CD)} \frac{\ab{CD25}\ab{CD61}\ab{8123}\ab{4567}}
 									           {\ab{CD81}\ab{CD12}\ab{CD23}\ab{CD45}\ab{CD56}\ab{CD67}}
\end{align}
does not have unit-leading singularities but contains one algebraic prefactor 
\begin{align}
\label{eq:gamma_prefactor_4mass_box}
 \gamma = \frac{\sqrt{(1-u_{1}-v_{1})^2-4u_{1}v_{1}}}{1-u_{1}-v_{1}}\,, \
  u_{1} = \frac{\ab{8123}\ab{4567}}{\ab{8145}\ab{2367}}\,, \
  v_{1} = \frac{\ab{2345}\ab{6781}}{\ab{8145}\ab{2367}}\,,
\end{align}
coming from the four-mass-box contribution when written in terms of a box expansion. This chiral hexagon integral has been discussed in~\cite{ArkaniHamed:2010gh} (see JHEP-version for corrected result) and serves as source term in the differential equation (\ref{eq:double_pent_8pt_b_diff_eq}). The knowledge of this integral allows us to match the inhomogeneous terms in our bootstrap inspired approach either as constraints on a general ansatz or as a cross check of the result.

%
\subsubsection*{Differential equations for \texorpdfstring{$\Omega^{(2)}_{8,b}$}{Omega(2)8b}}
%
For the second eight-point two-loop configuration, $\Omega^{(2)}_{8,b}$, the additional legs are attached to the central vertices where both loops meet.
Its unit-leading singularity chiral numerator for this integral is given by 
 \begin{align}
   N^{\text{dp}}_{8,b} = \ab{AB24}\ab{CD68} \overbrace{\ab{(123)\cap(345),(567)\cap(781)}}^{\equiv \bar{N}_a}\,,
 \end{align}
and it depends on the full nine-dimensional kinematic space of dual conformal eight-point scattering 

As for the lower-point cases, we use the mechanisms for the massless pentagons in order to derive four second-order differential operators for $\Omega^{(2)}_{8,b}$
 \begin{align*}
 \O_1\! =\! \bar{N} _aO_{23}O_{12} \bar{N}^{-1}_a\!\!\!\!\,, \hskip .3cm
 \O_2 \! =\!  \bar{N}_a O_{87}O_{18} \bar{N}^{-1}_a\!\!\!\!\,, \hskip .3cm
 \O_3 \! =\!  \bar{N}_a O_{43}O_{54} \bar{N}^{-1}_a\!\!\!\!\,, \hskip .3cm
 \O_4 \! =\!  \bar{N}_a O_{67}O_{56} \bar{N}^{-1}_a\,.
 \end{align*}
The differential equations are schematically given by,
\begin{align}
  D^{(2)}_2 \quad  \raisebox{-47.5pt}{\includegraphics[scale=.5]{./figures/2loop_8pt_dp_conf_a.pdf}} =  \quad  \raisebox{-47.5pt}{\includegraphics[scale=.5]{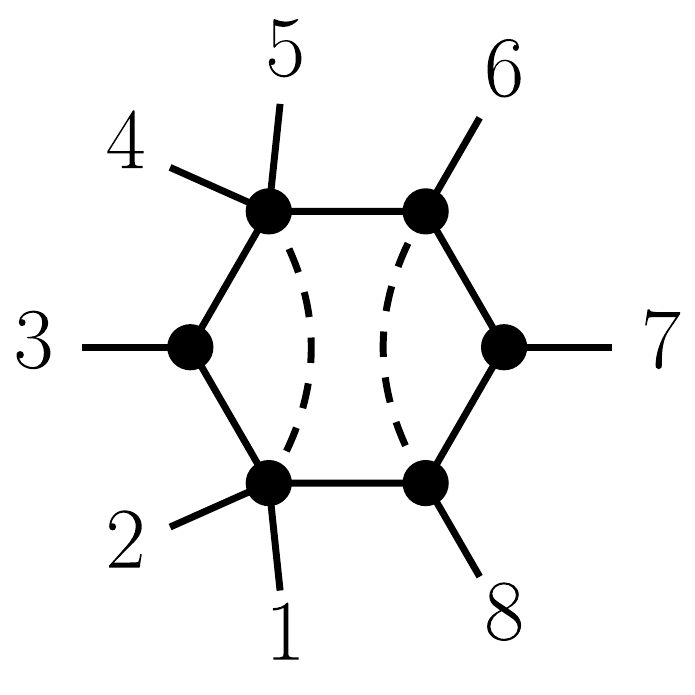}}
\end{align}
 In complete analogy to the discussion of $\Omega^{(2)}_{7,b}$, we focus on $\O_1$ (the remaining differential equations are related to $\O_1$ by symmetry)
\begin{align}
\O_1 \Omega^{(2)}_{8,b} = - \frac{\bar{N}_a}{\ab{2345}} 
					\int\limits_{(CD)} \frac{\ab{CD24}\ab{CD68}}
					{\ab{CD23}\ab{CD34}\ab{CD56}\ab{CD67}\ab{CD78}\ab{CD81}}\,,
\end{align}
where we define the one-loop chiral-hexagon integral that carries little-group weight,
$$
I^{(1),\chi\text{-h}}_{(12)(45)} \! = \!
\raisebox{-53pt}{
\includegraphics[scale=.55]{./figures/1loop_chiral_hex_12_45_massive.pdf}
}
\! = \! \int\limits_{(CD)} \frac{\ab{CD24}\ab{CD68}}
				   {\ab{CD23}\ab{CD34}\ab{CD56}\ab{CD67}\ab{CD78}\ab{CD81}}\,.
$$
In analogy to the seven-point example $I^{(1),\chi\text{-h}}_{(12)}$, we slightly abuse the notation for dashed-line numerator factors in the graph. The left dashed line denotes $\ab{AB24}$. This integral has not been explicitly written in the literature but it is easily obtained by a unitarity-inspired decomposition in terms of 15 (6 choose 4) boxes. It turns out that two boxes get zero coefficient in this expansion. The precise form of the answer is not so important at this point, we just note that  $I^{(1),\chi\text{-h}}_{(12)(45)}$ is not a pure integral but there are three linearly independent rational prefactors that can be chosen to be,
\begin{align*}
 \text{LS}_1 = \frac{1}{\ab{3567}\ab{3781}}\,,  
 \text{LS}_2 = \frac{\ab{1278}}{\ab{3781}\ab{23(567)\cap(781)}}\,,  
 \text{LS}_3 = \frac{\ab{4567}}{\ab{3567}\ab{34(781)\cap(567)}}\,.
\end{align*}
Unlike in the seven-point example (\ref{eq:chi_hex_12_massive_LS_rel}), these three leading singularities are independent and do not satisfy any further relations. In an ancillary file, we have given the result for this integral in the following form,
\begin{align}
 I^{(1),\chi\text{-h}}_{(12)(45)} = \sum^3_{i=1} \text{LS}_i \ \omega^{(2)}_i
\end{align}
where the $\omega^{(2)}_i$ are pure weight two symbols (at one loop due to the box-expansion, this result can easily be extended to the full function level) written in terms of our eight-point alphabet that we introduce later.

\section{From the dispersion integral to the symbol}
\label{subsubsec:integrating_symbols}

%
\subsection{Algorithm for integrating the dispersion relation}

%
%
In order to reconstruct the weight $n$ Feynman integral from its discontinuity at symbol level, we have to integrate the weight $(n-1)$ symbols of the discontinuity against the logarithmic factor $\frac{ds'}{s'-s}$ (or more complicated rational functions that first need to be partial fractioned in case subtraction terms have to be taken into account). To this end, we require an integration algorithm at symbol level. The algorithm sketched in the appendix of \cite{CaronHuot:2011kk} is similar in spirit to the integration algorithm for generalized polylogarithms that has been algorithmically implemented in e.g.~\cite{Panzer:2014caa}. In fact, the main properties of the symbol integration follow from the differentiation of hyperlogarithms. We found section 3.3 in Erik Panzer's PhD thesis \cite{Panzer:2015ida} especially helpful in this respect. 


Consider the following integral where the integrand converges both at zero and infinity (more generally one can consider an integral with finite integration endpoints),
\begin{align}
 I(y,x_i) = \int_0^\infty\!\!\! \dlog (x+y) F^{(n)}(x,x_i) \equiv  \int_0^\infty \frac{dx}{x+y} F^{(n)}(x,x_i) \,.
\end{align}
The first step in the symbol integration involves taking the total differential of the integral $I(y,x_i)$,
\begin{align}
 d I(y,x_i) = \left[dy \partial_y + dx_i \partial_{x_i}\right] I(y,x_i)\,. 
\end{align}
In the general case where the boundaries also depend on the parameters, one would add differentials with respect to the integration boundaries and add the corresponding boundary terms. Acting with the partial differentials on the integrand, we find,
\begin{align}
 d  I(y,x_i)  = - dy \int_0^\infty \!\!\! \frac{dx}{(x+y)^2} F^{(n)}(x,x_i) + dx_i \int_0^\infty \frac{dx}{(x+y)} \partial_{x_i}F^{(n)}(x,x_i)\,.
\end{align}
The first term we write as a derivative with respect to the integration variable $x$,
\begin{align}
d  I(y,x_i)  = + dy \int_0^\infty\!\!\!  dx \left[\partial_x\frac{1}{(x+y)}\right] F^{(n)}(x,x_i) + dx_i \int_0^\infty\!\!\!  \frac{dx}{(x+y)} \partial_{x_i}F^{(n)}(x,x_i)\,
\end{align}
and integrate by parts,
\begin{align}
 d  I(y,x_i)  =  dy \frac{F^{(n)}(x,x_i)}{x+y}\Big|^{x=\infty}_{x=0}\!\!\! 
 		   - dy \int_0^\infty\!\!\!  \frac{dx}{(x+y)} \partial_xF^{(n)}(x,x_i) 
		   + dx_i \int_0^\infty\!\!\!  \frac{dx}{(x+y)} \partial_{x_i} F^{(n)}(x,x_i) \,.
\end{align}
Since the $F^{(n)}(x,x_i)$ are defined as iterated integrals, we know their respective differentials,
\begin{align}
\begin{split}
\partial_x      F^{(n)}(x,x_i) &\!=\! \sum_{j} F^{(n-1)}_j(x,x_i) \frac{\log(x+\beta_j)}{\partial x } \\
\partial_{x_i} F^{(n)}(x,x_i) &\!=\! \sum_{j} F^{(n-1)}_j(x,x_i) \frac{\log(x+\beta_j)}{\partial \beta_j}  \left(\frac{\partial \beta_j}{\partial x_i} \right) 
					 \! +\! \sum_{j'} H^{(n-1)}_{j'} (x,x_i) \frac{\log f_{j'}}{\partial x_i}
\end{split}
\end{align}
Taking the boundary term at $x=\infty$ to vanish, the differential of the integral becomes,
\begin{align}
\begin{split}
 d  I(y,x_i)  = - \dlog y F^{(n)}(0,x_i) & - \sum_j dy \int_0^\infty\!\!\!  \frac{dx}{(x+y)(x+\beta_j)} F^{(n-1)}_j \\
 							 & + \sum_j \underbrace{dx_i   \left(\frac{\partial \beta_j}{\partial x_i} \right)}_{=d\beta_j}  \int_0^\infty\!\!\!  \frac{dx}{(x+y)(x+\beta_j)} F^{(n-1)}_j  \\
							 & + \sum_{j'} \underbrace{dx_i \frac{\partial \log f_{j'}}{\partial x_i}}_{=\dlog f_{j'}} \int_0^\infty\!\!\!  \frac{dx}{(x+y)} H^{(n-1)}_{j'}
\end{split}
\end{align}
In the first two terms, we need to linearize the denominators by partial fractioning the product,
\begin{align}
\frac{1}{(x+y)(x+\beta_j)} = \frac{1}{(y-\beta_j)} \left[\frac{1}{(x+\beta_j)}-\frac{1}{(x+y)}\right]\,.
\end{align}
Putting everything together, we find,
\begin{align}
\begin{split}
 dI =&  -\dlog y \ F^{(n)}(0,x_i) +  \sum_{j'} \dlog f_{j'} \int_0^\infty\!\!\!  \frac{dx}{(x+y)} H^{(n-1)}_{j'} \\
       &  -\sum_j \left[\frac{d y}{(y-\beta_j)} - \frac{d \beta_j}{(y-\beta_j)}\right]  \int_0^\infty\!  \left[\frac{dx}{(x+\beta_j)}-\frac{dx}{(x+y)}\right] F^{(n-1)}_j \,.
\end{split}
\end{align}
Combining the respective terms in the brackets back into $\dlog$-forms, we recover the expressions in \cite{CaronHuot:2011kk},
\begin{align}
\begin{split}
  dI(y,x_i) =& -\dlog y \ F^{(n)}(0,x_i)  +  \sum_{j'}  \dlog f_{j'} \int_0^\infty\!\!\!  \frac{dx}{(x+y)} H^{(n-1)}_{j'} (x,x_i) \\
        & +\sum_j \dlog(y-\beta_j) \int_0^\infty\!\!\! \dlog\left(\!\frac{x+y}{x+\beta_j}\!\right) F^{(n-1)}_j(x,x_i)
\end{split}
\end{align}
The structure of the symbol integration algorithm presented above also sheds some light on
how additional letters can appear.
Starting from the inhomogeneous term of the differential equation that only contains the four-mass box algebraic letters, the integration algorithm introduces a number of partial fractions with respect to the remaining letters. These partial fractions then introduce further combinations of the form $a_i \pm \sqrt{\Delta}$. 
%
%
\subsection{Example: Dispersion representation of the three-mass pentagon integral}
%
%
In order to both test our implementation of the symbol integration algorithm described in the previous subsection as well as elaborating on the dispersion representation of Feynman integrals, we are going to discuss a simple one-loop example with a slightly more complicated kinematic dependence than the three-mass triangle integral studied in \cite{Abreu:2014cla}. To this end, we are writing a dispersion representation of the three-mass dual conformal pentagon integral,
\begin{align}
\widetilde{\Psi}^{(1)}(u,v,w)=
\raisebox{-45pt}{
\includegraphics[scale=.4]{./figures/1loop_pent_massive}} 
\end{align} 
which is related to the un-normalized pentagon integral by the following factors
\begin{align}
\label{eq:3mass_pentagon_normalizations}
\tw{F}^{(1)}_{3m} = \int \frac{d^4 y (y,x_\ast)}{(y,1)(y,3)(y,4)(y,6)(y,7)} = \frac{(1,x_\ast)}{(1,4)(1,6)(3,7)} \frac{\tw{\Psi}^{(1)}(u,v,w)}{(1-u-v+uvw)}\,,
\end{align}
where $(i,j)\equiv x^2_{ij}=(x_i-x_j)^2$ and the point $x_\ast$ corresponds to one of the quadruple-cut solutions to $(y,3)=(y,4)=(y,6)=(y,7)=0$. Here we follow the conventions of \cite{Drummond:2010cz} and introduce the dual conformal cross ratios,
\begin{align}
 u \!=\! \frac{\ab{8123}\ab{3467}}{\ab{8134}\ab{2367}}\! \sim\! \frac{1}{\sab{6781}}\,, \
 v \!=\!  \frac{\ab{8167}\ab{2356}}{\ab{8156}\ab{2367}}\! \sim\! \frac{1}{\sab{6781}}\,, \
 w \!=\! \frac{\ab{2367}\ab{3456}}{\ab{2356}\ab{3467}}\! \sim\!  \sab{6781}\,.
\end{align}
The pentagon integral is of course known \cite{Drummond:2010cz}, but we are going to use it as an illustrative example of our setup,
\begin{align}
\widetilde{\Psi}^{(1)}(u,v,w)\! =\! \log u\log v\!+\!\Li_2(1\sm u)\!+\!\Li_2(1\sm v)\!+\!\Li_2(1\sm w) \sm \Li_2(1\sm uw) \sm \Li_2(1\sm vw)\,.
\end{align}
Here, we would like to take the discontinuity in the $\sab{7812}$-channel of the integral. In the form written here, all three dual-conformal cross-ratios $u,\ v,$ and $w$ depend on this Mandelstam invariant so that we should change variables to isolate $\sab{7812}$ in a single cross ratio. This can be accomplished by simply rescaling $u$ and $v$ by $w$, $\tw{u}= u w$ and $\tw{v}= v w$ which isolates $\sab{7812}$ in $w$. It is now easy to take the discontinuity in w and we find,
\begin{align}
 \underset{w}{\text{Disc}}\ \tw{\Psi}^{(1)}(\tw{u},\tw{v},w) = \raisebox{-55pt}{\includegraphics[scale=.5]{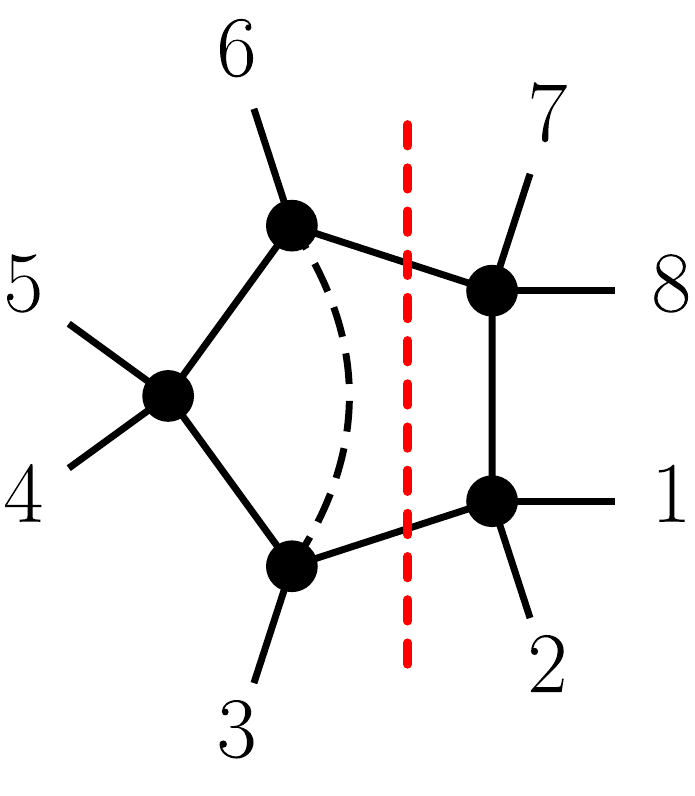}}
 =   \log\left[\frac{(w-\tw{v})(w-\tw{u})}{(w-1) \tw{u}\ \tw{v}}\right]\,.
\end{align}
Equipped with the discontinuity, we are now in the position to write the dispersion integral in the $w$-channel which represents the dispersion in $\sab{7812}$. In a massless theory, the branch cut is from zero to infinity along the real positive axis. This leads to the following one-fold integral representation for $\tw{\Psi}^{(1)}(\tw{u},\tw{v},w)$,
\begin{align}
\label{eq:dispersion_3mass_pentagon_naive}
  \tw{\Psi}^{(1)}(\tw{u},\tw{v},w)\ ``\!=\!'' \int \limits_0^\infty \frac{d w'}{w'-w}\  \underset{w'}{\text{Disc}}\ \tw{\Psi}^{(1)}(\tw{u},\tw{v},w') \quad (\text{naive})
\end{align}
where we wrote the equal sign in quotation marks and indicated the ``naive'' equality of both sides. This simple case illustrates our earlier comments on the need for subtraction terms in order to make the integral well-behaved at the upper integration boundary at infinity. As is, (\ref{eq:dispersion_3mass_pentagon_naive}) diverges at infinity, so that we are forced to introduce a subtraction term. For this integral, one of the natural choices for a subtraction term comes from looking at the normalization factor $(1-u-v+uvw)$ that is required in the numerator in order to define the pure function $\tw{\Psi}^{(1)}(u,v,w)$ from $\tw{F}^{(1)}_{3m}$ in (\ref{eq:3mass_pentagon_normalizations}). Since the normalization vanishes for the nonsingular point $w_1  = (u+v-1)/(uv)$, this is a natural subtraction point for the dispersion relation, since $\tw{\Psi}^{(1)}(u,v,w_1) =0$. Written in terms of our rescaled cross-ratios that make the discontinuity channel $\sab{7812}$ manifest, the subtracted dispersion relation now reads,
\begin{align}
\label{eq:dispersion_3mass_pentagon_subtracted}
 \tw{\Psi}^{(1)}(\tw{u},\tw{v},w) = \int \limits_0^\infty \dlog\left[ \frac{w'\sm w}{w' \sm \tw{u} (1\sm\tw{v})\sm\tw{v}}\right] \  \log\left[\frac{(w\sm\tw{v})(w\sm\tw{u})}{(w\sm1) \tw{u}\ \tw{v}}\right]  +  \overbrace{\tw{\Psi}^{(1)}(\tw{u},\tw{v},w^\ast) }^{=0}\,,
\end{align}
where $w^\ast\! =\! \tw{u} (1-\tw{v})-\tw{v}$ corresponds to the zero in the normalization when written in terms of the rescaled cross-ratios. Using the symbol integration algorithm described in the previous subsection, it is easy to check that we recover the full symbol of $\tw{\Psi}^{(1)}(u,v,w)$ from the subtracted dispersion relation in (\ref{eq:dispersion_3mass_pentagon_subtracted}).

%
\bibliographystyle{JHEP}
\phantomsection         
\bibliography{refs}

\providecommand{\href}[2]{#2}\begingroup\raggedright\begin{thebibliography}{10}

\bibitem{Bern:1994zx}
Z.~Bern, L.~J. Dixon, D.~C. Dunbar and D.~A. Kosower, \emph{{One loop n point
  gauge theory amplitudes, unitarity and collinear limits}},
  \href{https://doi.org/10.1016/0550-3213(94)90179-1}{\emph{Nucl. Phys.}
  {\bfseries B425} (1994) 217}
  [\href{https://arxiv.org/abs/hep-ph/9403226}{{\ttfamily hep-ph/9403226}}].

\bibitem{Bern:1994cg}
Z.~Bern, L.~J. Dixon, D.~C. Dunbar and D.~A. Kosower, \emph{{Fusing gauge
  theory tree amplitudes into loop amplitudes}},
  \href{https://doi.org/10.1016/0550-3213(94)00488-Z}{\emph{Nucl. Phys.}
  {\bfseries B435} (1995) 59}
  [\href{https://arxiv.org/abs/hep-ph/9409265}{{\ttfamily hep-ph/9409265}}].

\bibitem{Bern:2007ct}
Z.~Bern, J.~J.~M. Carrasco, H.~Johansson and D.~A. Kosower, \emph{{Maximally
  supersymmetric planar Yang-Mills amplitudes at five loops}},
  \href{https://doi.org/10.1103/PhysRevD.76.125020}{\emph{Phys. Rev.}
  {\bfseries D76} (2007) 125020}
  [\href{https://arxiv.org/abs/0705.1864}{{\ttfamily 0705.1864}}].

\bibitem{Vermaseren:1998uu}
J.~A.~M. Vermaseren, \emph{{Harmonic sums, Mellin transforms and integrals}},
  \href{https://doi.org/10.1142/S0217751X99001032}{\emph{Int. J. Mod. Phys.}
  {\bfseries A14} (1999) 2037}
  [\href{https://arxiv.org/abs/hep-ph/9806280}{{\ttfamily hep-ph/9806280}}].

\bibitem{Czakon:2005rk}
M.~Czakon, \emph{{Automatized analytic continuation of Mellin-Barnes
  integrals}}, \href{https://doi.org/10.1016/j.cpc.2006.07.002}{\emph{Comput.
  Phys. Commun.} {\bfseries 175} (2006) 559}
  [\href{https://arxiv.org/abs/hep-ph/0511200}{{\ttfamily hep-ph/0511200}}].

\bibitem{Smirnov:2009up}
A.~V. Smirnov and V.~A. Smirnov, \emph{{On the Resolution of Singularities of
  Multiple Mellin-Barnes Integrals}},
  \href{https://doi.org/10.1140/epjc/s10052-009-1039-6}{\emph{Eur. Phys. J.}
  {\bfseries C62} (2009) 445}
  [\href{https://arxiv.org/abs/0901.0386}{{\ttfamily 0901.0386}}].

\bibitem{Anastasiou:2013srw}
C.~Anastasiou, C.~Duhr, F.~Dulat and B.~Mistlberger, \emph{{Soft triple-real
  radiation for Higgs production at N3LO}},
  \href{https://doi.org/10.1007/JHEP07(2013)003}{\emph{JHEP} {\bfseries 07}
  (2013) 003} [\href{https://arxiv.org/abs/1302.4379}{{\ttfamily 1302.4379}}].

\bibitem{Henn:2014qga}
J.~M. Henn, \emph{{Lectures on differential equations for Feynman integrals}},
  \href{https://doi.org/10.1088/1751-8113/48/15/153001}{\emph{J. Phys.}
  {\bfseries A48} (2015) 153001}
  [\href{https://arxiv.org/abs/1412.2296}{{\ttfamily 1412.2296}}].

\bibitem{Kotikov:1990kg}
A.~V. Kotikov, \emph{{Differential equations method: New technique for massive
  Feynman diagrams calculation}},
  \href{https://doi.org/10.1016/0370-2693(91)90413-K}{\emph{Phys. Lett.}
  {\bfseries B254} (1991) 158}.

\bibitem{Remiddi:1997ny}
E.~Remiddi, \emph{{Differential equations for Feynman graph amplitudes}},
  {\emph{Nuovo Cim.} {\bfseries A110} (1997) 1435}
  [\href{https://arxiv.org/abs/hep-th/9711188}{{\ttfamily hep-th/9711188}}].

\bibitem{Gehrmann:1999as}
T.~Gehrmann and E.~Remiddi, \emph{{Differential equations for two loop four
  point functions}},
  \href{https://doi.org/10.1016/S0550-3213(00)00223-6}{\emph{Nucl. Phys.}
  {\bfseries B580} (2000) 485}
  [\href{https://arxiv.org/abs/hep-ph/9912329}{{\ttfamily hep-ph/9912329}}].

\bibitem{Argeri:2007up}
M.~Argeri and P.~Mastrolia, \emph{{Feynman Diagrams and Differential
  Equations}}, \href{https://doi.org/10.1142/S0217751X07037147}{\emph{Int. J.
  Mod. Phys.} {\bfseries A22} (2007) 4375}
  [\href{https://arxiv.org/abs/0707.4037}{{\ttfamily 0707.4037}}].

\bibitem{Henn:2013pwa}
J.~M. Henn, \emph{{Multiloop integrals in dimensional regularization made
  simple}}, \href{https://doi.org/10.1103/PhysRevLett.110.251601}{\emph{Phys.
  Rev. Lett.} {\bfseries 110} (2013) 251601}
  [\href{https://arxiv.org/abs/1304.1806}{{\ttfamily 1304.1806}}].

\bibitem{Goncharov:2001iea}
A.~B. Goncharov, \emph{{Multiple polylogarithms and mixed Tate motives}},
  \href{https://arxiv.org/abs/math/0103059}{{\ttfamily math/0103059}}.

\bibitem{Brown:2009ta}
F.~C.~S. Brown, \emph{{On the periods of some Feynman integrals}},
  \href{https://arxiv.org/abs/0910.0114}{{\ttfamily 0910.0114}}.

\bibitem{Panzer:2014gra}
E.~Panzer, \emph{{On hyperlogarithms and Feynman integrals with divergences and
  many scales}}, \href{https://doi.org/10.1007/JHEP03(2014)071}{\emph{JHEP}
  {\bfseries 03} (2014) 071} [\href{https://arxiv.org/abs/1401.4361}{{\ttfamily
  1401.4361}}].

\bibitem{Brown:2015qmm}
F.~Brown, \emph{{Periods and Feynman amplitudes}},  in \emph{{18th
  International Congress on Mathematical Physics (ICMP2015) Santiago de Chile,
  Chile, July 27-August 1, 2015}}, 2015,
  \href{https://arxiv.org/abs/1512.09265}{{\ttfamily 1512.09265}},
  \href{https://inspirehep.net/record/1411840/files/arXiv:1512.09265.pdf}{https://inspirehep.net/record/1411840/files/arXiv:1512.09265.pdf}.

\bibitem{ArkaniHamed:2010gh}
N.~Arkani-Hamed, J.~L. Bourjaily, F.~Cachazo and J.~Trnka, \emph{{Local
  Integrals for Planar Scattering Amplitudes}},
  \href{https://doi.org/10.1007/JHEP06(2012)125}{\emph{JHEP} {\bfseries 06}
  (2012) 125} [\href{https://arxiv.org/abs/1012.6032}{{\ttfamily 1012.6032}}].

\bibitem{Drummond:2006rz}
J.~M. Drummond, J.~Henn, V.~A. Smirnov and E.~Sokatchev, \emph{{Magic
  identities for conformal four-point integrals}},
  \href{https://doi.org/10.1088/1126-6708/2007/01/064}{\emph{JHEP} {\bfseries
  01} (2007) 064} [\href{https://arxiv.org/abs/hep-th/0607160}{{\ttfamily
  hep-th/0607160}}].

\bibitem{Alday:2007hr}
L.~F. Alday and J.~M. Maldacena, \emph{{Gluon scattering amplitudes at strong
  coupling}}, \href{https://doi.org/10.1088/1126-6708/2007/06/064}{\emph{JHEP}
  {\bfseries 06} (2007) 064} [\href{https://arxiv.org/abs/0705.0303}{{\ttfamily
  0705.0303}}].

\bibitem{Drummond:2008vq}
J.~M. Drummond, J.~Henn, G.~P. Korchemsky and E.~Sokatchev, \emph{{Dual
  superconformal symmetry of scattering amplitudes in N=4 super-Yang-Mills
  theory}}, \href{https://doi.org/10.1016/j.nuclphysb.2009.11.022}{\emph{Nucl.
  Phys.} {\bfseries B828} (2010) 317}
  [\href{https://arxiv.org/abs/0807.1095}{{\ttfamily 0807.1095}}].

\bibitem{Arkani-Hamed:2014bca}
N.~Arkani-Hamed, J.~L. Bourjaily, F.~Cachazo, A.~Postnikov and J.~Trnka,
  \emph{{On-Shell Structures of MHV Amplitudes Beyond the Planar Limit}},
  \href{https://doi.org/10.1007/JHEP06(2015)179}{\emph{JHEP} {\bfseries 06}
  (2015) 179} [\href{https://arxiv.org/abs/1412.8475}{{\ttfamily 1412.8475}}].

\bibitem{ArkaniHamed:2012nw}
N.~Arkani-Hamed, J.~L. Bourjaily, F.~Cachazo, A.~B. Goncharov, A.~Postnikov and
  J.~Trnka, \emph{{Grassmannian Geometry of Scattering Amplitudes}}. Cambridge
  University Press, 2016, [\href{https://arxiv.org/abs/1212.5605}{{\ttfamily
  1212.5605}}].

\bibitem{Arkani-Hamed:2013jha}
N.~Arkani-Hamed and J.~Trnka, \emph{{The Amplituhedron}},
  \href{https://doi.org/10.1007/JHEP10(2014)030}{\emph{JHEP} {\bfseries 10}
  (2014) 030} [\href{https://arxiv.org/abs/1312.2007}{{\ttfamily 1312.2007}}].

\bibitem{Arkani-Hamed:2013kca}
N.~Arkani-Hamed and J.~Trnka, \emph{{Into the Amplituhedron}},
  \href{https://doi.org/10.1007/JHEP12(2014)182}{\emph{JHEP} {\bfseries 12}
  (2014) 182} [\href{https://arxiv.org/abs/1312.7878}{{\ttfamily 1312.7878}}].

\bibitem{Britto:2004nc}
R.~Britto, F.~Cachazo and B.~Feng, \emph{{Generalized unitarity and one-loop
  amplitudes in N=4 super-Yang-Mills}},
  \href{https://doi.org/10.1016/j.nuclphysb.2005.07.014}{\emph{Nucl. Phys.}
  {\bfseries B725} (2005) 275}
  [\href{https://arxiv.org/abs/hep-th/0412103}{{\ttfamily hep-th/0412103}}].

\bibitem{Cachazo:2008vp}
F.~Cachazo, \emph{{Sharpening The Leading Singularity}},
  \href{https://arxiv.org/abs/0803.1988}{{\ttfamily 0803.1988}}.

\bibitem{Bern:2014kca}
Z.~Bern, E.~Herrmann, S.~Litsey, J.~Stankowicz and J.~Trnka, \emph{{Logarithmic
  Singularities and Maximally Supersymmetric Amplitudes}},
  \href{https://doi.org/10.1007/JHEP06(2015)202}{\emph{JHEP} {\bfseries 06}
  (2015) 202} [\href{https://arxiv.org/abs/1412.8584}{{\ttfamily 1412.8584}}].

\bibitem{Bern:2015ple}
Z.~Bern, E.~Herrmann, S.~Litsey, J.~Stankowicz and J.~Trnka, \emph{{Evidence
  for a Nonplanar Amplituhedron}},
  \href{https://doi.org/10.1007/JHEP06(2016)098}{\emph{JHEP} {\bfseries 06}
  (2016) 098} [\href{https://arxiv.org/abs/1512.08591}{{\ttfamily
  1512.08591}}].

\bibitem{Herrmann:2016qea}
E.~Herrmann and J.~Trnka, \emph{{Gravity On-shell Diagrams}},
  \href{https://doi.org/10.1007/JHEP11(2016)136}{\emph{JHEP} {\bfseries 11}
  (2016) 136} [\href{https://arxiv.org/abs/1604.03479}{{\ttfamily
  1604.03479}}].

\bibitem{Henn:2016jdu}
J.~M. Henn and B.~Mistlberger, \emph{{Four-Gluon Scattering at Three Loops,
  Infrared Structure, and the Regge Limit}},
  \href{https://doi.org/10.1103/PhysRevLett.117.171601}{\emph{Phys. Rev. Lett.}
  {\bfseries 117} (2016) 171601}
  [\href{https://arxiv.org/abs/1608.00850}{{\ttfamily 1608.00850}}].

\bibitem{Dixon:2011nj}
L.~J. Dixon, J.~M. Drummond and J.~M. Henn, \emph{{Analytic result for the
  two-loop six-point NMHV amplitude in N=4 super Yang-Mills theory}},
  \href{https://doi.org/10.1007/JHEP01(2012)024}{\emph{JHEP} {\bfseries 01}
  (2012) 024} [\href{https://arxiv.org/abs/1111.1704}{{\ttfamily 1111.1704}}].

\bibitem{Dixon:2011pw}
L.~J. Dixon, J.~M. Drummond and J.~M. Henn, \emph{{Bootstrapping the three-loop
  hexagon}}, \href{https://doi.org/10.1007/JHEP11(2011)023}{\emph{JHEP}
  {\bfseries 11} (2011) 023} [\href{https://arxiv.org/abs/1108.4461}{{\ttfamily
  1108.4461}}].

\bibitem{Dixon:2013eka}
L.~J. Dixon, J.~M. Drummond, M.~von Hippel and J.~Pennington, \emph{{Hexagon
  functions and the three-loop remainder function}},
  \href{https://doi.org/10.1007/JHEP12(2013)049}{\emph{JHEP} {\bfseries 12}
  (2013) 049} [\href{https://arxiv.org/abs/1308.2276}{{\ttfamily 1308.2276}}].

\bibitem{Golden:2014pua}
J.~Golden and M.~Spradlin, \emph{{A Cluster Bootstrap for Two-Loop MHV
  Amplitudes}}, \href{https://doi.org/10.1007/JHEP02(2015)002}{\emph{JHEP}
  {\bfseries 02} (2015) 002} [\href{https://arxiv.org/abs/1411.3289}{{\ttfamily
  1411.3289}}].

\bibitem{Dixon:2014iba}
L.~J. Dixon and M.~von Hippel, \emph{{Bootstrapping an NMHV amplitude through
  three loops}}, \href{https://doi.org/10.1007/JHEP10(2014)065}{\emph{JHEP}
  {\bfseries 10} (2014) 065} [\href{https://arxiv.org/abs/1408.1505}{{\ttfamily
  1408.1505}}].

\bibitem{Drummond:2014ffa}
J.~M. Drummond, G.~Papathanasiou and M.~Spradlin, \emph{{A Symbol of
  Uniqueness: The Cluster Bootstrap for the 3-Loop MHV Heptagon}},
  \href{https://doi.org/10.1007/JHEP03(2015)072}{\emph{JHEP} {\bfseries 03}
  (2015) 072} [\href{https://arxiv.org/abs/1412.3763}{{\ttfamily 1412.3763}}].

\bibitem{Dixon:2015iva}
L.~J. Dixon, M.~von Hippel and A.~J. McLeod, \emph{{The four-loop six-gluon
  NMHV ratio function}},
  \href{https://doi.org/10.1007/JHEP01(2016)053}{\emph{JHEP} {\bfseries 01}
  (2016) 053} [\href{https://arxiv.org/abs/1509.08127}{{\ttfamily
  1509.08127}}].

\bibitem{Caron-Huot:2016owq}
S.~Caron-Huot, L.~J. Dixon, A.~McLeod and M.~von Hippel, \emph{{Bootstrapping a
  Five-Loop Amplitude Using Steinmann Relations}},
  \href{https://doi.org/10.1103/PhysRevLett.117.241601}{\emph{Phys. Rev. Lett.}
  {\bfseries 117} (2016) 241601}
  [\href{https://arxiv.org/abs/1609.00669}{{\ttfamily 1609.00669}}].

\bibitem{Dixon:2016nkn}
L.~J. Dixon, J.~Drummond, T.~Harrington, A.~J. McLeod, G.~Papathanasiou and
  M.~Spradlin, \emph{{Heptagons from the Steinmann Cluster Bootstrap}},
  \href{https://doi.org/10.1007/JHEP02(2017)137}{\emph{JHEP} {\bfseries 02}
  (2017) 137} [\href{https://arxiv.org/abs/1612.08976}{{\ttfamily
  1612.08976}}].

\bibitem{Eden:1966dnq}
R.~J. Eden, P.~V. Landshoff, D.~I. Olive and J.~C. Polkinghorne, \emph{{The
  analytic S-matrix}}. Cambridge Univ. Press, Cambridge, 1966.

\bibitem{Goncharov:2010jf}
A.~B. Goncharov, M.~Spradlin, C.~Vergu and A.~Volovich, \emph{{Classical
  Polylogarithms for Amplitudes and Wilson Loops}},
  \href{https://doi.org/10.1103/PhysRevLett.105.151605}{\emph{Phys. Rev. Lett.}
  {\bfseries 105} (2010) 151605}
  [\href{https://arxiv.org/abs/1006.5703}{{\ttfamily 1006.5703}}].

\bibitem{Fomin:2001aa}
S.~Fomin and A.~Zelevinsky, \emph{{Cluster algebras I: Foundations}},
  \href{https://doi.org/10.1090/S0894-0347-01-00385-X}{\emph{J. Amer. Math.
  Soc.} {\bfseries 15} (2002) 497}
  [\href{https://arxiv.org/abs/0104151}{{\ttfamily 0104151}}].

\bibitem{Fock:2003aa}
V.~Fock and A.~Goncharov, \emph{{Cluster Ensembles, Quantization and the
  Dilogarithm}}, {\emph{Ann. Sci. L'Ecole Norm. Sup.} (2009) }
  [\href{https://arxiv.org/abs/0311245}{{\ttfamily 0311245}}].

\bibitem{Fock:2005aa}
V.~Fock and A.~Goncharov, \emph{{Cluster $\chi$-Varieties and Number Theory,
  Amalgamation and Poisson-Lie Groups}}, {\emph{Algebraic Geometry and Number
  Theory, Dedicated to Drinfeld's $50^{\text{th}} birthday$} (2006) 27}
  [\href{https://arxiv.org/abs/0508408}{{\ttfamily 0508408}}].

\bibitem{Golden:2013xva}
J.~Golden, A.~B. Goncharov, M.~Spradlin, C.~Vergu and A.~Volovich,
  \emph{{Motivic Amplitudes and Cluster Coordinates}},
  \href{https://doi.org/10.1007/JHEP01(2014)091}{\emph{JHEP} {\bfseries 01}
  (2014) 091} [\href{https://arxiv.org/abs/1305.1617}{{\ttfamily 1305.1617}}].

\bibitem{Golden:2014xqa}
J.~Golden, M.~F. Paulos, M.~Spradlin and A.~Volovich, \emph{{Cluster
  Polylogarithms for Scattering Amplitudes}},
  \href{https://doi.org/10.1088/1751-8113/47/47/474005}{\emph{J. Phys.}
  {\bfseries A47} (2014) 474005}
  [\href{https://arxiv.org/abs/1401.6446}{{\ttfamily 1401.6446}}].

\bibitem{Harrington:2015bdt}
T.~Harrington and M.~Spradlin, \emph{{Cluster Functions and Scattering
  Amplitudes for Six and Seven Points}},
  \href{https://doi.org/10.1007/JHEP07(2017)016}{\emph{JHEP} {\bfseries 07}
  (2017) 016} [\href{https://arxiv.org/abs/1512.07910}{{\ttfamily
  1512.07910}}].

\bibitem{Dennen:2015bet}
T.~Dennen, M.~Spradlin and A.~Volovich, \emph{{Landau Singularities and
  Symbology: One- and Two-loop MHV Amplitudes in SYM Theory}},
  \href{https://doi.org/10.1007/JHEP03(2016)069}{\emph{JHEP} {\bfseries 03}
  (2016) 069} [\href{https://arxiv.org/abs/1512.07909}{{\ttfamily
  1512.07909}}].

\bibitem{Dennen:2016mdk}
T.~Dennen, I.~Prlina, M.~Spradlin, S.~Stanojevic and A.~Volovich, \emph{{Landau
  Singularities from the Amplituhedron}},
  \href{https://doi.org/10.1007/JHEP06(2017)152}{\emph{JHEP} {\bfseries 06}
  (2017) 152} [\href{https://arxiv.org/abs/1612.02708}{{\ttfamily
  1612.02708}}].

\bibitem{Prlina:2017azl}
I.~Prlina, M.~Spradlin, J.~Stankowicz, S.~Stanojevic and A.~Volovich,
  \emph{{All-Helicity Symbol Alphabets from Unwound Amplituhedra}},
  \href{https://arxiv.org/abs/1711.11507}{{\ttfamily 1711.11507}}.

\bibitem{Prlina:2017tvx}
I.~Prlina, M.~Spradlin, J.~Stankowicz and S.~Stanojevic, \emph{{Boundaries of
  Amplituhedra and NMHV Symbol Alphabets at Two Loops}},
  \href{https://arxiv.org/abs/1712.08049}{{\ttfamily 1712.08049}}.

\bibitem{CaronHuot:2011ky}
S.~Caron-Huot, \emph{{Superconformal symmetry and two-loop amplitudes in planar
  N=4 super Yang-Mills}},
  \href{https://doi.org/10.1007/JHEP12(2011)066}{\emph{JHEP} {\bfseries 12}
  (2011) 066} [\href{https://arxiv.org/abs/1105.5606}{{\ttfamily 1105.5606}}].

\bibitem{Drummond:2017ssj}
J.~Drummond, J.~Foster and O.~Gurdogan, \emph{{Cluster adjacency properties of
  scattering amplitudes}},  \href{https://arxiv.org/abs/1710.10953}{{\ttfamily
  1710.10953}}.

\bibitem{Bourjaily:2018aeq}
J.~L. Bourjaily, A.~J. McLeod, M.~von Hippel and M.~Wilhelm,
  \emph{{Rationalizing Loop Integration}},
  \href{https://arxiv.org/abs/1805.10281}{{\ttfamily 1805.10281}}.

\bibitem{Drummond:2010cz}
J.~M. Drummond, J.~M. Henn and J.~Trnka, \emph{{New differential equations for
  on-shell loop integrals}},
  \href{https://doi.org/10.1007/JHEP04(2011)083}{\emph{JHEP} {\bfseries 04}
  (2011) 083} [\href{https://arxiv.org/abs/1010.3679}{{\ttfamily 1010.3679}}].

\bibitem{Caron-Huot:2018dsv}
S.~Caron-Huot, L.~J. Dixon, M.~von Hippel, A.~J. McLeod and G.~Papathanasiou,
  \emph{{The Double Pentaladder Integral to All Orders}},
  \href{https://arxiv.org/abs/1806.01361}{{\ttfamily 1806.01361}}.

\bibitem{Hodges:2009hk}
A.~Hodges, \emph{{Eliminating spurious poles from gauge-theoretic amplitudes}},
  \href{https://doi.org/10.1007/JHEP05(2013)135}{\emph{JHEP} {\bfseries 05}
  (2013) 135} [\href{https://arxiv.org/abs/0905.1473}{{\ttfamily 0905.1473}}].

\bibitem{Chicherin:2017dob}
D.~Chicherin, J.~Henn and V.~Mitev, \emph{{Bootstrapping pentagon functions}},
  \href{https://doi.org/10.1007/JHEP05(2018)164}{\emph{JHEP} {\bfseries 05}
  (2018) 164} [\href{https://arxiv.org/abs/1712.09610}{{\ttfamily
  1712.09610}}].

\bibitem{Duhr:2011zq}
C.~Duhr, H.~Gangl and J.~R. Rhodes, \emph{{From polygons and symbols to
  polylogarithmic functions}},
  \href{https://doi.org/10.1007/JHEP10(2012)075}{\emph{JHEP} {\bfseries 10}
  (2012) 075} [\href{https://arxiv.org/abs/1110.0458}{{\ttfamily 1110.0458}}].

\bibitem{Duhr:2014woa}
C.~Duhr, \emph{{Mathematical aspects of scattering amplitudes}},  in
  \emph{{Proceedings, Theoretical Advanced Study Institute in Elementary
  Particle Physics: Journeys Through the Precision Frontier: Amplitudes for
  Colliders (TASI 2014): Boulder, Colorado, June 2-27, 2014}}, pp.~419--476,
  2015, \href{https://arxiv.org/abs/1411.7538}{{\ttfamily 1411.7538}},
  \href{https://doi.org/10.1142/9789814678766_0010}{DOI}.

\bibitem{Steinmann}
O.~Steinmann, \emph{{Uber den Zusammenhang zwischen den Wightmanfunktionen und
  der retardierten Kommutatoren}}, {\emph{Helv. Physica Acta} {\bfseries 33}
  (1960) 257}.

\bibitem{Steinmann2}
O.~Steinmann, \emph{{Wightman-Funktionen und retardierten Kommutatoren. II}},
  {\emph{Helv. Physica Acta} {\bfseries 33} (1960) 347}.

\bibitem{Cahill:1973qp}
K.~E. Cahill and H.~P. Stapp, \emph{{OPTICAL THEOREMS AND STEINMANN
  RELATIONS}}, \href{https://doi.org/10.1016/0003-4916(75)90006-8}{\emph{Annals
  Phys.} {\bfseries 90} (1975) 438}.

\bibitem{Bartels:2008ce}
J.~Bartels, L.~N. Lipatov and A.~Sabio~Vera, \emph{{BFKL Pomeron, Reggeized
  gluons and Bern-Dixon-Smirnov amplitudes}},
  \href{https://doi.org/10.1103/PhysRevD.80.045002}{\emph{Phys. Rev.}
  {\bfseries D80} (2009) 045002}
  [\href{https://arxiv.org/abs/0802.2065}{{\ttfamily 0802.2065}}].

\bibitem{Bartels:2008sc}
J.~Bartels, L.~N. Lipatov and A.~Sabio~Vera, \emph{{N=4 supersymmetric Yang
  Mills scattering amplitudes at high energies: The Regge cut contribution}},
  \href{https://doi.org/10.1140/epjc/s10052-009-1218-5}{\emph{Eur. Phys. J.}
  {\bfseries C65} (2010) 587}
  [\href{https://arxiv.org/abs/0807.0894}{{\ttfamily 0807.0894}}].

\bibitem{Landau:1959fi}
L.~D. Landau, \emph{{On analytic properties of vertex parts in quantum field
  theory}}, \href{https://doi.org/10.1016/0029-5582(59)90154-3}{\emph{Nucl.
  Phys.} {\bfseries 13} (1959) 181}.

\bibitem{Remiddi:1981hn}
E.~Remiddi, \emph{{Dispersion Relations for Feynman Graphs}}, {\emph{Helv.
  Phys. Acta} {\bfseries 54} (1982) 364}.

\bibitem{Bauberger:1994hx}
S.~Bauberger and M.~Bohm, \emph{{Simple one-dimensional integral
  representations for two loop selfenergies: The Master diagram}},
  \href{https://doi.org/10.1016/0550-3213(95)00199-3}{\emph{Nucl. Phys.}
  {\bfseries B445} (1995) 25}
  [\href{https://arxiv.org/abs/hep-ph/9501201}{{\ttfamily hep-ph/9501201}}].

\bibitem{Remiddi:2016gno}
E.~Remiddi and L.~Tancredi, \emph{{Differential equations and dispersion
  relations for Feynman amplitudes. The two-loop massive sunrise and the kite
  integral}},
  \href{https://doi.org/10.1016/j.nuclphysb.2016.04.013}{\emph{Nucl. Phys.}
  {\bfseries B907} (2016) 400}
  [\href{https://arxiv.org/abs/1602.01481}{{\ttfamily 1602.01481}}].

\bibitem{Abreu:2014cla}
S.~Abreu, R.~Britto, C.~Duhr and E.~Gardi, \emph{{From multiple unitarity cuts
  to the coproduct of Feynman integrals}},
  \href{https://doi.org/10.1007/JHEP10(2014)125}{\emph{JHEP} {\bfseries 10}
  (2014) 125} [\href{https://arxiv.org/abs/1401.3546}{{\ttfamily 1401.3546}}].

\bibitem{Caron-Huot:2014lda}
S.~Caron-Huot and J.~M. Henn, \emph{{Iterative structure of finite loop
  integrals}}, \href{https://doi.org/10.1007/JHEP06(2014)114}{\emph{JHEP}
  {\bfseries 06} (2014) 114} [\href{https://arxiv.org/abs/1404.2922}{{\ttfamily
  1404.2922}}].

\bibitem{Cheung:2015cba}
C.~Cheung, C.-H. Shen and J.~Trnka, \emph{{Simple Recursion Relations for
  General Field Theories}},
  \href{https://doi.org/10.1007/JHEP06(2015)118}{\emph{JHEP} {\bfseries 06}
  (2015) 118} [\href{https://arxiv.org/abs/1502.05057}{{\ttfamily
  1502.05057}}].

\bibitem{Cheung:2015ota}
C.~Cheung, K.~Kampf, J.~Novotny, C.-H. Shen and J.~Trnka, \emph{{On-Shell
  Recursion Relations for Effective Field Theories}},
  \href{https://doi.org/10.1103/PhysRevLett.116.041601}{\emph{Phys. Rev. Lett.}
  {\bfseries 116} (2016) 041601}
  [\href{https://arxiv.org/abs/1509.03309}{{\ttfamily 1509.03309}}].

\bibitem{Cheung:2016drk}
C.~Cheung, K.~Kampf, J.~Novotny, C.-H. Shen and J.~Trnka, \emph{{A Periodic
  Table of Effective Field Theories}},
  \href{https://doi.org/10.1007/JHEP02(2017)020}{\emph{JHEP} {\bfseries 02}
  (2017) 020} [\href{https://arxiv.org/abs/1611.03137}{{\ttfamily
  1611.03137}}].

\bibitem{Chicherin:2018ubl}
D.~Chicherin, J.~M. Henn and E.~Sokatchev, \emph{{Amplitudes from
  superconformal Ward identities}},
  \href{https://arxiv.org/abs/1804.03571}{{\ttfamily 1804.03571}}.

\bibitem{Gehrmann:2015bfy}
T.~Gehrmann, J.~M. Henn and N.~A. Lo~Presti, \emph{{Analytic form of the
  two-loop planar five-gluon all-plus-helicity amplitude in QCD}},
  \href{https://doi.org/10.1103/PhysRevLett.116.189903,
  10.1103/PhysRevLett.116.062001}{\emph{Phys. Rev. Lett.} {\bfseries 116}
  (2016) 062001} [\href{https://arxiv.org/abs/1511.05409}{{\ttfamily
  1511.05409}}].

\bibitem{Bourjaily:2013mma}
J.~L. Bourjaily, S.~Caron-Huot and J.~Trnka, \emph{{Dual-Conformal
  Regularization of Infrared Loop Divergences and the Chiral Box Expansion}},
  \href{https://doi.org/10.1007/JHEP01(2015)001}{\emph{JHEP} {\bfseries 01}
  (2015) 001} [\href{https://arxiv.org/abs/1303.4734}{{\ttfamily 1303.4734}}].

\bibitem{CaronHuot:2011kk}
S.~Caron-Huot and S.~He, \emph{{Jumpstarting the All-Loop S-Matrix of Planar
  N=4 Super Yang-Mills}},
  \href{https://doi.org/10.1007/JHEP07(2012)174}{\emph{JHEP} {\bfseries 07}
  (2012) 174} [\href{https://arxiv.org/abs/1112.1060}{{\ttfamily 1112.1060}}].

\bibitem{Panzer:2014caa}
E.~Panzer, \emph{{Algorithms for the symbolic integration of hyperlogarithms
  with applications to Feynman integrals}},
  \href{https://doi.org/10.1016/j.cpc.2014.10.019}{\emph{Comput. Phys. Commun.}
  {\bfseries 188} (2015) 148}
  [\href{https://arxiv.org/abs/1403.3385}{{\ttfamily 1403.3385}}].

\bibitem{Panzer:2015ida}
E.~Panzer, \emph{{Feynman integrals and hyperlogarithms}}, Ph.D. thesis,
  Humboldt U., Berlin, Inst. Math., 2015.
\newblock \href{https://arxiv.org/abs/1506.07243}{{\ttfamily 1506.07243}}.

\end{thebibliography}\endgroup
            \clearpage
            
\end{fmffile}

\end{document}